\documentclass[twocolumn,tighten]{aastex62}
\usepackage{wasysym}

\newcommand{\sion}[2]{#1$\;$\textsc{#2}\relax}
\newcommand{\sub}[2]{\ifmmode #1_\mathrm{\scriptstyle #2} \else $#1_\mathrm{\scriptstyle #2}$\fi}
\newcommand{\ssub}[2]{\ifmmode #1_\mathrm{\scriptscriptstyle #2} \else $#1_\mathrm{\scriptscriptstyle #2}$\fi}
\newcommand{\pme}[2]{$^{+#1}_{-#2}$}
\newcommand{\pe}[1]{$^{+#1}$}
\newcommand{\me}[1]{$_{-#1}$}
\shorttitle{\textit{HST}/COS Observations of Outflows in Five Quasars}
\shortauthors{Miller et al.}

\begin{document}

\title{\textit{HST}/COS Observations of Quasar Outflows in the 500 -- 1050~\AA\ Rest Frame: VII\\Distances and Energetics for 11 Outflows in Five Quasars}
\altaffiliation{Based on observations with the NASA/ESA \\Hubble Space Telescope obtained at the Space Telescope \\Science Institute, which is operated by the Association \\of Universities for Research in Astronomy, Incorporated, \\under NASA contract NAS5-26555.}

\author[0000-0002-0730-2322]{Timothy R. Miller}
\affiliation{Department of Physics, Virginia Polytechnic Institute and State University, Blacksburg, VA 24061, USA}
\author[0000-0003-2991-4618]{Nahum Arav}
\affiliation{Department of Physics, Virginia Polytechnic Institute and State University, Blacksburg, VA 24061, USA}
\author[0000-0002-9217-7051]{Xinfeng Xu}
\affiliation{Department of Physics, Virginia Polytechnic Institute and State University, Blacksburg, VA 24061, USA}
\author[0000-0002-2180-8266]{Gerard A. Kriss}
\affiliation{Space Telescope Science Institute, 3700 San Martin Drive, Baltimore, MD 21218, USA}
\author{Rachel J. Plesha}
\affiliation{Space Telescope Science Institute, 3700 San Martin Drive, Baltimore, MD 21218, USA}



\begin{abstract}
From Hubble Space Telescope/Cosmic Origins Spectrograph spectra of five quasars, 16 outflows are detected. For 11 outflows, we are able to constrain their distances to the central source ($R$) and their energetics. In instances of multiple electron number density determinations (used in the calculation of $R$) for the same outflow, the values are consistent within errors. For the 11 outflows, eight have measurements for $R$ (between 10 and 1000 pc), one has a lower limit, another has an upper limit, and the last has a range in $R$. There are two outflows that have enough kinetic luminosity to be major contributors to active galactic nucleus feedback. The outflowing mass is found primarily in a very high-ionization phase, which is probed using troughs from, e.g., \sion{Ne}{viii}, \sion{Na}{ix}, \sion{Mg}{x}, and \sion{Si}{xii}. Such ions connect the physical conditions of these ultraviolet outflows to the X-ray warm absorber outflows seen in nearby Seyfert galaxies. The ion \sion{Cl}{vii} and several new transitions from \sion{Ne}{v} have been detected for the first time.
\end{abstract}

\section{Introduction}\label{sec:int}
Outflowing material from the centers of quasars can manifest as blueshifted absorption troughs in the rest frame of quasar spectra. Upward of 40\% \cite[][]{hew03,dai08,gan08,kni08} of the quasar population contains absorption outflows. These outflows are candidates for major feedback processes within active galactic nuclei (AGNs) as detailed in section 1 of \citet[][hereafter Paper I]{ara20} and references therein. 

The kinetic luminosity ($\dot{\sub{E}{K}}$) of these outflows is the key quantity needed to assess the potential they have to produce the feedback processes. $\dot{\sub{E}{K}}$ is linearly dependent on the distance from the central source ($R$), which can be inferred by simultaneously determining the electron number density (\sub{n}{e}) and ionization parameter (\sub{U}{H}) of the outflow \cite[see section 7.1 of][]{ara18}. Around 30 such distances are currently found within the literature using this method (e.g., Paper I and references within their section 4.2.3). Accretion disk wind models predict these outflows to reside around $R$ = 0.03 pc \cite[e.g.,][]{mur95,pro00,pro04}, but the range found in the literature is between parsecs to tens of kiloparsecs.

The aforementioned feedback potential is determined by the ratio of the kinetic luminosity with respect to the Eddington luminosity. \cite{hop10} and \cite{sca04} require ratios exceeding 0.5\% and 5\%, respectively, for sufficient feedback. Of the outflow systems, 10 are currently known in the literature to meet at least one of these criteria. \cite[][]{moe09,ara13,bor13,cha15b,xu19,xu20a,xu20c,mil20a}.

The data analyzed here are part of a spectroscopic survey taken during Cycle 24 (GO-14777, PI: N. Arav). There are 10 quasars in total with redshifts around 1. The survey aimed to probe the 500-1050~\AA\ rest-frame wavelength range (EUV500). The goal was to identify and measure diagnostic troughs \cite[examples listed in][]{ara13} capable of yielding outflow properties such as \sub{n}{e} as well as troughs that arise from very high-ionization potential ions (e.g. \sion{Ne}{viii} and \sion{Mg}{x}), like those typically seen in X-ray warm absorbers \cite[e.g.,][]{rey97,kaa00,cre03,kaa14}. Such very high-ionization potential ions could establish a connection between X-ray warm absorbers and ultraviolet (UV) AGN outflows \cite[][]{ara13}. A previous publication \cite[][]{mil18} details the results for the lowest redshift object where its highest-ionization potential ion is \sion{O}{vi}.

This paper is part of a series of publications describing the results
of Hubble Space Telescope (HST) program GO-14777. \\
Paper I summarizes the results
for the individual objects and discusses their importance to various
aspects of quasar outflow research. \\
Paper II \cite[][]{xu20a} gives the full
analysis for four outflows detected in SDSS J1042+1646, including the
largest $\dot{E}_k$ ($10^{47}$~erg~s$^{-1}$) outflow measured to date
at $R=800$~pc, and another outflow at $R=15$~pc. \\
Paper III \cite[][]{mil20a} analyzes four outflows
detected in 2MASS J1051+1247, which show remarkable similarities, are
situated at $R\sim200$~pc, and have a combined $\dot{E}_k=10^{46}$ erg
s$^{-1}$. \\
Paper IV \cite[][]{xu20b} presents the largest
velocity shift and acceleration measured to date in a broad absorption line (BAL) outflow.  \\
Paper V \cite[][]{mil20b} analyzes two outflows
detected in PKS J0352-0711, including one outflow at $R=500$~pc and a second
outflow at $R=10$~pc that shows an ionization potential-dependent
velocity shift for troughs from different ions. \\
Paper VI \cite[][]{xu20c} analyzes two outflows
detected in SDSS J0755+2306, including one at $R=220$~pc with
$\dot{E}_k=10^{46}$~erg~s$^{-1}$. \\
Paper VII is this work.

This paper is structured as follows. Section~\ref{sec:od} describes the observations taken by the HST/Cosmic Origins Spectrograph \cite[COS;][]{gre12} for the remaining five quasars of the survey. The spectral fitting for the unabsorbed continuum and emission lines is also discussed. Section~\ref{sec:da} details the ionic column density and electron number density measurements in addition to the photoionization modeling. Comments on the individual outflows are given in Section~\ref{sec:indobj}. The physical properties, distances, and energetics of each outflow are given in section~\ref{sec:rd}. A discussion of these results is in Section~\ref{sec:ds} with a summary and conclusions in section~\ref{sec:sc}. In this paper, we adopt a cosmology with $h = 0.696$, $\Omega_m = 0.286$, and $\Omega_\Lambda = 0.714$, and we use Ned Wright's Javascript Cosmology Calculator website \cite[][]{wri06}.

\section{Observations, Data Reduction, and Spectral Fitting}
\label{sec:od}
Table~\ref{tab:obs} contains the details of each observation as well as the coordinates, redshifts, and \textit{E}(\textit{B-V}) values for the five quasars: SDSS J093602.10+200542.9, 7C 163119.39+393037.00, UM 425, 2MASS J14362129+0727208, and VV2006 J132957.2+540506. \cite{mil18} details the data processing steps, and each spectrum was corrected for Galactic extinction with the corresponding \textit{E}(\textit{B-V}) values \cite[]{sch11}. Figure~\ref{fig:spectrum} shows the dereddened, one-dimensional spectrum in black with errors in gray for each quasar. Individual outflow systems and corresponding absorption troughs are delineated S1, S2, S3, etc., in order of increasing absolute centroid velocities as summarized in Table~\ref{tab:out}, which also contains the widths of the widest trough for each system. The slanted, dark green lines mark intervening hydrogen absorption systems identified by at least two hydrogen Lyman line transitions. The trough labels combine multiple transitions with wavelength separations less than 0.5~\AA\ into a single transition. Table 3 of Paper II provides a list of transition atomic data. 

\begin{deluxetable*}{lccccc}[h!]
	\tablecaption{HST/COS observations, coordinates, redshifts, and Galactic \textit{E}(\textit{B-V}) values for the five quasars.\label{tab:obs}}
	\tablewidth{0pt}
	\tabletypesize{\footnotesize}
	\tablehead{
		\colhead{Object} & \colhead{SDSS J0936+2005} & \colhead{UM 425} & \colhead{VV2006 J1329+5405} & \colhead{2MASS J1436+0727} & \colhead{7C 1631+3930}
	}
	\startdata
	J2000 R.A. & 09:36:02.11 & 11:23:20.73 & 13:29:57.15 & 14:36:21.30 & 16:33:02.10 \\
	J2000 Decl. & +20:05:42.90 & +01:37:47.50 & +54:05:05.90 & +07:27:20.89 & +39:24:27.4 \\
	Redshift (\textit{z}) & 1.1832 & 1.4720 & 0.9496 & 0.8944 & 1.0246 \\
	Galactic \textit{E}(\textit{B-V}) & 0.0267 & 0.0313 & 0.0128 & 0.0244 & 0.0070 \\
	\tableline
	& \multicolumn{5}{c}{HST/COS Grating G130M} \\
	Observation date & 2017 Nov 21 & 2017 Nov 15 & 2017 Sep 30 & 2017 Jul 17 & 2017 May 13 \\
	Exposure time (s) & 4360 & 3870 & 3690 & 4130 & 4200 \\
	Observed range (\AA) & 1155--1460 & 1130--1470 & 1130--1470 & 1130--1470 & 1130--1470 \\
	Rest-frame range (\AA) & 520--670 & 460--595 & 580--755 & 600--775 & 560--725 \\
	\tableline
	& \multicolumn{5}{c}{HST/COS Grating G160M} \\
	Observation date & 2017 Nov 22 & 2017 Nov 15--16 & 2017 Sep 30 & 2017 Jul 17 & 2017 May 14 \\
	Exposure time (s) & 4660 & 4660 & 4660 & 4660 & 5200 \\
	Observed range (\AA) & 1405--1800 & 1405--1800 & 1405--1800 & 1405--1800 & 1380--1765 \\
	Rest-frame range (\AA) & 645--825 & 570--730 & 720--920 & 740--950 & 695--890 \\
	\enddata
\end{deluxetable*}

\begin{deluxetable}{ccc}
	\tablecaption{Detected Outflows in each quasar.\label{tab:out}}
	\tablewidth{0pt}
	\tabletypesize{\footnotesize}
	\tablehead{
		\colhead{Outflow System} & \colhead{Centroid Velocity } & \colhead{FWHM}\\ 
		& \colhead{(km~s$^{-1}$)} & \colhead{(km~s$^{-1}$)} 
	}
	\startdata
	\multicolumn{3}{c}{SDSS J0936+2005}\\
	\tableline
	S1 & --7960 & 200\\
	S2 & --8200 & 400\\
	S3 & --9300 & 350\\
	\tableline
	\multicolumn{3}{c}{UM 425}\\
	\tableline
	S1 & --1640 & 400\\
	S2 & --1980 & 200\\
	S3 & --2200 & 200\\
	S4 & --9420 & 300\\
	\tableline
	\multicolumn{3}{c}{VV2006 J1329+5405}\\
	\tableline
	S1 & --11600 & 1450\\
	S2 & --12900 & 600\\
	\tableline
	\multicolumn{3}{c}{2MASS J1436+0727}\\
	\tableline
	S1 & --14400 & 300\\
	\tableline
	\multicolumn{3}{c}{7C 1631+3930}\\
	\tableline
	S1 & --1010 & 200\\
	S2 & --1430 & 550\\
	S3 & --5300 & 150\\
	S4 & --5770 & 250\\
	S5 & --6150 & 150\\
	S6 & --7210 & 200\\
	\enddata
\end{deluxetable}

We fit the unabsorbed line and continuum emission in the same way as in \cite{mil18} and \cite{xu18}. Specifically, a power law was used to fit the continuum emission while Gaussian profiles were used to fit line emission features, constrained by the red side of each line. Each emission line had the Gaussian centroid fixed at the rest-frame wavelength. For regions where a partial Lyman limit exists, the continuum was manually fitted with a cubic spline blueward of each limit, accounting for the decreased flux. The adopted, unabsorbed emission model is shown as the solid red contour in Figure~\ref{fig:spectrum} for each quasar.

\section{Data Analysis}
\label{sec:da}
\subsection{Ionic Column Density}
\label{sec:cd}
From the analysis steps in Paper II, we first followed step 1 of the Synthetic Spectral Simulation (SSS) method to determine which troughs yield reliable ionic column density measurements (\sub{N}{ion}). The two methods used to obtain each measurement are the apparent optical depth (AOD) and partial covering (PC) methods (see e.g., \cite{sav91} and \cite{ara08} for each method, respectively). The AOD method is used to measure \sub{N}{ion} from a single ionic transition, typically yielding lower or upper limits. The PC method uses two ionic transitions from the same ion and energy level to yield an \sub{N}{ion} measurement for one ionic energy state. 

The total column density of each ion (listed in Table~\ref{tab:col} for each outflow) is the sum of all ionic energy states. In Table~\ref{tab:col}, red denotes upper limits while blue marks lower limits. The ratio of the adopted column densities to the best-fit, model predicted column densities are in the last column (see Section~\ref{sec:photmod} and Figure~\ref{fig:sol}). Ratios greater than one are expected for upper limits and vice versa for lower limits. 

The same criteria from Paper II are used to determine measurements, upper limits, and lower limits. All PC determined \sub{N}{ion} are treated as measurements, and \sub{N}{ion} for troughs where both 0.05 $<$ $\tau_{max}$ $<$ 0.5 and where troughs from ions of similar ionization potential have $\tau_{max}$ $>$ 2 are also treated as measurements. \sub{N}{ion} measured from regions where no trough is identified (a maximum optical depth, $\tau_{max}$, less than 0.05) are upper limits.  Our adopted values are the PC values when available and AOD values otherwise. The adopted errors include a systematic error (20\% of the adopted value), which is added in quadrature to the corresponding AOD/PC errors (see Table~\ref{tab:col}). This systematic error accounts for uncertainties in the unabsorbed emission model \cite[e.g.,][]{mil18,xu18}. 

\subsection{Photoionization Modeling}
\label{sec:photmod}
The ionization equilibrium of a quasar outflow is dominated by photoionization. Therefore, the outflow is characterized by its total hydrogen column density (\sub{N}{H}) and \sub{U}{H}. These two values together are the photoionization solution for each outflow, where the best-fit photoionization solution is determined from grids of photoionization models generated by the code Cloudy \cite[][version c17.00]{fer17}. Each component of an outflow system is modeled with one \sub{N}{H} and \sub{U}{H}. The grid containing the best-fit solution assumes one of two metallicities and the spectral energy distribution (SED) HE0238 SED \cite[][]{ara13}, which has a similar rest-frame wavelength range to our objects. The two metallicities are solar, Z$_{\astrosun}$, from \citet[][]{gre10} and super-solar, Z = 4.68Z$_{\astrosun}$, from Paper V. Only the solutions for VV2006 J1329+5405 used the super-solar value (see section~\ref{obj1329}).

To determine the best-fit solution, we follow the remaining steps of the SSS method as outlined here. We obtain preliminary values for \sub{N}{H} and \sub{U}{H} from the measured \sub{N}{ion} that are known to be uncontaminated (a subset of the values in Table~\ref{tab:col}). This preliminary photoionization solution and subsequent solutions are found through $\chi^2$-minimization, corresponding to the minimum of the merit function given in equation 4 of \cite{bor12}. Then a model spectra (see section 3.3 of Paper II for construction details) containing all outflows for each quasar is created and compared to the observed spectra. New upper/lower limits and measurements are assessed and used to further constrain the values of \sub{N}{H} and \sub{U}{H}. A best-fit photoionization solution is generated after a few iterations of this process. Figure~\ref{fig:sol} shows these best-fit photoionization solutions for each outflow, which were constrained by all total \sub{N}{ion} values in Table~\ref{tab:col}. 

The colored contours for individual ions in Figure~\ref{fig:sol} show the \sub{N}{H} and \sub{U}{H} values where the model \sub{N}{ion} are within 1$\sigma$ of the observed values, assuming the HE0238 SED and solar metallicity (super-solar metallicity for VV2006 1329+5405). Solid contours represent \sub{N}{ion} measurements while dotted and dashed lines indicate upper and lower limits, respectively. The adopted best-fit solution is the solid black dots and 1$\sigma$ error ellipses. A two phase photoionization solution \cite[e.g.,][Paper II; Paper III; Paper V]{ara13}, comprised of a high-ionization phase (HP) and very high-ionization phase (VHP), is needed for 10 of the 16 outflow systems. The remaining six require only a single phase solution (one VHP). The values for all \sub{N}{H} and \sub{U}{H} determinations are given in Table~\ref{tab:res}. Comments on individual objects can be found in section~\ref{sec:indobj}.



\subsection{Electron Number Density Determination}
\label{sec:ed}
All of the excited-state troughs shown in Figure~\ref{fig:spectrum} become populated primarily through electron collisions, where the collision frequency and energy transferred depend on \sub{n}{e} and the electron temperature. Therefore, the relative populations between an excited-state and a ground state from the same ion can be used to calculate \sub{n}{e} \cite[e.g.,][]{dek01,ham01,dek02,kor08}. To calculate the necessary population ratios, which is equal to the column density ratios, we used the CHIANTI 8.0.7 database \cite[]{der97,lan13} as was done in previous works \cite[e.g.,][]{bor12b,ara13,ara15,ara18,cha15b}. Separate discussions follow for each quasar, with the results shown in Figure~\ref{fig:dens} where the theoretical column density ratios as a function of \sub{n}{e} are shown with the black contours for each population ratio at the temperature of the first outflow listed in each panel. The top axis shows the distance scale obtained from equation~\ref{eq:R} for this outflow as well. The measured column density ratios with uncertainties for each outflow system are overlaid, and the individual column densities are listed in Table~\ref{tab:col}.

\section{Comments on Individual Outflows}\label{sec:indobj}
\subsection{SDSS J0936+2005}\label{obj0936}
\textit{S1}: The photoionization solution consists of two phases. A single phase solution at the intersection of \sion{Ar}{viii} and \sion{Mg}{x} overpredicts the \sub{N}{ion} of \sion{Ne}{viii} by a factor of 10, requiring this two phase solution. The VHP is constrained primarily by the \sub{N}{ion} measurements of \sion{Ne}{viii} and \sion{Mg}{x} as well as an upper limit on \sion{Al}{xi} while the HP has only upper and lower limit constraints. The VHP contains four times more hydrogen column density than the HP. 

The only detected troughs with useful \sub{n}{e} diagnostics for S1 are from \sion{Ne}{v} and \sion{Ne}{vi}. The \sion{Ne}{v*} 569.83~\AA\ trough is deeper than the \sion{Ne}{v} 568.42~\AA\ trough. Conversely, the \sion{Ne}{vi*} 562.81~\AA\ trough is shallower than the \sion{Ne}{vi} 558.60~\AA\ trough. These facts already predict that the \sub{n}{e} of S1 is above the critical density for the energy level associated with the \sion{Ne}{v*} 569.83~\AA\ transition (413~cm$^{-1}$) and below that of the \sion{Ne}{vi*} 562.81~\AA\ transition (1307~cm$^{-1}$). 

The majority of the \sub{N}{ion} for \sion{Ne}{v} and \sion{Ne}{vi} come from the HP. Therefore, the total \sub{N}{ion} of each can be constrained from the HP solution in the same way as was done in section 3.3 of Paper III. Specifically, the deeper trough is assumed to contain the missing column density needed for the total to match the value constrained by the photoionization solution. Paper III showed that the shallower trough \sub{N}{ion} does not increase significantly when considering maximum saturation effects, making this assumption acceptable. From there, a measured ratio from the \sub{N}{ion} of \sion{Ne}{vi*} and \sion{Ne}{vi} is obtained. We are left with a lower limit for \sion{Ne}{v}. This happens because there is another excited energy level of \sion{Ne}{v} (1111~cm$^{-1}$) that would also be populated, but the transitions needed to probe the level are blended with the geocoronal Ly$\alpha$ emission line. Since we do not know the relative trough depths of the excited levels, we cannot determine if the observed \sion{Ne}{v*} 569.83~\AA\ trough is the deepest. Therefore, the adopted value for \sub{n}{e} is from the \sion{Ne}{vi} diagnostic, which is also consistent within errors with the lower limit determined from the \sion{Ne}{v} diagnostic. 

\textit{S2}: A two phase photoionization solution is needed since a single phase solution at the intersection of \sion{Ca}{vi} and \sion{Al}{xi} overpredicts the column densities of \sion{Ca}{x} (by a factor of 100) and \sion{Ar}{viii} (by a factor of 10). The HP is bounded by the \sub{N}{ion} measurement of \sion{Ar}{viii} and the upper and lower limits of \sion{S}{iv} and \sion{Ca}{vi}, respectively. The VHP is constrained by the column density measurements of \sion{Na}{ix} and \sion{Al}{xi} along with the upper limit from \sion{Ca}{x}. The \sub{N}{ion} from \sion{Al}{xi} 550~\AA\ is treated as a measurement even though we measure only a single trough since it is well defined and much shallower than the \sion{Mg}{x} troughs, which have a similar ionization potential to \sion{Al}{xi}. We assume since the ionization potentials are similar that they should be produced in the same regions of the outflow and have the same covering solutions. The VHP carries 90\% of the total hydrogen column density. 

The troughs from the excited transitions of \sion{O}{iv} are either heavily blended or show one-to-one trough depths with ground-state transitions, making them unsuitable for \sub{n}{e} measurements. However, the troughs from excited levels of \sion{Ne}{v} and \sion{Ne}{vi} can yield useful measurements. Like S1, the excited-state trough of \sion{Ne}{v} is deeper than its corresponding ground-state trough while the opposite is true for the \sion{Ne}{vi} troughs. The majority of the \sub{N}{ion} for \sion{Ne}{v} and \sion{Ne}{vi} also comes from the HP. Therefore, the same method used for S1 to determine \sub{n}{e} is used for S2. Again, the adopted value for \sub{n}{e} is from the \sion{Ne}{vi} diagnostic.

\textit{S3}: The two phase solution contains only upper and lower limits where the HP is constrained by \sion{O}{iii}, \sion{N}{iv}, and \sion{Ne}{v} while the VHP is bounded by \sion{Mg}{x}, \sion{Ne}{viii}, and \sion{Ne}{vi}. Choosing a single phase solution where \sion{N}{iv} crosses \sion{Ne}{viii} overpredicts the upper limit of \sion{Ne}{v} by a factor of 15. The VHP contains over 99\% of the total hydrogen column density.

The only useful \sub{n}{e} diagnostic comes from the transitions of \sion{Ne}{vi}, where the majority of the \sub{N}{ion} resides in the VHP. The same process as was used for S1 and S2 yields the \sub{n}{e} measurement for S3.

\subsection{UM 425}\label{obj425}
\textit{S1}: A well constrained two phase photoionization solution is determined from the \sub{N}{ion} constraints. Like S2 of SDSS J0936+2005, the \sub{N}{ion} of \sion{Al}{xi} is treated as a measurement since the 550~\AA\ trough is shallow compared to the \sion{Mg}{X} and \sion{Si}{xii} troughs (all three have similar ionization potentials and therefore have similar assumed origins and covering solutions within the outflow) and the trough is well defined. About 98\% of the total \sub{N}{H} is contained within the VHP.

The \sion{O}{iv} and \sion{O}{iv*} transitions are quite blended, preventing usable \sub{n}{e} estimates. The \sion{S}{iv} and \sion{S}{iv*} troughs are also only upper limits. This leaves the transitions from \sion{Ne}{v} and \sion{Ne}{vi}. The \sion{Ne}{v*} 482.99~\AA\ trough is not well defined, and the \sion{Ne}{v*} 481.35~\AA\ is blended with an intervening system and other absorption, leading to unreliable \sub{N}{ion} measurements for both troughs. However, An upper limit and measurement for \sub{n}{e} is obtained from the \sion{Ne}{v*} 572.30~\AA\ and \sion{Ne}{v*} 569.83~\AA\ \sub{N}{ion}, respectively, along with the ground-state column density of \sion{Ne}{v}, which is constrained by the HP since it carries the majority of the \sub{N}{ion} for \sion{Ne}{v}. For the \sion{Ne}{vi} troughs, the \sion{Ne}{vi*} 562.81~\AA\ trough is severely blending with an intervening system, preventing an \sub{n}{e} estimate. Therefore, the \sub{n}{e} measurement using the \sion{Ne}{v*} 569.83~\AA\ diagnostic is adopted as the best value.

\textit{S2}: There is a two phase solution with an unbounded VHP. Unlike S1, the \sion{Al}{xi} 550~\AA\ trough is not well defined so we do not take it as a measurement nor a lower limit since it is uncertain if the absorption arises from intervening systems. The VHP carries at least 93\% of the total \sub{N}{H}.

Like S1, the \sion{O}{iv} and \sion{O}{iv*} transitions are blended, and the \sion{S}{iv} and \sion{S}{iv*} transitions are upper limits. Therefore, neither species is usable for \sub{n}{e} estimates. The \sion{Ne}{v*} 481.35~\AA\ and 572.30~\AA\ as well as the \sion{Ne}{vi*} 562.81~\AA\ troughs do not yield accurate \sub{N}{ion} as they are blended with intervening systems and other absorption. However, S2 does have two consistent \sub{n}{e} measurements where the ground-state column density is constrained by the HP: one from the \sub{N}{ion} ratio of \sion{Ne}{v*} 569.83~\AA\ and the ground-state and another from the \sub{N}{ion} ratio of \sion{Ne}{v*} 482.99~\AA\ and the ground state. Since the \sion{Ne}{v*} 569.83~\AA\ trough is better defined than the \sion{Ne}{v*} 482.99~\AA\ trough and lies within a portion of the spectrum with less unaccounted for absorption that may lead one to question the identification of the absorption trough, we choose the \sub{n}{e} value determined from the \sion{Ne}{v*} 569.83~\AA\ and ground-state diagnostic as the best value.

\textit{S3}: There is an unbounded VHP and bounded HP. Like in the case of S2, the \sion{Al}{xi} 550~\AA\ trough is unreliable as a measurement or a lower limit. The errors allow for both phases to carry equal amounts of the total hydrogen column density.

The \sion{O}{iv} and \sion{O}{iv*} transitions as well as the \sion{S}{iv} and \sion{S}{iv*} transitions are upper limits and therefore cannot yield \sub{n}{e} estimates. The only excited trough that is not severely blended is the \sion{Ne}{v*} 572.30~\AA\ transition. From the \sub{N}{ion} ratio of \sion{Ne}{v*} 572.30~\AA\ and the ground state (constrained by the HP), we determine an \sub{n}{e} measurement.

\textit{S4}: A two phase solution exists with a bounded HP and VHP. The \sion{Al}{xi} 550\AA\ trough is treated as a lower limit since it has a similar depth to the \sion{Mg}{x} and \sion{Si}{xii} troughs. The VHP contains the majority of the total \sub{N}{H} at 90\%. 

The only useful \sub{n}{e} estimates arise from the \sion{Ne}{v*} 482.99~\AA, \sion{Ne}{v*} 569.83~\AA\ (blended on the blue side with an intervening system), and \sion{Ne}{vi*} 562.81~\AA\ troughs. The first two yield electron number density lower limits since they are deeper than the \sion{Ne}{v} 480.42~\AA\ (blended on the red side with an intervening system), resulting in ratios that are consistent with the theoretical limit. For \sion{Ne}{vi}, the \sion{Ne}{vi*} 562.81~\AA\ trough is blended with unidentified absorption, preventing an \sub{n}{e} estimate. The lower limit obtained from the \sion{Ne}{v*} 482.99~\AA\ diagnostic is adopted as the best value since it provides a tighter constraint compared to the other lower limit. 

\subsection{VV2006 J1329+5405}\label{obj1329}
\textit{S1}: The photoionization solution uses the super-solar value since solar values yield a solution that overpredicts the hydrogen \sub{N}{ion} upper limit by more than a factor of five. The two phase solution has an unbounded VHP, which carries around 90\% of the total \sub{N}{H}.

Two lower limits and one upper limit on \sub{n}{e} are determined for S1. First, the \sion{O}{iv*} 790.11~\AA\ trough is free from major contamination or blending, but can only yield a lower limit since it shows a one to one trough depth with the uncontaminated red side of the \sion{O}{iv} 787.71~\AA\ and blue side of the \sion{O}{iv} 608.40~\AA\ troughs. Therefore, using the HP to constrain the total \sub{N}{ion} of \sion{O}{iv}, a lower limit on \sub{n}{e} is obtained. Second, the \sion{O}{v*} multiplet near 730~\AA\ in the quasar rest frame is quite blended, but the red half of the \sion{O}{v*} 762.00~\AA\ trough is not blended. However, the trough is likely saturated as it shows a one to one trough depth with the \sion{O}{v*} 760.45~\AA\ trough, whose transition is from the same energy level and has a larger oscillator strength. Therefore, a lower limit to the \sub{N}{ion} of \sion{O}{v*} 762.00~\AA\ trough is estimated by doubling the \sub{N}{ion} measured from the red half of the trough. Using the HP solution to constrain the total \sub{N}{ion} of \sion{O}{v} along with this lower limit yields the other lower limit to \sub{n}{e}. Third, an upper limit to the \sion{N}{iv*} 923.22~\AA\ trough is measured, and the total \sub{N}{ion} for \sion{N}{iv} is constrained by the HP. Together, an \sub{n}{e} upper limit is calculated. The lower limit \sub{n}{e} determined from the \sion{O}{v} diagnostic and upper limit from the \sion{N}{iv} diagnostic are adopted as the best range since they provide the tightest constraints on \sub{n}{e}.

\textit{S2}: For consistency, S2 also used the super-solar value, but a solar metallicity solution is equally viable. It contains a single VHP with the possibility of an HP if its log(\sub{U}{H}) is less than 0 and total hydrogen column density is below the \sion{O}{iv} upper limit.

The \sion{O}{v*} multiplet is highly blended with intervening systems. Similarly, the \sion{O}{iv*} 790.11~\AA\ trough is blended with troughs from S1. Therefore, \sub{n}{e} cannot be constrained, leaving the distance unknown.

\subsection{2MASS J1436+0727}\label{obj1436}
\textit{S1}: The only detected trough is \sion{Ne}{viii}, yielding a bounded VHP solution with the aide of upper limit constraints from \sion{H}{i} and \sion{O}{v}. An HP solution could exist for \sub{N}{H} values below the \sion{O}{v} upper limit and \sub{U}{H} smaller than the VHP. The outflow has no \sub{N}{ion} lower limits or measurements from ground-state transitions with possible excited levels nor any excited level transitions, rendering an \sub{n}{e} estimate indeterminable. 

\subsection{7C 1631+3930}\label{obj1631}
\textit{S1}: A single phase (VHP) is detected. An HP could exist following similar requirements of previously discussed single phase outflows.

Only upper limits to excited-state \sub{N}{ion} are measured, which at best will result in \sub{n}{e} upper limits. These \sub{N}{ion} upper limits are typically accompanied by upper limits on the ground-state \sub{N}{ion} as well, yielding no constraint on \sub{n}{e}. However, the VHP solution contains a large amount of \sion{Ne}{vi} that would be detected if the spectrum covered the \sion{Ne}{vi} 558.60~\AA\ trough. Therefore, an upper limit to \sub{n}{e} is determined from the upper limit \sub{N}{ion} of the \sion{Ne}{vi*} 562.81~\AA\ and the ground-state \sub{N}{ion} of \sion{Ne}{vi} (determined by subtracting the upper limit from the total column density of \sion{Ne}{vi} constrained by the VHP). 

\textit{S2}: A VHP and an HP are needed following similar arguments as was shown for the outflows in SDSS J0936+2005. Like S3 of SDSS J0936+2005, the overwhelming majority (98\%) of the hydrogen column density is contained within the VHP.

For \sub{n}{e} determinations, S2 has an upper limit \sub{N}{ion} for the \sion{Ne}{v*} 572.30~\AA\ trough and a measurement for the \sion{Ne}{v*} 569.83~\AA\ trough since this trough is much shallower than the \sion{Ne}{v} 568.42~\AA\ trough. The HP also contains the majority of the \sion{Ne}{v} column density, yielding an \sub{n}{e} upper limit from the \sion{Ne}{v*} 572.30~\AA\ and ground-state column density ratio and a measurement from the \sub{N}{ion} ratio of \sion{Ne}{v*} 569.83~\AA\ and the ground state. The adopted \sub{n}{e} is the value determined by the \sion{Ne}{v*} 569.83~\AA\ and ground-state diagnostic.

\textit{S3}: Similar to S1, only a VHP is needed to explain the detected troughs, and an HP could exist following similar requirements as discussed previously. The VHP solution is unbounded, allowing solutions beyond log(\sub{U}{H}) = 2 and log(\sub{N}{H}) = 23.5. There are not any useful \sub{n}{e} diagnostics for this outflow.

\textit{S4}: Like S2, a two phase solution explains the observed column density constraints, and the VHP contains 96\% of the total \sub{N}{H}. Similar to S1, an \sub{n}{e} measurement is obtained from the \sub{N}{ion} ratio of the observed \sion{Ne}{v*} 572.30~\AA\ trough and the total \sub{N}{ion} constrained by the HP (the ground-state trough is outside the observed wavelength range like in the case of S1). 

\textit{S5} and \textit{S6}: A single VHP is detected in both outflows with possible HPs existing below each \sion{N}{iv} upper limit and log(\sub{U}{H}) $<$ 0. Neither outflow has useful \sub{n}{e} diagnostics, leaving their distances unconstrained.

\section{Results}
\label{sec:rd}
\subsection{Outflow Distance, Energetics, and Properties}
The ionization parameter relates the distance of each outflow from the central source to other measurable properties: 
\begin{equation}
\label{eq:R}
\sub{U}{H} = \frac{\sub{Q}{H}}{4\pi R^2\ssub{n}{H} c}
\end{equation}
where \ssub{n}{H}~is the hydrogen number density ($\sub{n}{e} \approx 1.2\ssub{n}{H}$ for highly ionized plasma), $R$ is the distance from the central source, $c$ is the speed of light, and $\sub{Q}{H}$ is the incident ionizing photon rate of hydrogen. Taking the HE0238 SED and integrating for energies above 1 Ryd yields the $\sub{Q}{H}$ values in Table~\ref{tab:res}, which also contains the distances of each outflow. This integration process includes the methodology of \cite{hog99} for calculating the luminosity distance and applying the \textit{k}-correction to the observed flux. 


From \cite{bor12}, the average mass flow rate and kinetic luminosity over a dynamical timescale (\textit{R/v}) for a partially filled, thin shell outflow are given by
\begin{equation}
\label{eq:M}
\dot{M}\simeq 4\pi \Omega R N_H \mu m_p v 
\end{equation}
and
\begin{equation}
\label{eq:E}
\dot{\sub{E}{K}}\simeq \frac{1}{2} \dot{M} v^2
\end{equation}
where $\Omega$ = 0.4\pme{0.14}{0.14} is the global covering factor \cite[the fraction of quasars with observed \sion{Ne}{viii} mini-BAL outflows;][]{muz13}, $R$ is the distance from the central source, $\mu = 1.4$ is the mean atomic mass per proton, \sub{N}{H} is the hydrogen column density, $m_p$ is the proton mass, and $v$ is the outflow velocity. Table~\ref{tab:res} contains the calculated energetics for each outflow assuming the HP and VHP have the same global covering factor, and the hydrogen column density value used for the energetics is the sum of the \sub{N}{H} from each phase.

\startlongtable
\begin{deluxetable}{ccccc}
	\tablecaption{Total Ionic Column Densities\label{tab:col}}
	\tabletypesize{\footnotesize}
	\tablehead{
		\colhead{Ion} & \colhead{AOD\tablenotemark{a}} & \colhead{PC\tablenotemark{a}} & \colhead{Adopted\tablenotemark{b}} & \colhead{\scriptsize$\frac{\textnormal{Adopted}}{\textnormal{Best Model}}$\tablenotemark{c}} \\ 
		& \colhead{($10^{12} $cm$^{-2}$)} & \colhead{($10^{12} $cm$^{-2}$)} & \colhead{($10^{12} $cm$^{-2}$)} & 
	}
	\startdata
	\multicolumn{5}{c}{\textbf{SDSS J0936+2005}}\\
	\tableline
	\multicolumn{5}{c}{S1: v = -7960~km~s$^{-1}$}\\
	\tableline
	\sion{N}{iv} & 48\pme{12}{10} & \nodata & \color{blue}\textgreater48\me{14} & \textgreater0.99\me{0.29}\\
	\sion{O}{iv} & 250\pme{70}{60} & \nodata & \color{red}\textless250\pe{80} & \textless0.99\pe{0.34}\\
	\sion{O}{v} & 220\pme{20}{10} & \nodata & \color{blue}\textgreater220\me{50} & \textgreater0.06\me{0.01}\\
	\sion{Ne}{v} & 550\pme{80}{70} & \nodata & \color{blue}\textgreater550\me{130} & \textgreater0.67\me{0.16}\\
	\sion{Ne}{v*}\tablenotemark{d} & 390\pme{70}{50} & \nodata & \color{blue}\textgreater390\me{100} & \nodata\\
	\sion{Ne}{v}\tablenotemark{e} & 160\pme{50}{40} & \nodata & 160\pme{60}{50} & \nodata\\	
	\sion{Ne}{vi} & 2000\pme{120}{100} & \nodata & \color{blue}\textgreater2000\me{420} & \textgreater0.57\me{0.12}\\
	\sion{Ne}{vi*}\tablenotemark{f} & 730\pme{70}{70} & \nodata & 730\pme{160}{160} & \nodata\\
	\sion{Ne}{vi}\tablenotemark{e} & 1290\pme{90}{80} & \nodata & 6690\pme{3810}{2040} & \nodata\\
	\sion{Ne}{viii} & 1900\pme{240}{190} & 2200\pme{470}{230} & 2200\pme{650}{510} & 1.00\pme{0.24}{0.22}\\
	\sion{Na}{ix} & 140\pme{30}{30} & \nodata & \color{red}\textless140\pe{50} & \textless2.32\pe{0.73}\\
	\sion{Mg}{x} & 1500\pme{140}{130} & 2400\pme{440}{240} & 2400\pme{650}{530} & 1.00\pme{0.27}{0.22}\\
	\sion{Al}{xi} & 190\pme{60}{70} & \nodata & \color{red}\textless190\pe{70} & \textless1.04\pe{0.40}\\
	\sion{S}{iv} & 10\pme{2}{3} & \nodata & \color{red}\textless10\pe{3} & \textless84.2\pe{25.2}\\
	\sion{Ar}{vi} & 43\pme{15}{11} & \nodata & \color{red}\textless43\pe{17} & \textless3.53\pe{1.39}\\
	\sion{Ar}{vii} & 16\pme{4}{3} & \nodata & \color{red}\textless16\pe{5} & \textless1.66\pe{0.52}\\
	\sion{Ar}{viii} & 25\pme{11}{11} & \nodata & \color{red}\textless25\pe{12} & \textless1.08\pe{0.52}\\
	\sion{Ca}{vi} & 110\pme{70}{50} & \nodata & \color{red}\textless110\pe{80} & \textless3.87\pe{2.78}\\
	\sion{Ca}{viii} & 160\pme{70}{50} & \nodata & \color{red}\textless160\pe{80} & \textless4.15\pe{1.89}\\
	\sion{Ca}{x} & 74\pme{3}{34} & \nodata & \color{red}\textless74\pe{15} & \textless6.14\pe{1.24}\\
	\tableline
	\multicolumn{5}{c}{S2: v = -8200~km~s$^{-1}$}\\
	\tableline
	\sion{N}{iv} & 200\pme{20}{20} & \nodata & \color{blue}\textgreater200\me{40} & \textgreater0.09\me{0.02}\\
	\sion{O}{iv} & 1500\pme{140}{120} & \nodata & \color{blue}\textgreater1500\me{320} & \textgreater0.09\me{0.02}\\
	\sion{O}{v} & 710\pme{40}{30} & \nodata & \color{blue}\textgreater710\me{150} & \textgreater0.01\me{0.002}\\
	\sion{Ne}{v} & 3200\pme{160}{130} & \nodata & \color{blue}\textgreater3200\me{650} & \textgreater0.15\me{0.03}\\
	\sion{Ne}{v*}\tablenotemark{d} & 1960\pme{120}{100} & \nodata & \color{blue}\textgreater1960\me{410} & \nodata\\
	\sion{Ne}{v}\tablenotemark{e} & 1240\pme{100}{80} & \nodata & 1240\pme{270}{260} & \nodata\\	
	\sion{Ne}{vi} & 5900\pme{250}{200} & \nodata & \color{blue}\textgreater5900\me{1200} & \textgreater0.15\me{0.03}\\
	\sion{Ne}{vi*}\tablenotemark{f} & 2420\pme{130}{110} & \nodata & 2420\pme{500}{500} & \nodata\\
	\sion{Ne}{vi}\tablenotemark{e} & 3450\pme{220}{160} & \nodata & 29200\pme{16200}{11700} & \nodata\\
	\sion{Ne}{viii} & 6800\pme{570}{400} & \nodata & \color{blue}\textgreater6800\me{1400} & \textgreater0.12\me{0.02}\\
	\sion{Na}{ix} & 1700\pme{160}{140} & 1900\pme{160}{110} & 1900\pme{420}{400} & 1.46\pme{0.32}{0.30}\\
	\sion{Mg}{x} & 6200\pme{290}{250} & \nodata & \color{blue}\textgreater6200\me{1300} & \textgreater0.26\me{0.05}\\
	\sion{Al}{xi} & 530\pme{100}{90} & \nodata & 530\pme{140}{140} & 0.77\pme{0.21}{0.20}\\
	\sion{S}{iv} & 23\pme{4}{4} & \nodata & \color{red}\textless23\pe{6} & \textless0.98\pe{0.25}\\
	\sion{Ar}{vi} & 91\pme{20}{17} & \nodata & \color{blue}\textgreater91\me{25} & \textgreater0.17\me{0.05}\\
	\sion{Ar}{vii} & 88\pme{6}{5} & \nodata & \color{blue}\textgreater88\me{18} & \textgreater0.53\me{0.11}\\
	\sion{Ar}{viii} & 210\pme{30}{30} & 240\pme{40}{30} & 240\pme{60}{60} & 0.93\pme{0.24}{0.23}\\
	\sion{Ca}{vi} & 680\pme{90}{90} & \nodata & \color{blue}\textgreater680\me{160} & \textgreater1.18\me{0.28}\\
	\sion{Ca}{vii} & 1300\pme{190}{170} & \nodata & \color{red}\textless1300\pe{320} & \textless1.88\pe{0.46}\\
	\sion{Ca}{viii} & 210\pme{50}{50} & \nodata & \color{red}\textless210\pe{70} & \textless1.16\pe{0.39}\\
	\sion{Ca}{x} & 61\pme{39}{27} & \nodata & \color{red}\textless61\pe{41} & \textless0.48\pe{0.32}\\
	\tableline
	\multicolumn{5}{c}{S3: v = -9300~km~s$^{-1}$}\\
	\tableline
	\sion{N}{iv} & 93\pme{16}{14} & \nodata & \color{blue}\textgreater93\me{23} & \textgreater1.08\me{0.27}\\
	\sion{O}{iii} & 110\pme{30}{30} & \nodata & \color{red}\textless110\pe{40} & \textless0.96\pe{0.34}\\
	\sion{O}{v} & 300\pme{20}{20} & \nodata & \color{blue}\textgreater300\me{60} & \textgreater0.37\me{0.08}\\
	\sion{Ne}{v} & 120\pme{60}{40} & \nodata & \color{red}\textless120\pe{60} & \textless0.84\pe{0.43}\\
	\sion{Ne}{vi} & 1300\pme{120}{110} & \nodata & \color{blue}\textgreater1300\me{280} & \textgreater1.06\me{0.23}\\
	\sion{Ne}{vi*}\tablenotemark{f} & 530\pme{70}{70} & \nodata & 530\pme{130}{130} & \nodata\\
	\sion{Ne}{vi}\tablenotemark{e} & 790\pme{90}{80} & \nodata & 790\pme{870}{180} & \nodata\\
	\sion{Ne}{viii} & 2300\pme{300}{240} & \nodata & \color{blue}\textgreater2300\me{530} & \textgreater0.95\me{0.21}\\
	\sion{Na}{ix} & 48\pme{23}{22} & \nodata & \color{red}\textless48\pe{25} & \textless1.93\pe{1.01}\\
	\sion{Mg}{x} & 350\pme{100}{90} & \nodata & \color{red}\textless350\pe{120} & \textless0.99\pe{0.34}\\
	\sion{Al}{xi} & 160\pme{60}{70} & \nodata & \color{red}\textless160\pe{70} & \textless34.3\pe{15.5}\\
	\sion{S}{iv} & 24\pme{3.6}{3.6} & \nodata & \color{red}\textless24\pe{6.0} & \textless1.79\pe{0.45}\\
	\sion{Ar}{vi} & 46\pme{19}{15} & \nodata & \color{red}\textless46\pe{21} & \textless67.9\pe{31.0}\\
	\sion{Ar}{vii} & 9.2\pme{3.7}{2.9} & \nodata & \color{red}\textless9.2\pe{4.1} & \textless10.0\pe{4.47}\\
	\sion{Ar}{viii} & 25\pme{9.0}{10} & \nodata & \color{red}\textless25\pe{10} & \textless8.62\pe{3.55}\\
	\tableline
	\multicolumn{5}{c}{\textbf{UM 425}}\\
	\tableline
	\multicolumn{5}{c}{S1: v = -1640~km~s$^{-1}$}\\
	\tableline
	\sion{O}{v} & 1300\pme{71}{44} & \nodata & \color{blue}\textgreater1300\me{260} & \textgreater0.09\me{0.02}\\
	\sion{Ne}{iv} & 140\pme{26}{24} & \nodata & \color{red}\textless140\pe{38} & \textless0.99\pe{0.27}\\
	\sion{Ne}{v} & 3500\pme{89}{81} & \nodata & \color{blue}\textgreater3500\me{710} & \textgreater1.03\me{0.21}\\
	\sion{Ne}{v*}\tablenotemark{d} & 350\pme{30}{30} & \nodata & 350\pme{80}{80} & \nodata\\
	\sion{Ne}{v*}\tablenotemark{g} & 80\pme{20}{20} & \nodata & \color{red}\textless80\pe{30} & \nodata\\
	\sion{Ne}{v}\tablenotemark{e} & 3180\pme{80}{70} & \nodata & 3180\pme{710}{710} & \nodata\\	
	\sion{Ne}{vi} & 7300\pme{420}{270} & \nodata & \color{blue}\textgreater7300\me{1500} & \textgreater0.74\me{0.15}\\
	\sion{Ne}{vii} & 12000\pme{0}{0} & \nodata & \color{red}\textless12000\pe{4000} & \textless0.94\pe{0.31}\\
	\sion{Na}{ix} & 2400\pme{260}{210} & 3900\pme{1400}{600} & 3900\pme{1600}{980} & 1.12\pme{0.46}{0.28}\\
	\sion{Mg}{x} & 18000\pme{1000}{670} & \nodata & \color{blue}\textgreater18000\me{3600} & \textgreater0.12\me{0.02}\\
	\sion{Al}{xi} & 1290\pme{100}{100} & \nodata & 1290\pme{280}{280} & 1.00\pme{0.22}{0.22}\\
	\sion{Si}{xii} & 7300\pme{270}{220} & \nodata & \color{blue}\textgreater7300\me{1500} & \textgreater0.05\me{0.01}\\
	\sion{S}{iv} & 23\pme{19}{14} & \nodata & \color{red}\textless23\pe{20} & \textless21.4\pe{18.2}\\
	\sion{Ca}{viii} & 100\pme{45}{32} & \nodata & \color{red}\textless100\pe{49} & \textless1.47\pe{0.66}\\
	\tableline
	\multicolumn{5}{c}{S2: v = -1980~km~s$^{-1}$}\\
	\tableline
	\sion{O}{v} & 330\pme{20}{16} & \nodata & \color{blue}\textgreater330\me{68} & \textgreater0.04\me{0.01}\\
	\sion{Ne}{iv} & 180\pme{51}{44} & \nodata & \color{red}\textless180\pe{63} & \textless1.31\pe{0.45}\\
	\sion{Ne}{v} & 1900\pme{71}{64} & \nodata & \color{blue}\textgreater1900\me{380} & \textgreater1.01\me{0.20}\\
	\sion{Ne}{v*}\tablenotemark{d} & 500\pme{30}{30} & \nodata & 500\pme{110}{110} & \nodata\\
	\sion{Ne}{v*}\tablenotemark{g} & 270\pme{50}{40} & \nodata & 270\pme{70}{70} & \nodata\\
	\sion{Ne}{v}\tablenotemark{e} & 1080\pme{40}{40} & \nodata & 1080\pme{220}{220} & \nodata\\	
	\sion{Ne}{vi} & 2000\pme{140}{120} & \nodata & \color{blue}\textgreater2000\me{420} & \textgreater0.39\me{0.08}\\
	\sion{Na}{ix} & 280\pme{78}{68} & \nodata & \color{red}\textless280\pe{96} & \textless1.31\pe{0.45}\\
	\sion{Mg}{x} & 3300\pme{190}{170} & \nodata & \color{blue}\textgreater3300\me{680} & \textgreater0.94\me{0.19}\\
	\sion{S}{iv} & 13\pme{3.7}{4.2} & \nodata & \color{red}\textless13\pe{4.5} & \textless5.83\pe{2.02}\\
	\sion{Ar}{v} & 25\pme{6.9}{11} & \nodata & \color{red}\textless25\pe{8.5} & \textless1.02\pe{0.35}\\
	\sion{Ar}{vii} & 14\pme{2.1}{2.0} & \nodata & \color{red}\textless14\pe{3.5} & \textless1.00\pe{0.25}\\
	\sion{Ca}{vii} & 140\pme{66}{71} & \nodata & \color{red}\textless140\pe{71} & \textless2.25\pe{1.16}\\
	\sion{Ca}{x} & 36\pme{13}{13} & \nodata & \color{red}\textless36\pe{15} & \textless1.01\pe{0.41}\\
	\tableline
	\multicolumn{5}{c}{S3: v = -2200~km~s$^{-1}$}\\
	\tableline
	\sion{O}{v} & 360\pme{19}{16} & \nodata & \color{blue}\textgreater360\me{74} & \textgreater0.10\me{0.02}\\
	\sion{Ne}{iv} & 260\pme{66}{57} & \nodata & \color{red}\textless260\pe{84} & \textless6.56\pe{2.09}\\
	\sion{Ne}{v} & 830\pme{51}{48} & \nodata & \color{blue}\textgreater830\me{170} & \textgreater1.00\me{0.21}\\
	\sion{Ne}{v*}\tablenotemark{g} & 230\pme{30}{30} & \nodata & 230\pme{60}{50} & \nodata\\
	\sion{Ne}{v}\tablenotemark{e} & 600\pme{40}{40} & \nodata & 600\pme{130}{130} & \nodata\\	
	\sion{Ne}{vi} & 1900\pme{110}{90} & \nodata & \color{blue}\textgreater1900\me{400} & \textgreater0.78\me{0.16}\\
	\sion{Na}{vi} & 140\pme{63}{48} & \nodata & \color{red}\textless140\pe{69} & \textless3.10\pe{1.58}\\
	\sion{Na}{ix} & 320\pme{78}{70} & \nodata & \color{red}\textless320\pe{100} & \textless1.31\pe{0.41}\\
	\sion{Mg}{x} & 4000\pme{200}{180} & \nodata & \color{blue}\textgreater4000\me{830} & \textgreater0.66\me{0.14}\\
	\sion{Si}{xii} & 1100\pme{96}{88} & \nodata & \color{blue}\textgreater1100\me{240} & \textgreater0.97\me{0.21}\\
	\sion{Ar}{vi} & 20\pme{19.2}{9.8} & \nodata & \color{red}\textless20\pe{20} & \textless1.05\pe{1.02}\\
	\sion{Ar}{vii} & 15\pme{2.6}{2.4} & \nodata & \color{red}\textless15\pe{4.0} & \textless1.80\pe{0.47}\\
	\sion{Ca}{x} & 85\pme{15}{15} & \nodata & \color{red}\textless85\pe{23} & \textless9.71\pe{2.63}\\
	\tableline
	\multicolumn{5}{c}{S4: v = -9420~km~s$^{-1}$}\\
	\tableline
	\sion{O}{iv} & 250\pme{45}{42} & \nodata & \color{red}\textless250\pe{66} & \textless0.91\pe{0.25}\\
	\sion{O}{v} & 330\pme{16}{15} & \nodata & \color{blue}\textgreater330\me{67} & \textgreater0.04\me{0.01}\\
	\sion{Ne}{v} & 1400\pme{160}{130} & \nodata & \color{blue}\textgreater1400\me{320} & \textgreater1.10\me{0.24}\\
	\sion{Ne}{v*}\tablenotemark{d} & 510\pme{60}{60} & \nodata & \color{blue}\textgreater510\me{120} & \nodata\\
	\sion{Ne}{v*}\tablenotemark{g} & 770\pme{130}{100} & \nodata & \color{blue}\textgreater770\me{70} & \nodata\\
	\sion{Ne}{v}\tablenotemark{e} & 170\pme{80}{50} & \nodata & 170\pme{90}{60} & \nodata\\	
	\sion{Ne}{vi} & 3800\pme{160}{140} & \nodata & \color{blue}\textgreater3800\me{780} & \textgreater0.27\me{0.05}\\
	\sion{Na}{ix} & 650\pme{120}{82} & \nodata & \color{blue}\textgreater650\me{150} & \textgreater1.00\me{0.24}\\
	\sion{Mg}{x} & 2500\pme{79}{73} & \nodata & \color{blue}\textgreater2500\me{510} & \textgreater0.08\me{0.02}\\
	\sion{Al}{xi} & 2000\pme{180}{160} & \nodata & \color{blue}\textgreater2000\me{420} & \textgreater0.48\me{0.10}\\
	\sion{Si}{xii} & 7400\pme{310}{280} & \nodata & \color{blue}\textgreater7400\me{1500} & \textgreater0.12\me{0.02}\\
	\sion{P}{xiii} & 870\pme{320}{310} & \nodata & \color{red}\textless870\pe{370} & \textless1.79\pe{0.77}\\
	\sion{Ar}{viii} & 190\pme{41}{39} & \nodata & \color{red}\textless190\pe{57} & \textless3.86\pe{1.13}\\
	\sion{Ca}{x} & 150\pme{52}{47} & \nodata & \color{red}\textless150\pe{60} & \textless0.97\pe{0.38}\\
	\tableline
	\multicolumn{5}{c}{\textbf{2MASS J1436+0727}}\\
	\tableline
	\multicolumn{5}{c}{S1: v = -14400~km~s$^{-1}$}\\
	\tableline
	\sion{H}{i} & 590\pme{160}{190} & \nodata & \color{red}\textless590\pe{200} & \textless6.30\pe{2.14}\\
	\sion{O}{v} & 110\pme{21}{19} & \nodata & \color{red}\textless110\pe{30} & \textless6.64\pe{1.81}\\
	\sion{Ne}{viii} & 2400\pme{97}{91} & 2900\pme{150}{120} & 2900\pme{590}{580} & 1.00\pme{0.20}{0.20}\\
	\sion{Na}{ix} & 1600\pme{170}{130} & \nodata & \color{red}\textless1600\pe{350} & \textless10.6\pe{2.31}\\
	\tableline
	\multicolumn{5}{c}{\textbf{VV2006 J1329+5405}}\\
	\tableline
	\multicolumn{5}{c}{S1: v = -11600~km~s$^{-1}$}\\
	\tableline
	\sion{H}{i} & 5100\pme{1400}{1200} & \nodata & \color{red}\textless5100\pe{1700} & \textless1.23\pe{0.42}\\
	\sion{N}{iv} & 560\pme{12}{11} & \nodata & \color{blue}\textgreater560\me{110} & \textgreater0.18\me{0.04}\\
	\sion{N}{iv*}\tablenotemark{h} & 190\pme{50}{40} & \nodata & \color{red}\textless190\pe{60} & \nodata\\
	\sion{N}{iv}\tablenotemark{e} & 560\pme{12}{11} & \nodata & 3020\pme{870}{1510} & \nodata\\
	\sion{O}{iv} & 3600\pme{100}{95} & \nodata & \color{blue}\textgreater3600\me{720} & \textgreater0.42\me{0.08}\\
	\sion{O}{iv*}\tablenotemark{i} & 1790\pme{70}{70} & \nodata & \color{blue}\textgreater1790\me{370} & \nodata\\
	\sion{O}{iv}\tablenotemark{e} & 1770\pme{70}{70} & \nodata & 6160\pme{4660}{4390} & \nodata\\	
	\sion{O}{v} & 940\pme{51}{43} & \nodata & \color{blue}\textgreater940\me{200} & \textgreater0.01\me{0.002}\\
	\sion{O}{v*}\tablenotemark{j} & 1000\pme{100}{90} & \nodata & \color{blue}\textgreater1000\me{220} & \nodata\\
	\sion{O}{v}\tablenotemark{e} & 940\pme{51}{43} & \nodata & 87100\pme{54200}{44400} & \nodata\\
	\sion{Ne}{viii} & 7300\pme{96}{88} & \nodata & \color{blue}\textgreater7300\me{1500} & \textgreater0.02\me{0.004}\\
	\sion{Na}{ix} & 3700\pme{130}{120} & \nodata & \color{blue}\textgreater3700\me{760} & \textgreater0.98\me{0.20}\\
	\sion{Mg}{x} & 11000\pme{880}{660} & \nodata & \color{blue}\textgreater11000\me{2400} & \textgreater0.16\me{0.03}\\
	\sion{S}{iv} & 28\pme{5.2}{4.9} & \nodata & \color{red}\textless28\pe{7.6} & \textless1.00\pe{0.29}\\
	\sion{S}{v} & 93\pme{5.2}{4.8} & \nodata & \color{blue}\textgreater93\me{19} & \textgreater0.33\me{0.07}\\
	\sion{S}{vi} & 390\pme{130}{120} & \nodata & \color{blue}\textgreater390\me{140} & \textgreater0.19\me{0.07}\\
	\sion{Cl}{vii} & 190\pme{13}{12} & \nodata & \color{blue}\textgreater190\me{40} & \textgreater1.05\me{0.22}\\
	\sion{Ar}{vi} & 1100\pme{65}{62} & \nodata & \color{blue}\textgreater1100\me{230} & \textgreater0.52\me{0.11}\\
	\sion{Ar}{viii} & 1100\pme{45}{41} & \nodata & \color{blue}\textgreater1100\me{230} & \textgreater0.30\me{0.06}\\
	\sion{K}{ix} & 210\pme{50}{40} & \nodata & \color{red}\textless210\pe{66} & \textless1.00\pe{0.31}\\
	\sion{Ca}{vi} & 2700\pme{450}{360} & \nodata & \color{red}\textless2700\pe{700} & \textless0.98\pe{0.25}\\
	\tableline
	\multicolumn{5}{c}{S2: v = -12900~km~s$^{-1}$}\\
	\tableline
	\sion{H}{i} & 9100\pme{1500}{1900} & \nodata & \color{red}\textless9100\pe{2300} & \textless127\pe{32.5}\\
	\sion{O}{iv} & 460\pme{42}{39} & \nodata & \color{red}\textless460\pe{100} & \textless7.30\pe{1.60}\\
	\sion{O}{v} & 290\pme{33}{28} & \nodata & \color{blue}\textgreater290\me{64} & \textgreater0.17\me{0.04}\\
	\sion{Ne}{viii} & 3400\pme{59}{56} & \nodata & \color{blue}\textgreater3400\me{680} & 1.00\me{0.20}\\
	\sion{Na}{ix} & 220\pme{57}{48} & \nodata & \color{red}\textless220\pe{71} & \textless10.4\pe{3.37}\\
	\sion{Mg}{x} & 2200\pme{630}{350} & \nodata & \color{red}\textless2200\pe{680} & \textless6.61\pe{2.07}\\
	\sion{S}{v} & 34\pme{2.9}{2.7} & \nodata & \color{red}\textless34\pe{7.5} & \textless55.8\pe{12.1}\\
	\sion{S}{vi} & 110\pme{18}{18} & \nodata & \color{red}\textless110\pe{29} & \textless29.0\pe{7.44}\\
	\tableline
	\multicolumn{5}{c}{\textbf{7C 1631+3930}}\\
	\tableline
	\multicolumn{5}{c}{S1: v = -1010~km~s$^{-1}$}\\
	\tableline
	\sion{O}{iv} & 170\pme{40}{40} & \nodata & \color{red}\textless170\pe{50} & \textless29.6\pe{8.71}\\
	\sion{O}{v} & 310\pme{10}{10} & \nodata & \color{blue}\textgreater310\me{60} & \textgreater1.01\me{0.20}\\
	\sion{Ne}{v} & 350\pme{70}{70} & \nodata & \color{red}\textless350\pe{100} & \textless9.08\pe{2.60}\\
	\sion{Ne}{vi*}\tablenotemark{f} & 160\pme{70}{70} & \nodata & \color{red}\textless160\pe{70} & \nodata\\
	\sion{Ne}{vi}\tablenotemark{e} & \nodata & \nodata & 780\pme{650}{190} & \nodata\\
	\sion{Ne}{viii} & 4100\pme{170}{140} & 5400\pme{1600}{340} & 5400\pme{2000}{1100} & 0.97\pme{0.35}{0.20}\\
	\sion{Mg}{x} & 2200\pme{110}{100} & 2500\pme{130}{110} & 2500\pme{520}{520} & 1.01\pme{0.21}{0.21}\\
	\sion{Ar}{vii} & 8.4\pme{3.3}{3.3} & \nodata & \color{red}\textless8.4\pe{3.7} & \textless34.5\pe{15.2}\\
	\sion{Ar}{viii} & 32\pme{9.0}{13} & \nodata & \color{red}\textless32\pe{11} & \textless34.8\pe{12.0}\\
	\tableline
	\multicolumn{5}{c}{S2: v = -1430~km~s$^{-1}$}\\
	\tableline
	\sion{N}{iv} & 270\pme{10}{10} & \nodata & \color{blue}\textgreater270\me{55} & \textgreater0.92\me{0.21}\\
	\sion{O}{iii} & 51\pme{12}{22} & \nodata & \color{red}\textless51\pe{16} & \textless0.98\pe{0.31}\\
	\sion{O}{iv} & 1200\pme{50}{50} & \nodata & \color{blue}\textgreater1200\me{250} & \textgreater0.51\me{0.11}\\
	\sion{O}{v} & 1800\pme{210}{40} & \nodata & \color{blue}\textgreater1800\me{360} & \textgreater0.29\me{0.06}\\
	\sion{Ne}{v} & 1400\pme{100}{90} & \nodata & \color{blue}\textgreater1400\me{300} & \textgreater1.05\me{0.23}\\
	\sion{Ne}{v*}\tablenotemark{d} & 250\pme{40}{40} & \nodata & 250\pme{70}{60} & \nodata\\
	\sion{Ne}{v*}\tablenotemark{g} & 130\pme{40}{30} & \nodata & \color{red}\textless130\pe{50} & \nodata\\
	\sion{Ne}{v}\tablenotemark{e} & 1050\pme{80}{70} & \nodata & 1050\pme{230}{220} & \nodata\\	
	\sion{Ne}{vi} & 1900\pme{240}{200} & \nodata & \color{blue}\textgreater1900\me{420} & \textgreater0.72\me{0.16}\\
	\sion{Ne}{viii} & 12100\pme{400}{300} & 14100\pme{2400}{500} & 14100\pme{3700}{2900} & 1.00\pme{0.26}{0.21}\\
	\sion{Na}{ix} & 670\pme{180}{220} & \nodata & \color{red}\textless670\pe{260} & \textless1.73\pe{0.67}\\
	\sion{Mg}{x} & 6500\pme{200}{180} & 8100\pme{200}{180} & 8100\pme{1600}{1600} & 1.02\pme{0.20}{0.20}\\
	\sion{S}{iii} & 26\pme{7.8}{6.7} & \nodata & \color{red}\textless26\pe{9.4} & \textless55.6\pe{20.1}\\
	\sion{S}{iv} & 35\pme{6.1}{5.2} & \nodata & \color{red}\textless35\pe{9.3} & \textless4.00\pe{1.06}\\
	\sion{S}{v} & 18\pme{2.0}{1.9} & \nodata & \color{blue}\textgreater18\me{4.0} & \textgreater0.90\me{0.20}\\
	\sion{Ar}{vii} & 3.8\pme{2.8}{1.9} & \nodata & \color{red}\textless3.8\pe{2.8} & \textless0.91\pe{0.67}\\
	\sion{Ar}{viii} & 30\pme{14}{12} & \nodata & \color{red}\textless30\pe{15} & \textless5.10\pe{2.55}\\
	\tableline
	\multicolumn{5}{c}{S3: v = -5300~km~s$^{-1}$}\\
	\tableline
	\sion{N}{iv} & 48\pme{9.2}{8.5} & \nodata & \color{red}\textless48\pe{13} & \textless12.3\pe{3.32}\\
	\sion{O}{iv} & 120\pme{31}{34} & \nodata & \color{red}\textless120\pe{40} & \textless6.52\pe{2.21}\\
	\sion{Ne}{viii} & 770\pme{40}{37} & \nodata & \color{blue}\textgreater770\me{160} & \textgreater0.05\me{0.01}\\
	\sion{Na}{ix} & 340\pme{59}{54} & \nodata & \color{blue}\textgreater340\me{90} & \textgreater1.00\me{0.26}\\
	\sion{Mg}{x} & 1500\pme{120}{110} & \nodata & \color{blue}\textgreater1500\me{330} & \textgreater0.25\me{0.05}\\
	\sion{Si}{xi} & 96000\pme{1000}{58000} & \nodata & \color{red}\textless96000\pe{19000} & \textless62.6\pe{12.5}\\
	\sion{Ar}{vii} & 5.1\pme{3.1}{2.4} & \nodata & \color{red}\textless5.7\pe{3.3} & \textless6.74\pe{4.36}\\
	\sion{Ar}{viii} & 32\pme{15}{15} & \nodata & \color{red}\textless32\pe{16} & \textless11.10\pe{6.45}\\
	\sion{Ca}{x} & 46\pme{32}{18} & \nodata & \color{red}\textless46\pe{33} & \textless1.37\pe{0.98}\\
	\tableline
	\multicolumn{5}{c}{S4: v = -5770~km~s$^{-1}$}\\
	\tableline
	\sion{N}{iv} & 240\pme{13}{11} & \nodata & \color{blue}\textgreater240\me{49} & \textgreater0.34\me{0.07}\\
	\sion{O}{iii} & 75\pme{39}{29} & \nodata & \color{red}\textless75\pe{42} & \textless1.04\pe{0.58}\\
	\sion{O}{iv} & 660\pme{59}{53} & \nodata & \color{blue}\textgreater660\me{140} & \textgreater0.13\me{0.03}\\
	\sion{O}{v} & 570\pme{24}{20} & \nodata & \color{blue}\textgreater570\me{120} & \textgreater0.02\me{0.004}\\
	\sion{Ne}{v*}\tablenotemark{g} & 250\pme{80}{50} & \nodata & 250\pme{90}{70} & \nodata\\
	\sion{Ne}{v}\tablenotemark{e} & \nodata & \nodata & 3550\pme{7400}{2680} & \nodata\\	
	\sion{Ne}{viii} & 6000\pme{190}{160} & \nodata & \color{blue}\textgreater6000\me{1200} & \textgreater0.13\me{0.03}\\
	\sion{Na}{ix} & 600\pme{84}{77} & \nodata & \color{blue}\textgreater600\me{140} & \textgreater1.53\me{0.36}\\
	\sion{Mg}{x} & 3400\pme{170}{150} & 4300\pme{270}{210} & 4300\pme{900}{890} & 0.81\pme{0.17}{0.17}\\
	\sion{S}{iv} & 28\pme{5.3}{3.5} & \nodata & \color{red}\textless28\pe{7.8} & \textless2.59\pe{0.71}\\
	\sion{S}{v} & 28\pme{3.4}{3.2} & \nodata & \color{blue}\textgreater28\me{6.4} & \textgreater0.74\me{0.17}\\
	\sion{Ar}{vii} & 39\pme{4.9}{4.8} & \nodata & \color{blue}\textgreater39\me{9.3} & \textgreater1.02\me{0.24}\\
	\sion{Ar}{viii} & 96\pme{23}{20} & \nodata & \color{red}\textless96\pe{30} & \textless0.95\pe{0.30}\\
	\sion{Ca}{x} & 540\pme{70}{57} & \nodata & \color{red}\textless540\pe{130} & \textless1.45\pe{0.35}\\
	\tableline
	\multicolumn{5}{c}{S5: v = -6150~km~s$^{-1}$}\\
	\tableline
	\sion{N}{iv} & 9.3\pme{2.5}{4.4} & \nodata & \color{red}\textless9.3\pe{3.1} & \textless1.80\pe{0.60}\\
	\sion{O}{v} & 76\pme{7.3}{7.2} & \nodata & \color{blue}\textgreater76\me{17} & \textgreater0.06\me{0.01}\\
	\sion{Ne}{viii} & 2200\pme{98}{88} & \nodata & \color{blue}\textgreater2200\me{440} & \textgreater0.11\me{0.02}\\
	\sion{Na}{ix} & 450\pme{64}{57} & \nodata & \color{blue}\textgreater450\me{110} & \textgreater1.00\me{0.24}\\
	\sion{Mg}{x} & 890\pme{110}{98} & \nodata & \color{blue}\textgreater890\me{200} & \textgreater0.11\me{0.03}\\
	\sion{Si}{xi} & 130000\pme{1000}{70000} & \nodata & \color{red}\textless130000\pe{26000} & \textless66.6\pe{13.3}\\
	\sion{Ca}{x} & 49\pme{31}{26} & \nodata & \color{red}\textless49\pe{33} & \textless1.10\pe{0.73}\\
	\tableline
	\multicolumn{5}{c}{S6: v = -7210~km~s$^{-1}$}\\
	\tableline
	\sion{N}{iv} & 25\pme{5.1}{4.8} & \nodata & \color{red}\textless25\pe{7.0} & \textless126\pe{35.2}\\
	\sion{O}{iv} & 210\pme{32}{29} & \nodata & \color{red}\textless210\pe{52} & \textless240\pe{59.9}\\
	\sion{Ne}{viii} & 1100\pme{58}{52} & \nodata & \color{blue}\textgreater1100\me{230} & \textgreater1.00\me{0.20}\\
	\sion{Na}{ix} & 130\pme{44}{39} & \nodata & \color{red}\textless130\pe{50} & \textless3.79\pe{1.49}\\
	\sion{Mg}{x} & 680\pme{79}{71} & \nodata & \color{blue}\textgreater680\me{150} & \textgreater1.00\me{0.23}\\
	\enddata
	\tablecomments{	\tablenotetext{a}{Sum of all \sub{N}{ion} from excited and ground states for a given ion in \\each outflow system using the AOD and PC methods.}
		\tablenotetext{b}{The adopted values in blue are lower limits, in red are upper limits, \\and in black are measurements.}
		\tablenotetext{c}{The ratio of the adopted values to the column densities from the \\best-fit Cloudy model.}
		\tablenotetext{d}{\sion{Ne}{v} 413 cm$^{-1}$ energy level.}
		\tablenotetext{e}{0 cm$^{-1}$ energy level of the respective ion.}
		\tablenotetext{f}{\sion{Ne}{vi} 1307 cm$^{-1}$ energy level.}
		\tablenotetext{g}{\sion{Ne}{v} 1111 cm$^{-1}$ energy level.}
		\tablenotetext{h}{\sion{N}{iv} 67416 cm$^{-1}$ energy level.}
		\tablenotetext{i}{\sion{O}{iv} 386 cm$^{-1}$ energy level.}	
		\tablenotetext{j}{\sion{O}{v} 82385 cm$^{-1}$ energy level.\\}		
		(This table is available in machine-readable form.)}
\end{deluxetable}

\begin{figure*}
	\includegraphics[scale=1.0]{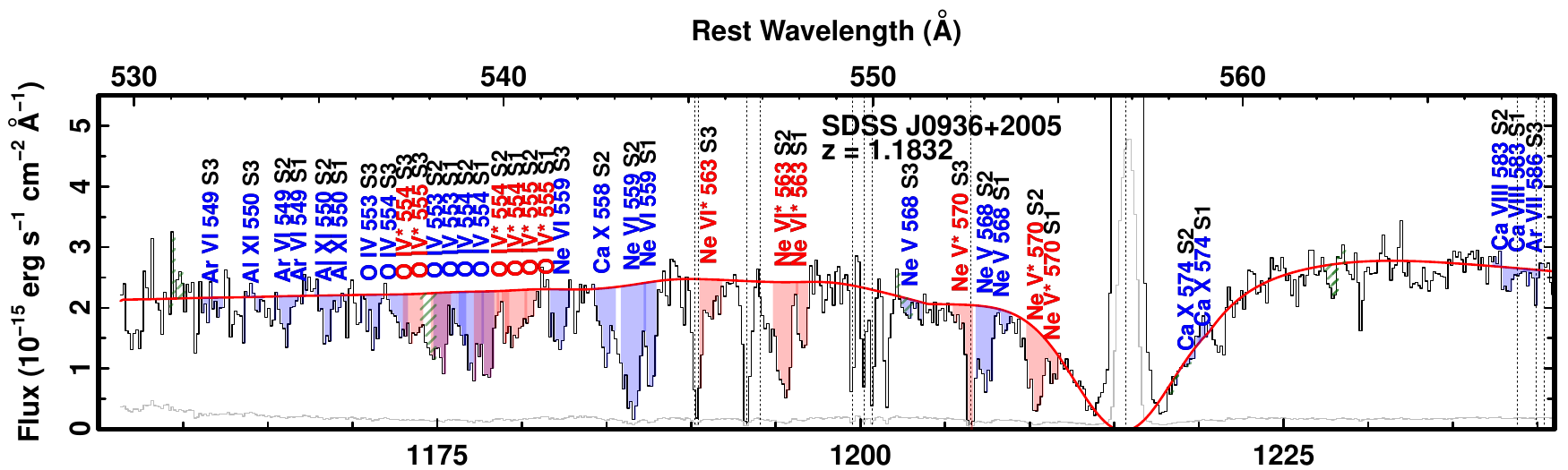}
	\includegraphics[scale=1.0]{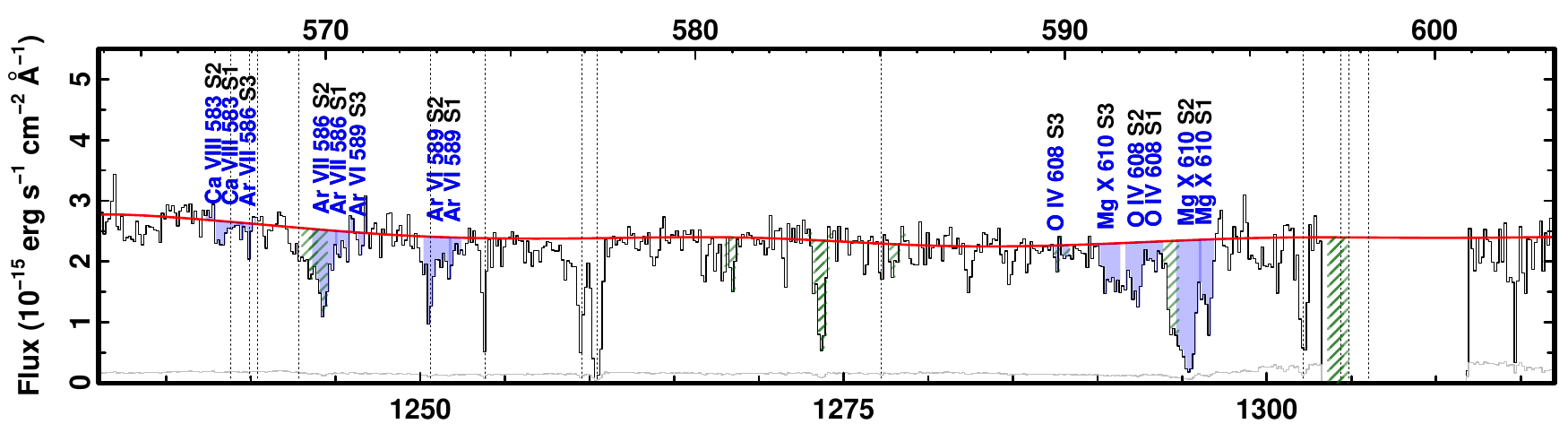}
	\includegraphics[scale=1.0]{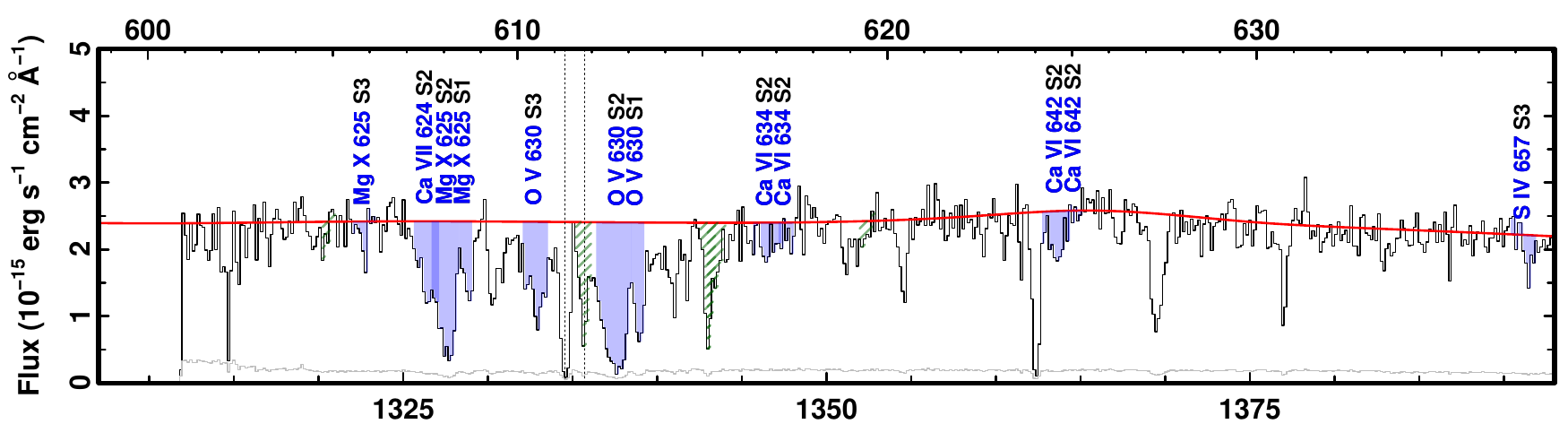}
	\includegraphics[scale=1.0]{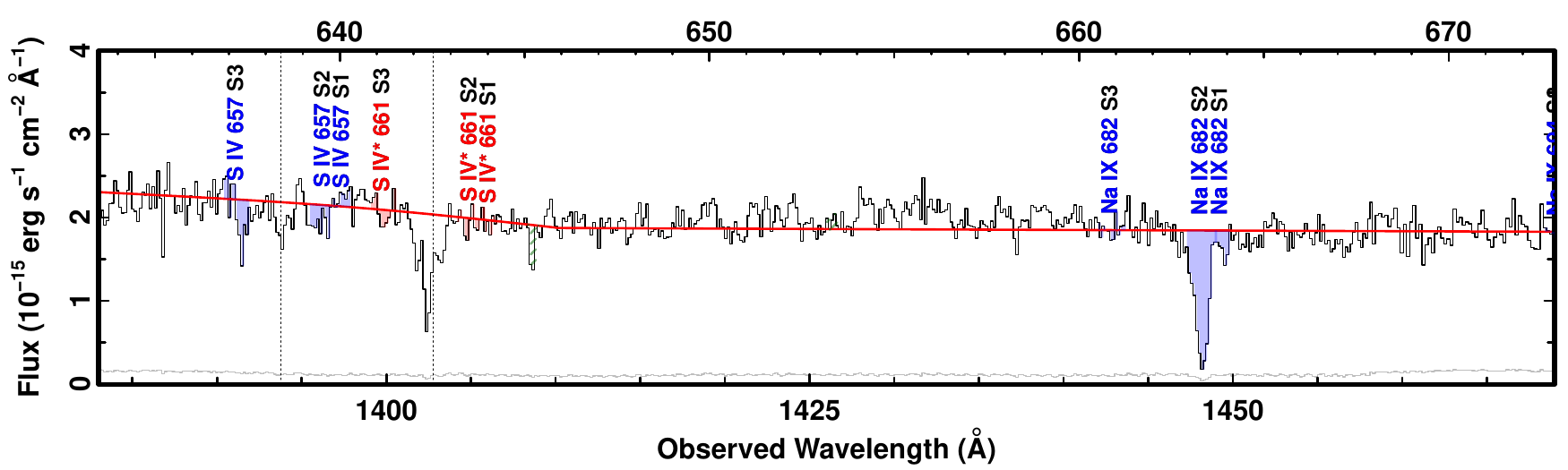}
	\caption{\footnotesize{The dereddened HST/COS spectrum (in black) with errors (in gray) for the five quasars. The main absorption troughs are labeled for all outflow systems as well as regions used for upper limit measurements (see Table~\ref{tab:out}). Blue shaded regions mark transitions from ground absorption lines, and red shaded regions are for excited ones. Absorption troughs from intervening systems are the slanted dark green shaded regions, and the vertical dashed lines mark Galactic absorption and geocoronal emission features. The red contour traces the unabsorbed emission model for each quasar. Overlapping sections of troughs have regions of darker blues and reds as well as mixtures of blue and red.}}
\label{fig:spectrum}	
\end{figure*}
\begin{figure*}
	\includegraphics[scale=1.0]{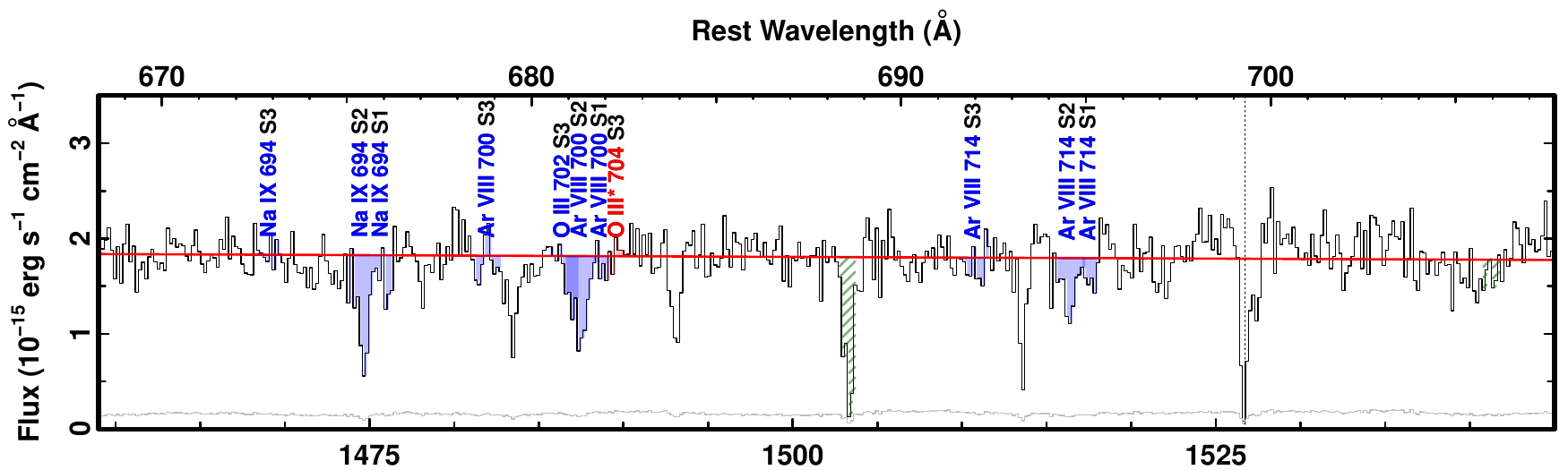}
	\includegraphics[scale=1.0]{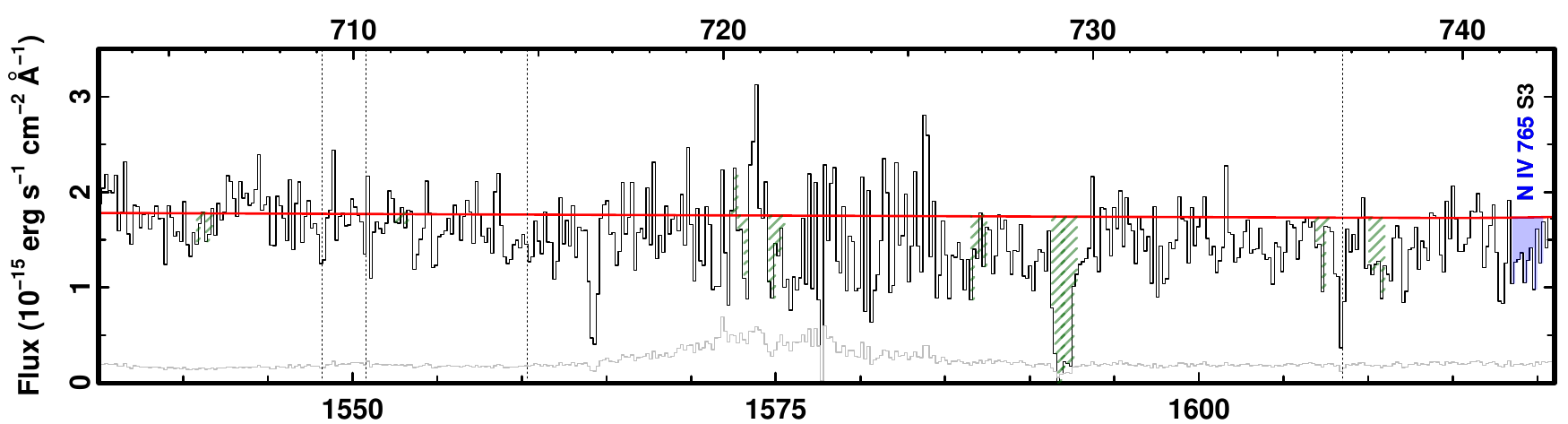}
	\includegraphics[scale=1.0]{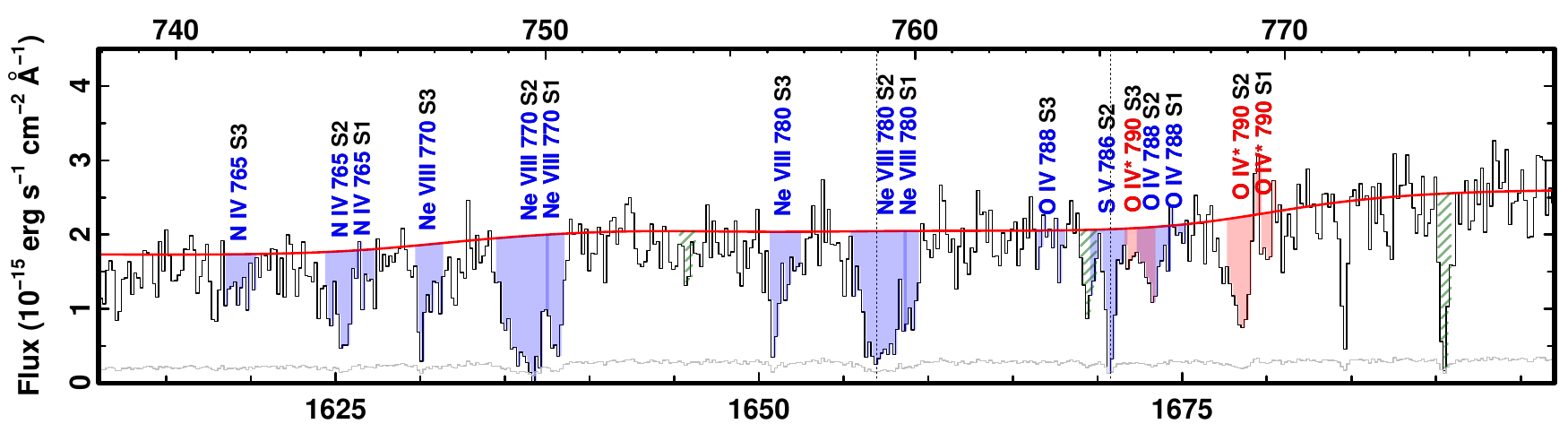}
	\includegraphics[scale=1.0]{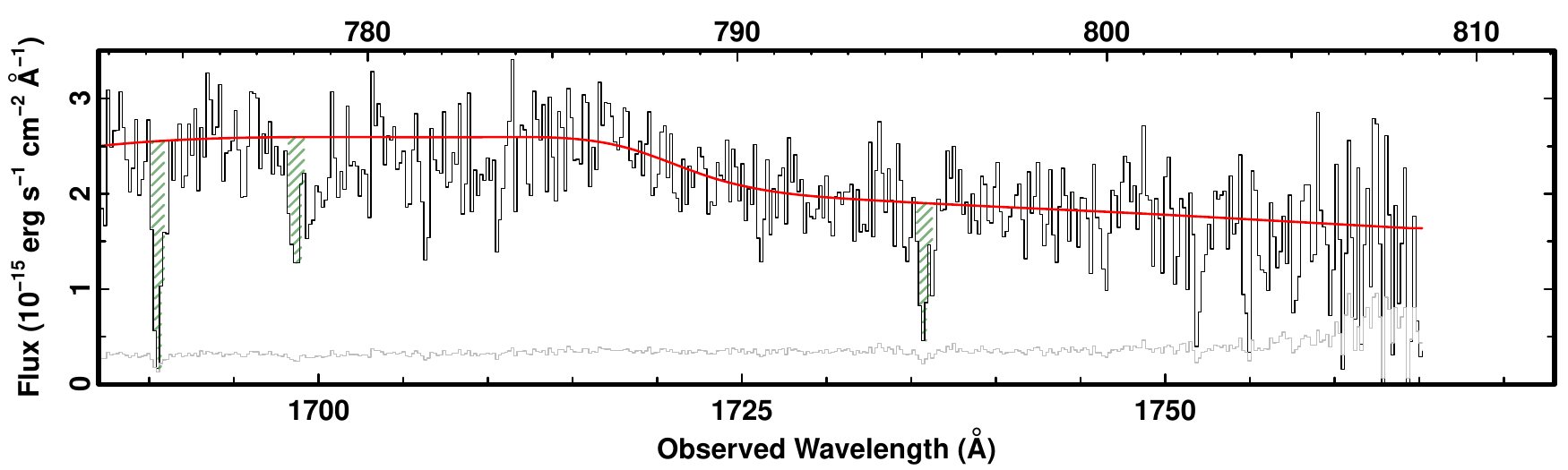}
	\figurenum{1}
	\caption{(Continued.)}
\end{figure*}
\begin{figure*}
	\includegraphics[scale=1.0]{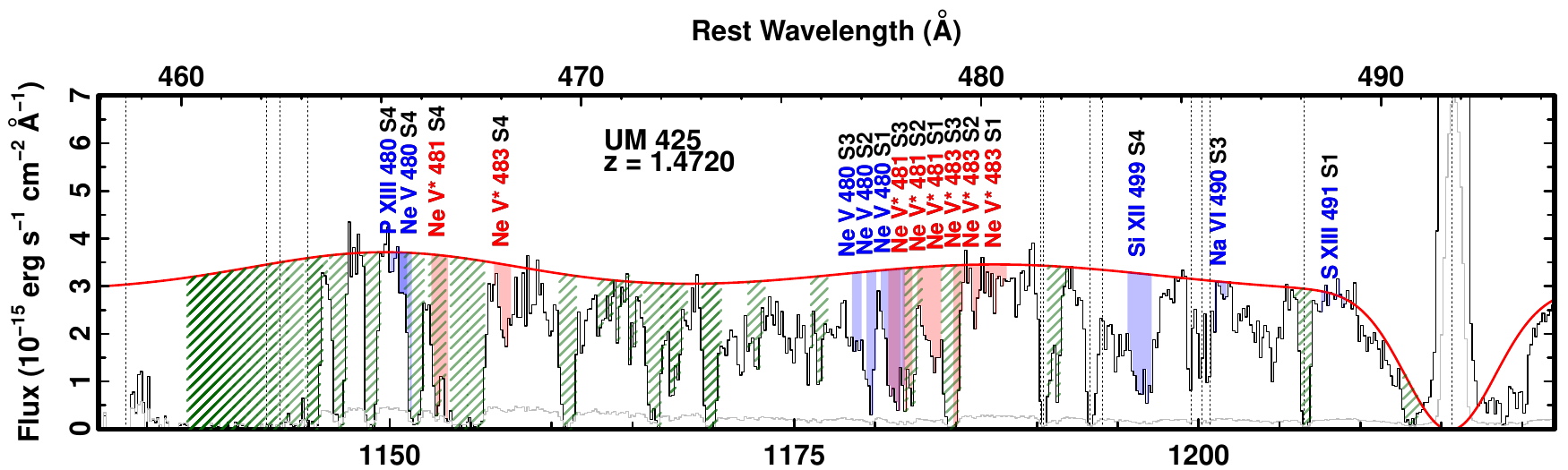}
	\includegraphics[scale=1.0]{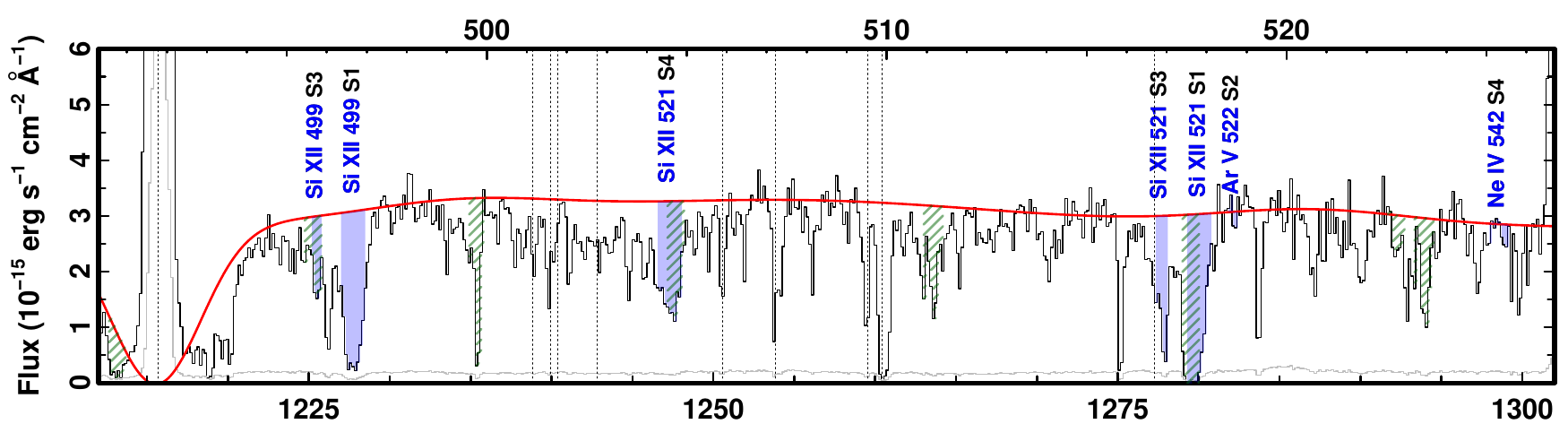}
	\includegraphics[scale=1.0]{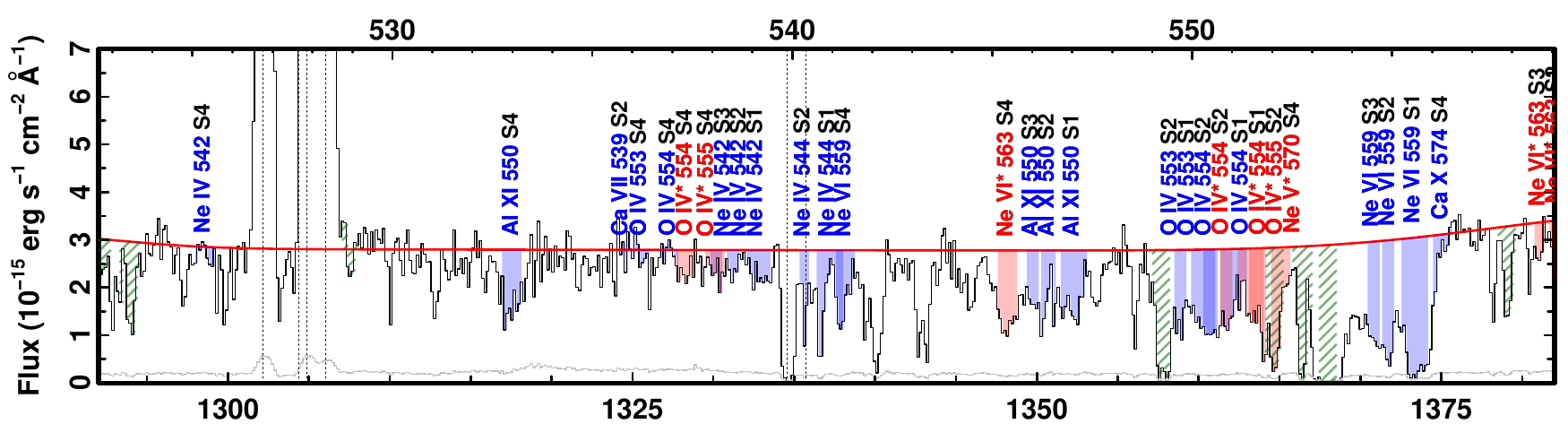}
	\includegraphics[scale=1.0]{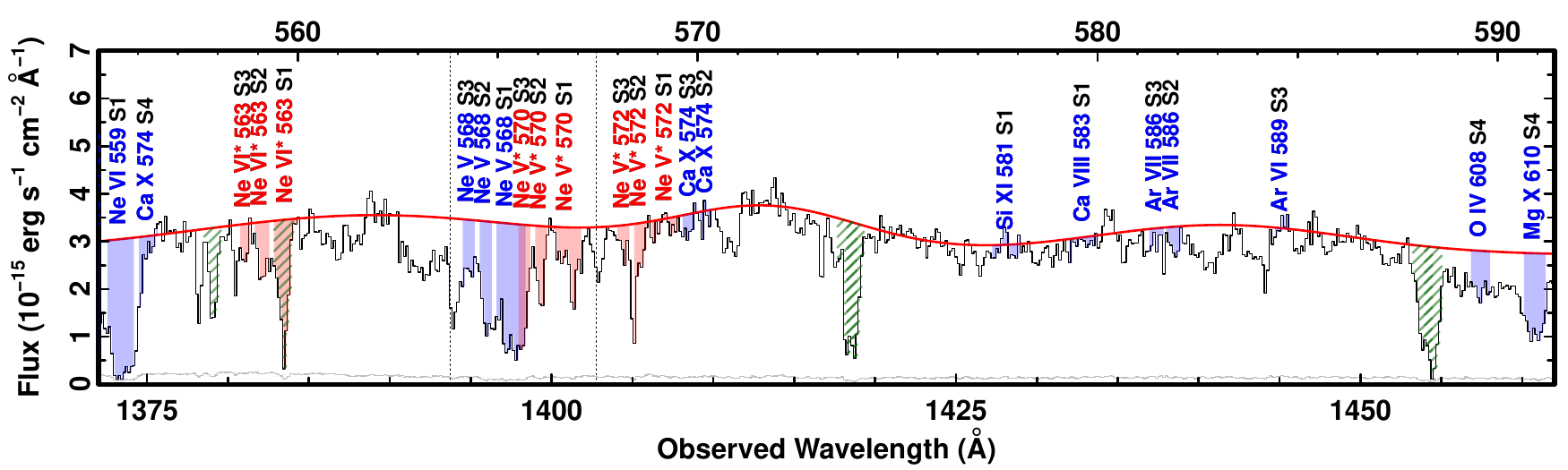}
	\figurenum{1}
	\caption{(Continued.)}
\end{figure*}
\begin{figure*}
	\includegraphics[scale=1.0]{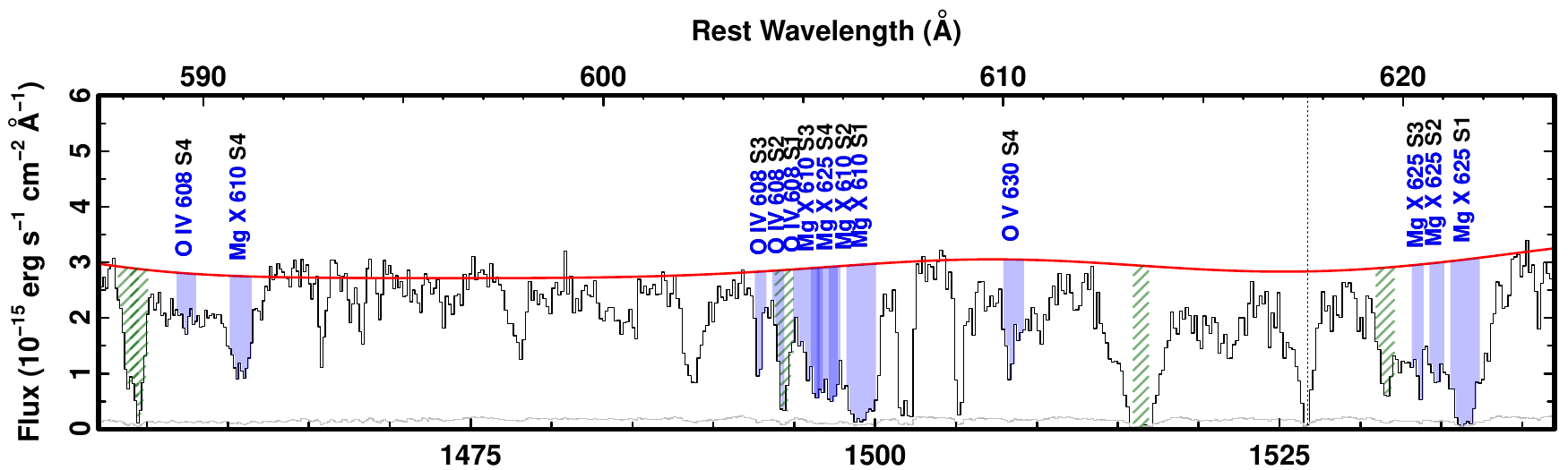}
	\includegraphics[scale=1.0]{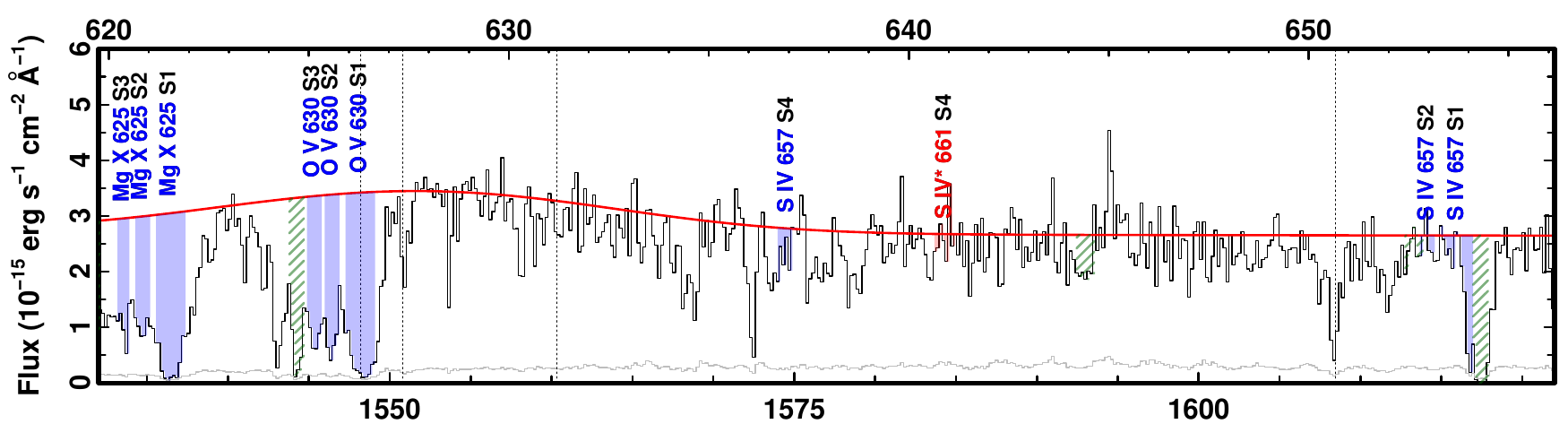}
	\includegraphics[scale=1.0]{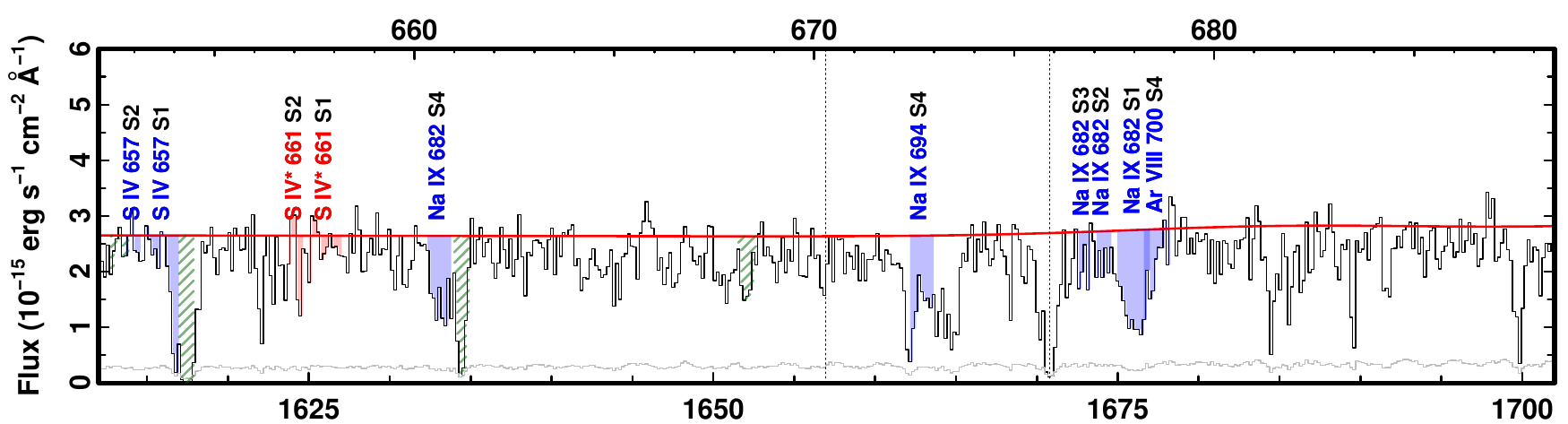}
	\includegraphics[scale=1.0]{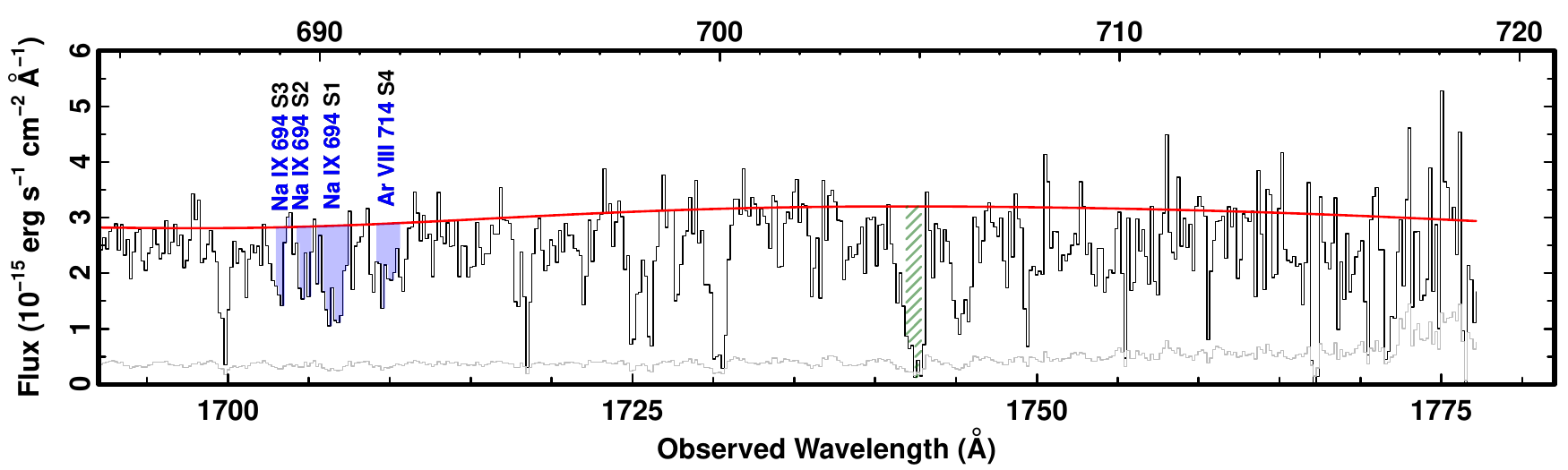}
	\figurenum{1}
	\caption{(Continued.)}
\end{figure*}
\begin{figure*}
	\includegraphics[scale=1.0]{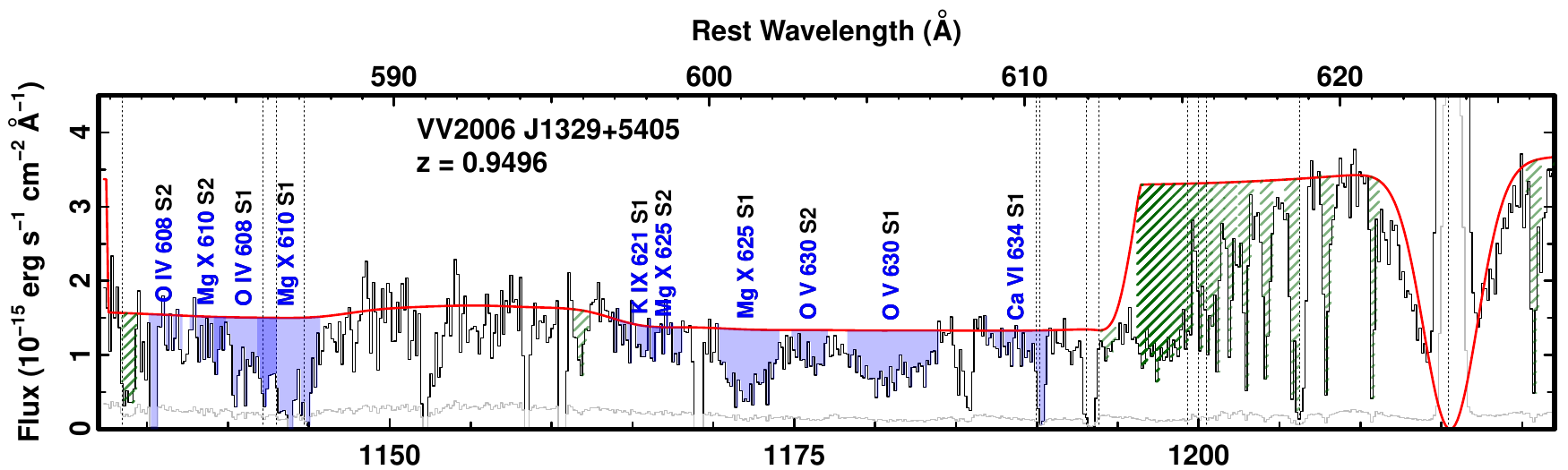}
	\includegraphics[scale=1.0]{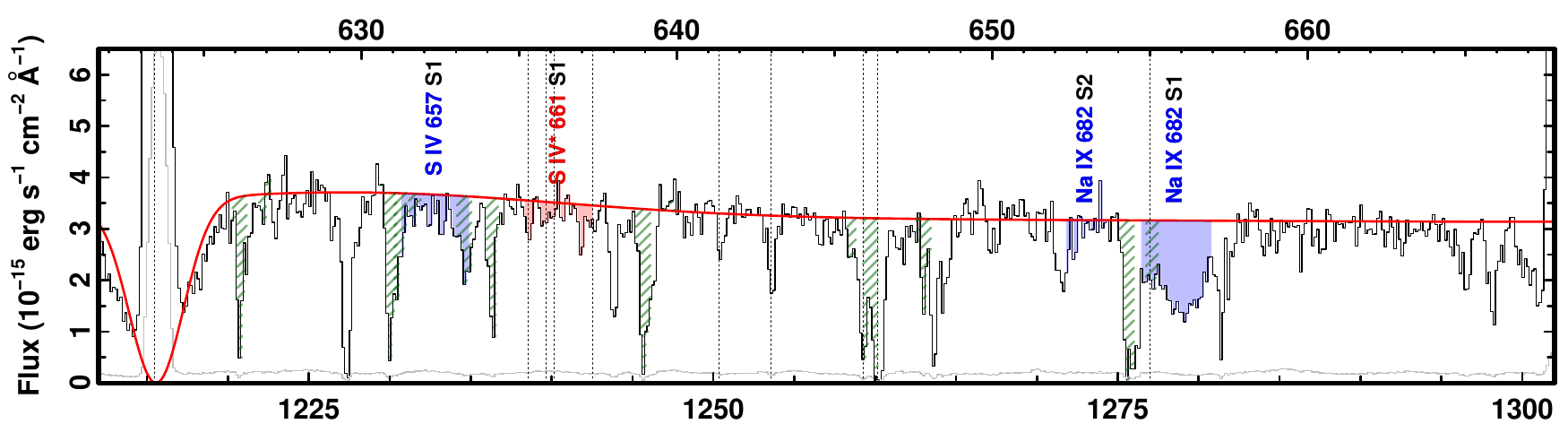}
	\includegraphics[scale=1.0]{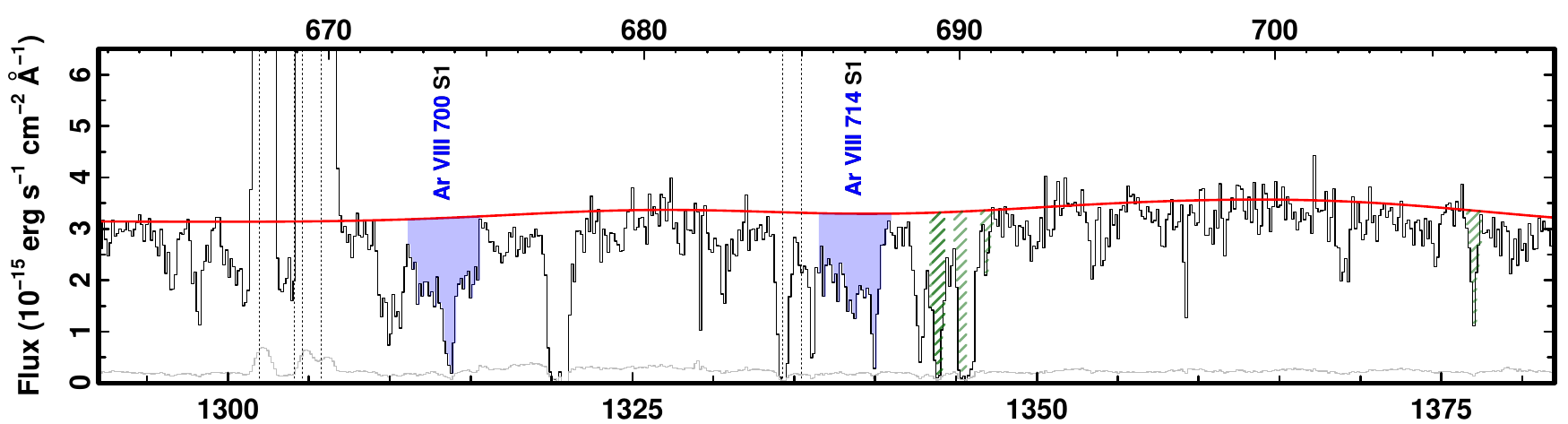}
	\includegraphics[scale=1.0]{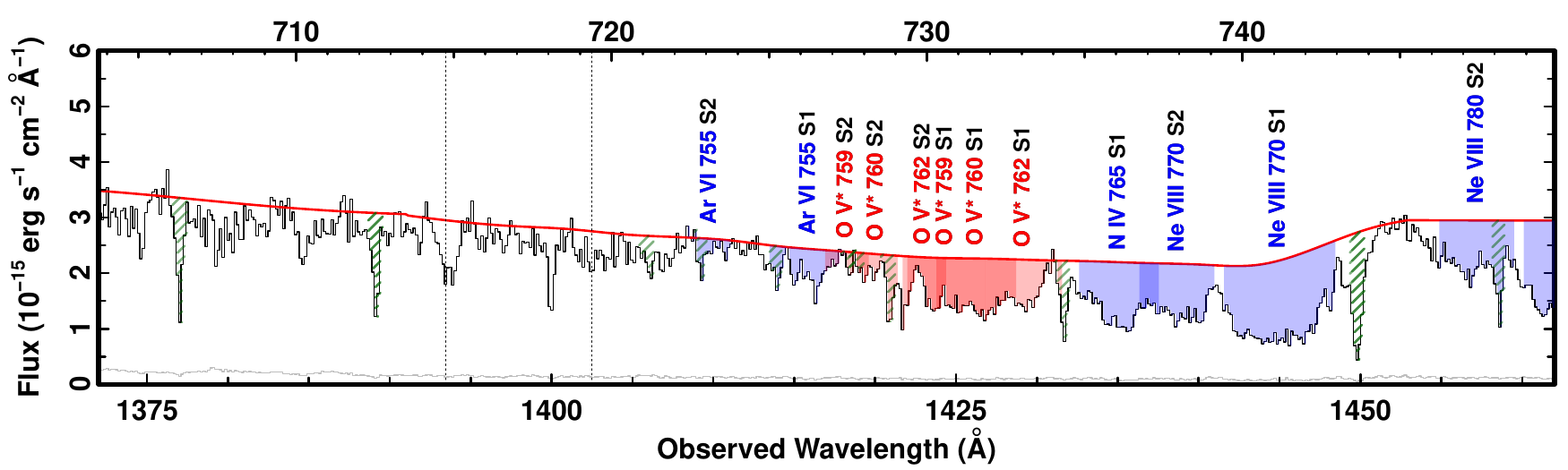}
	\figurenum{1}
	\caption{(Continued.)}
\end{figure*}
\begin{figure*}
	\includegraphics[scale=1.0]{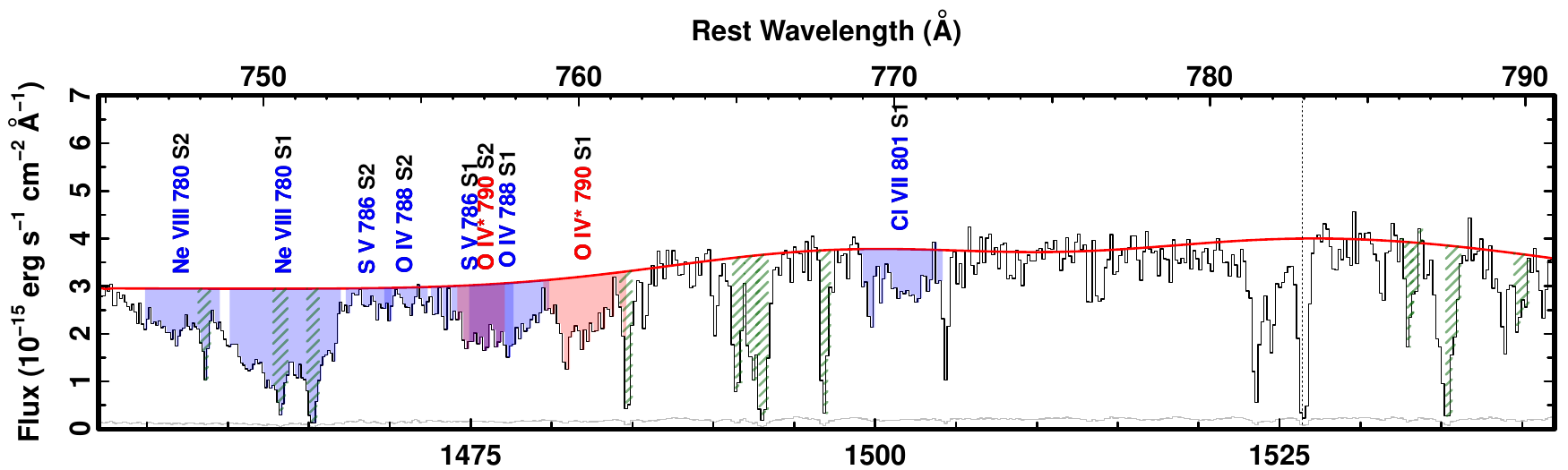}
	\includegraphics[scale=1.0]{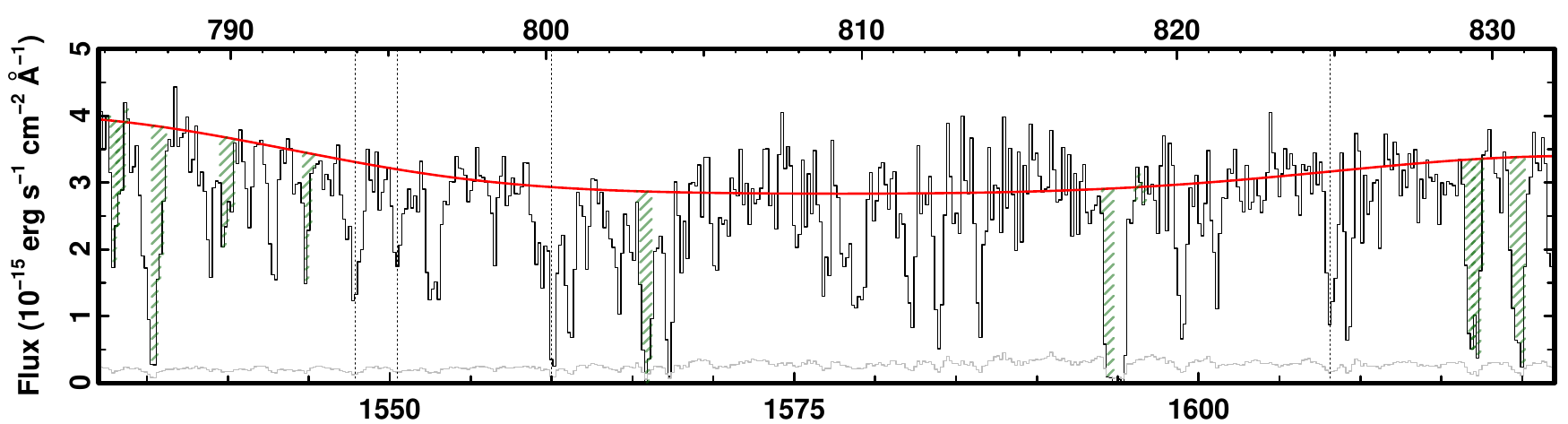}
	\includegraphics[scale=1.0]{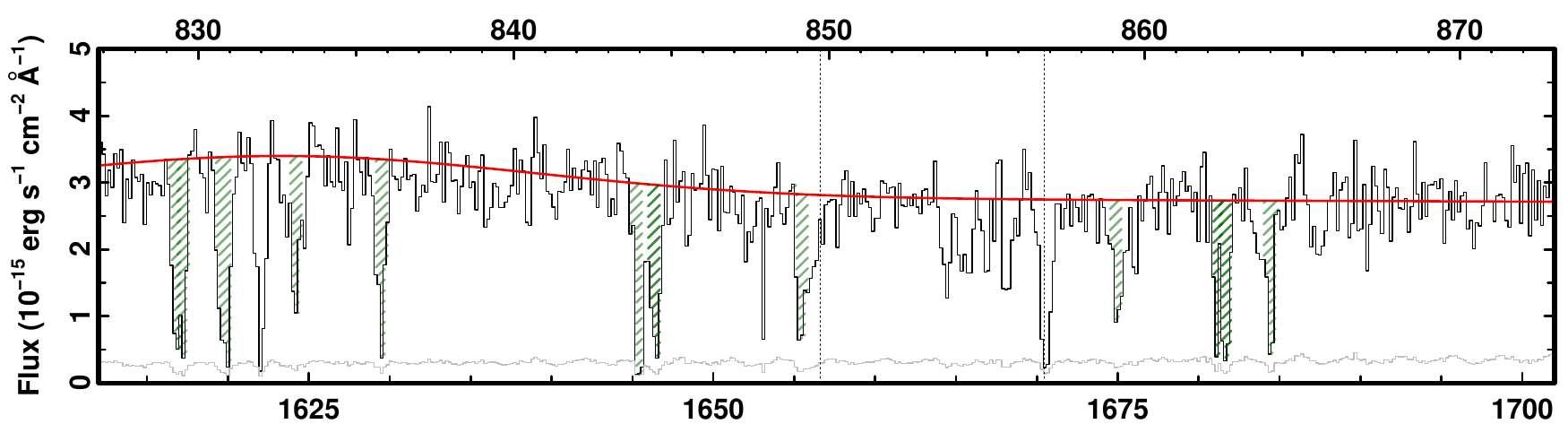}
	\includegraphics[scale=1.0]{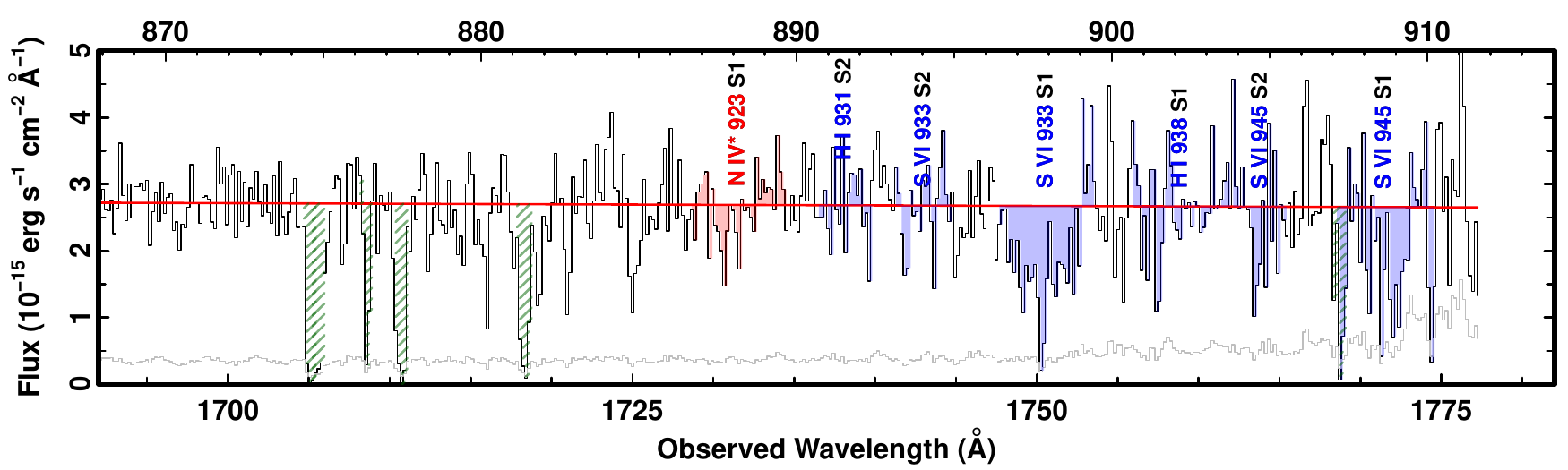}
	\figurenum{1}
	\caption{(Continued.)}
\end{figure*}
\begin{figure*}
	\includegraphics[scale=1.0]{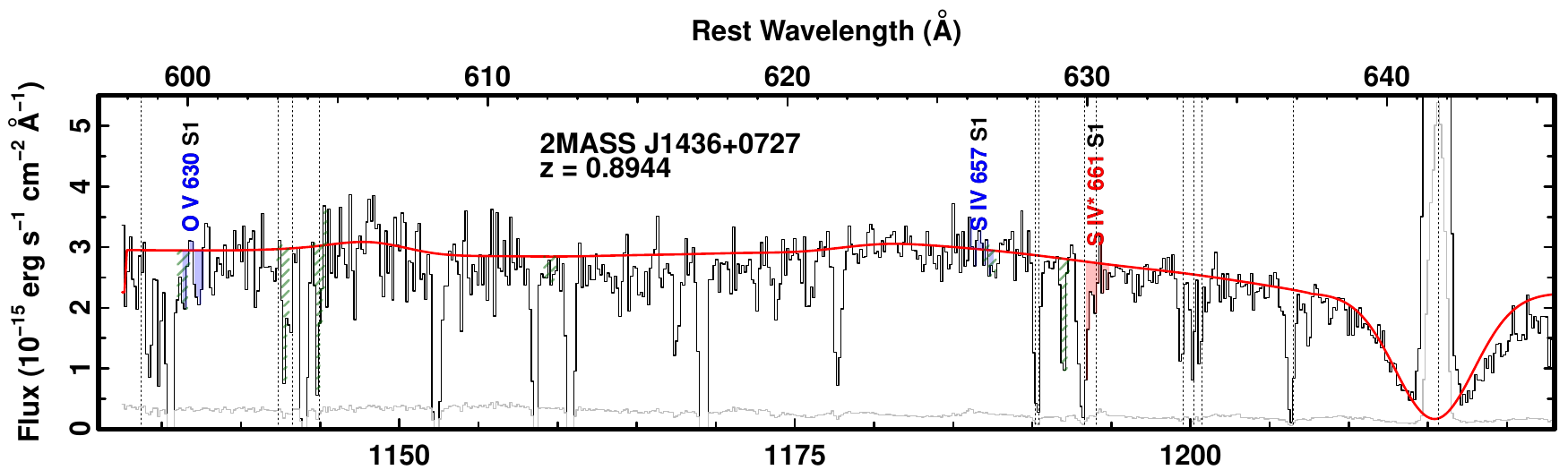}
	\includegraphics[scale=1.0]{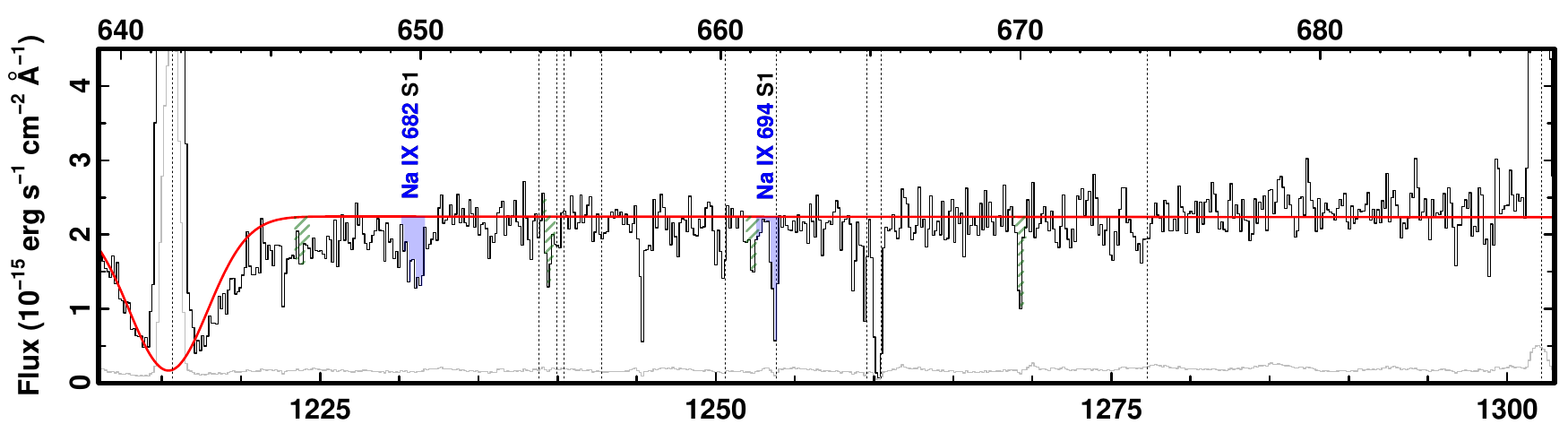}
	\includegraphics[scale=1.0]{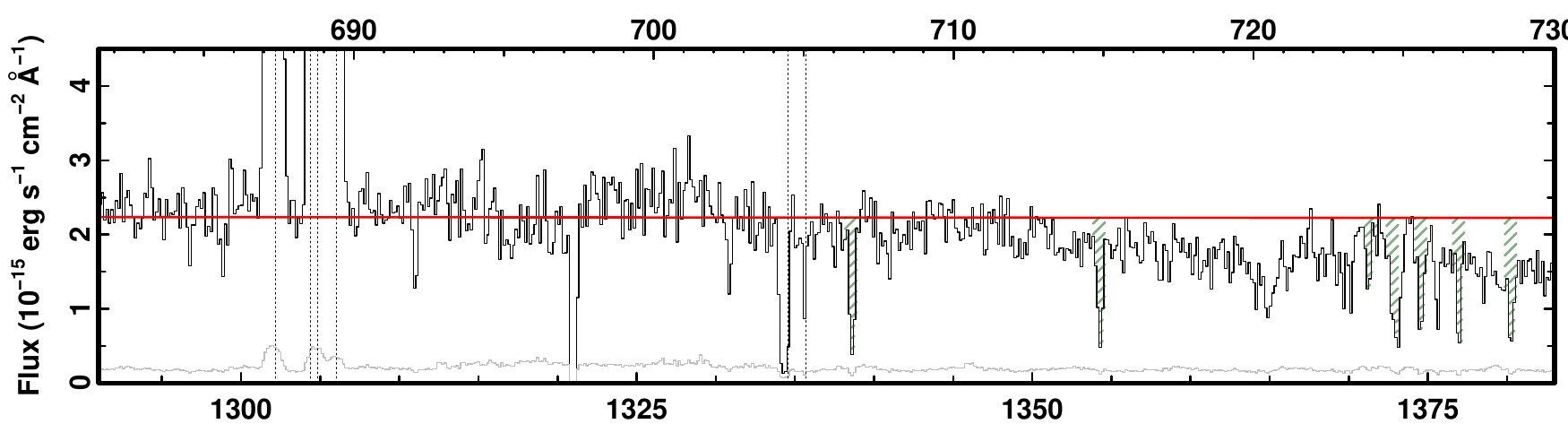}
	\includegraphics[scale=1.0]{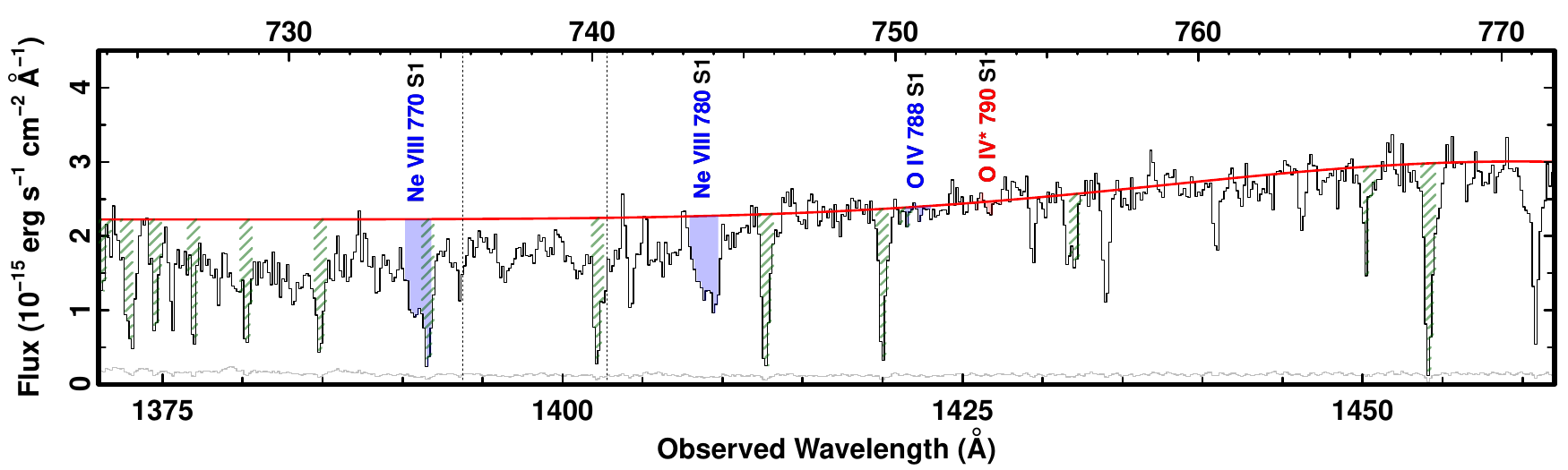}
	\figurenum{1}
	\caption{(Continued.)}
\end{figure*}
\begin{figure*}
	\includegraphics[scale=1.0]{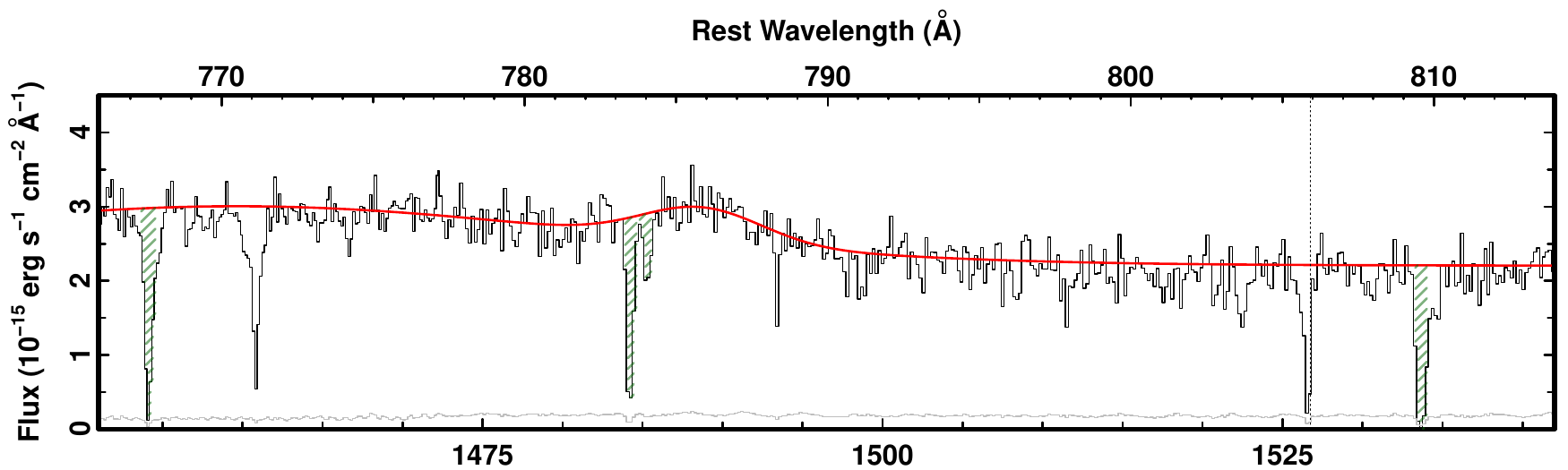}
	\includegraphics[scale=1.0]{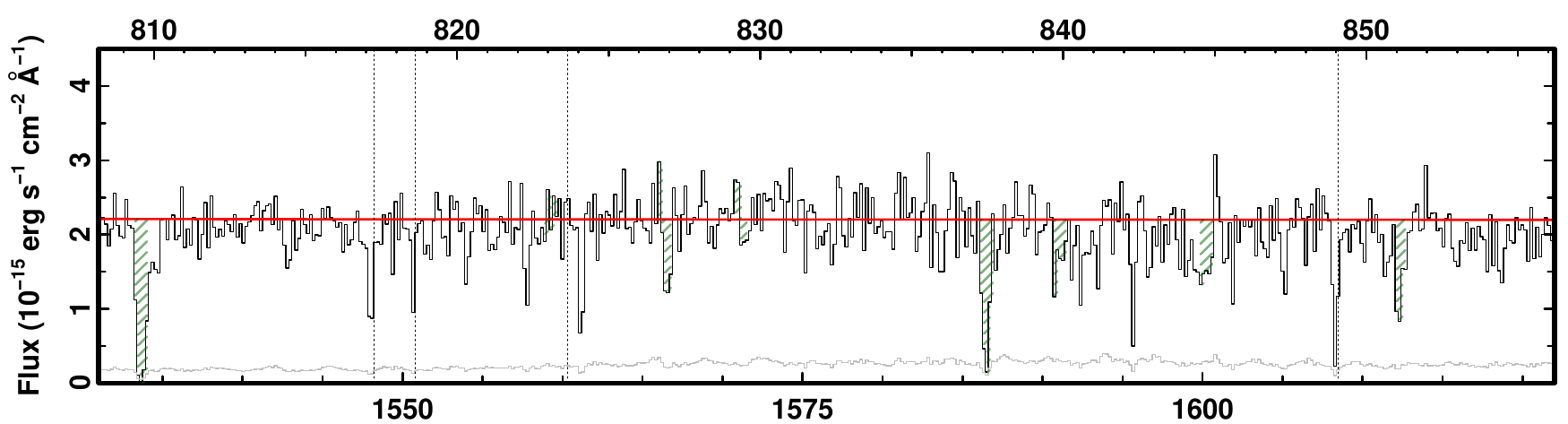}
	\includegraphics[scale=1.0]{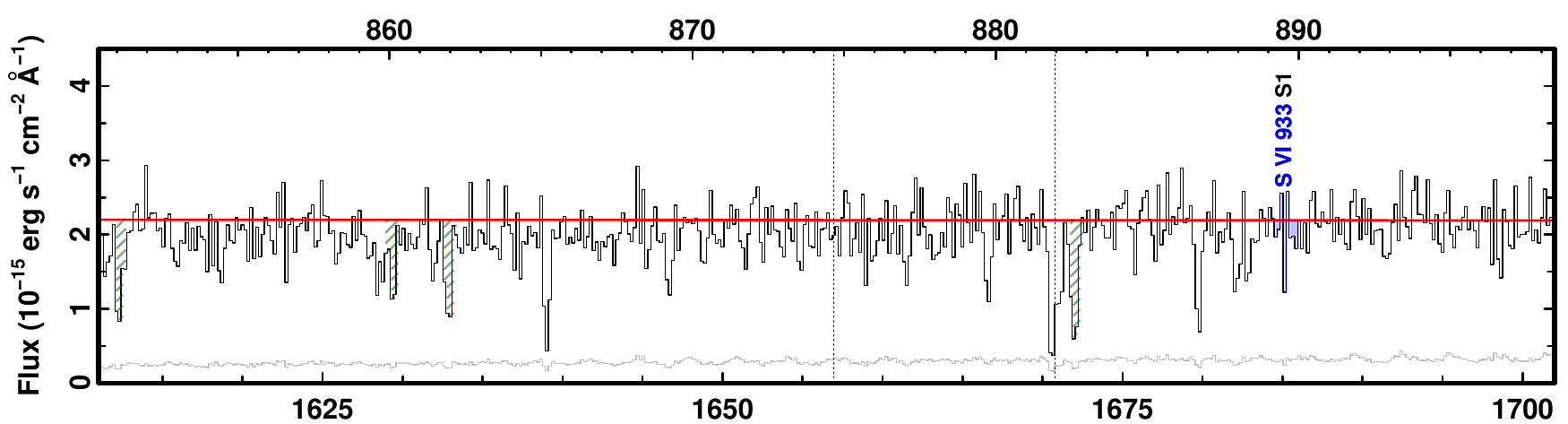}
	\includegraphics[scale=1.0]{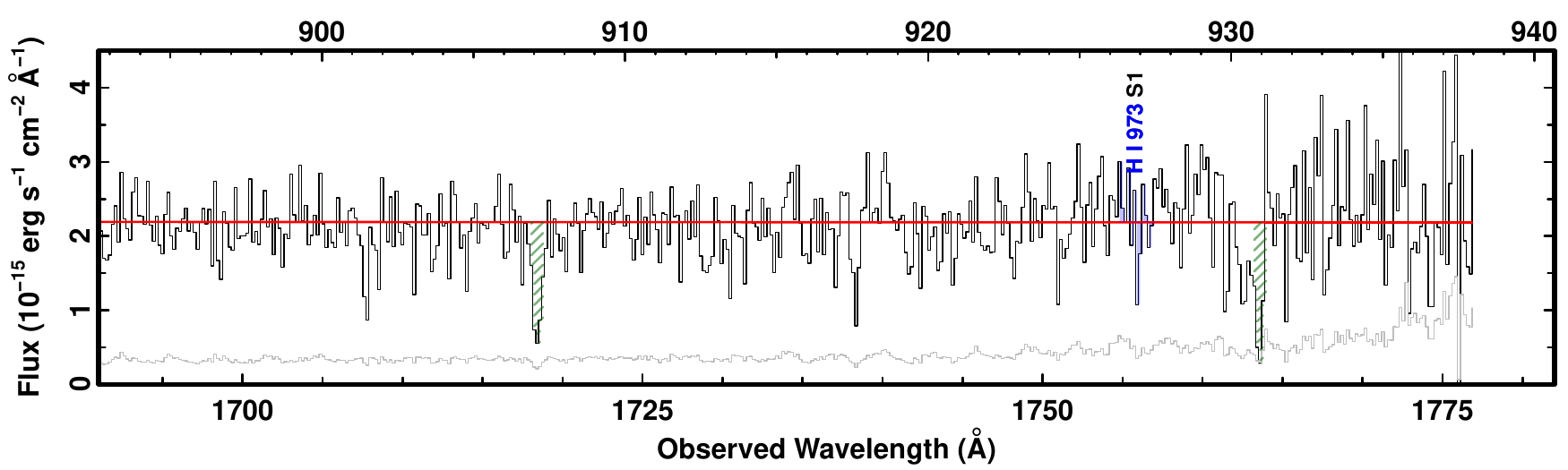}
	\figurenum{1}
	\caption{(Continued.)}
\end{figure*}
\begin{figure*}
	\includegraphics[scale=1.0]{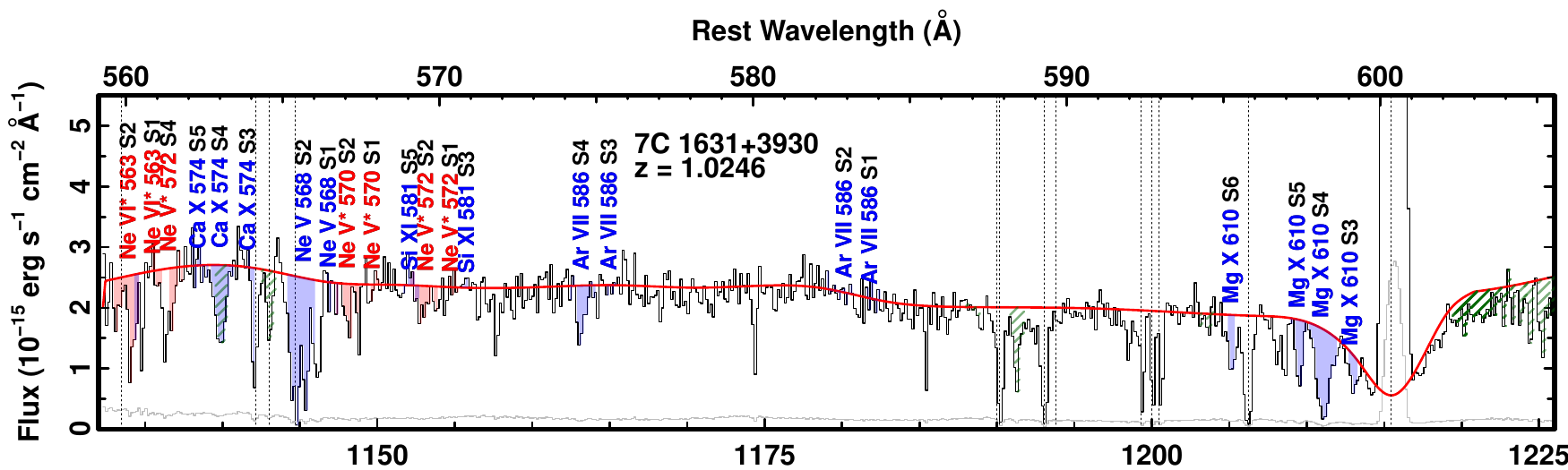}
	\includegraphics[scale=1.0]{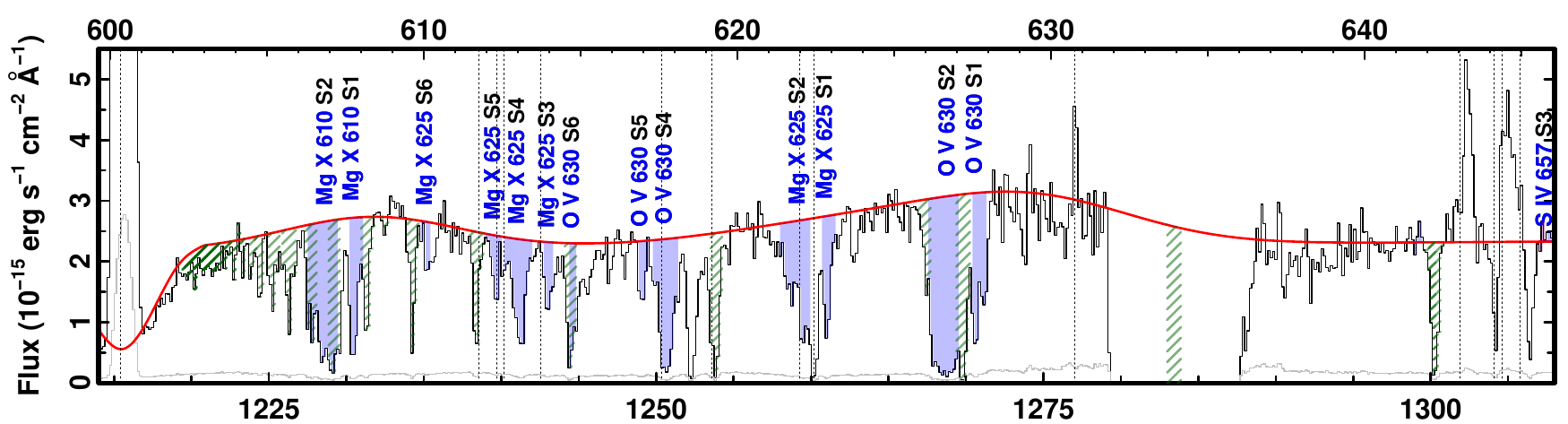}
	\includegraphics[scale=1.0]{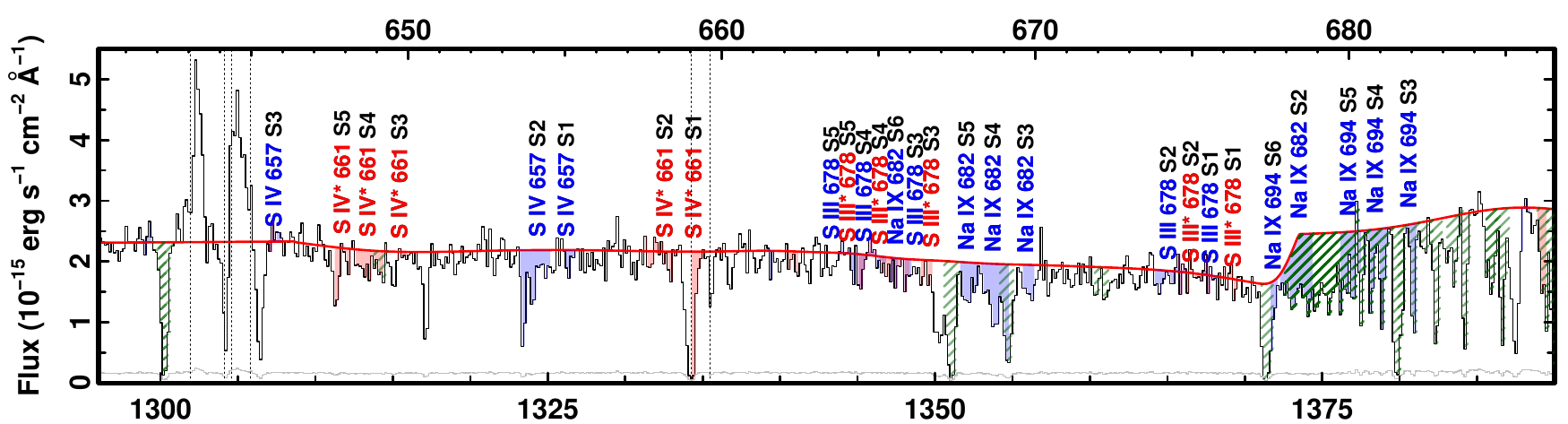}
	\includegraphics[scale=1.0]{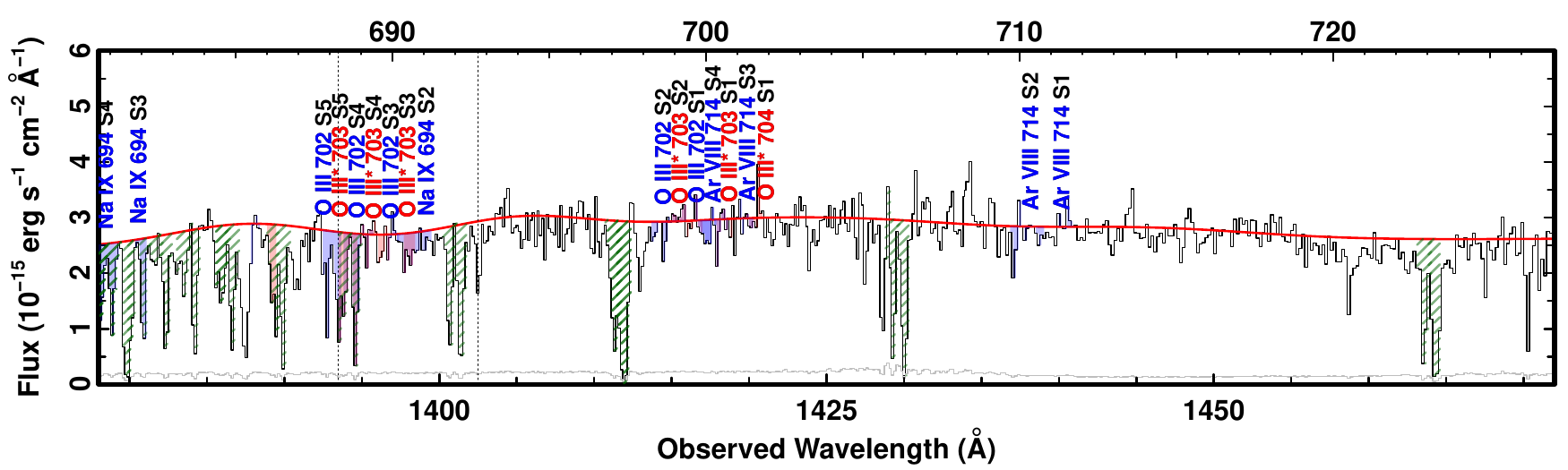}
	\figurenum{1}
	\caption{(Continued.)}
\end{figure*}
\begin{figure*}
	\includegraphics[scale=1.0]{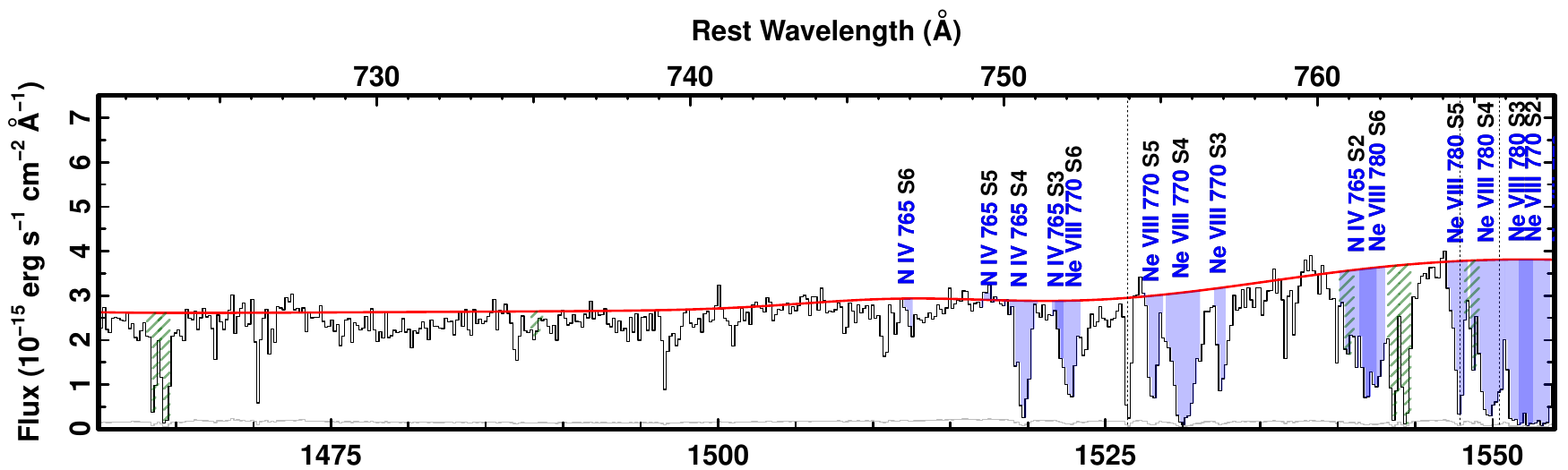}
	\includegraphics[scale=1.0]{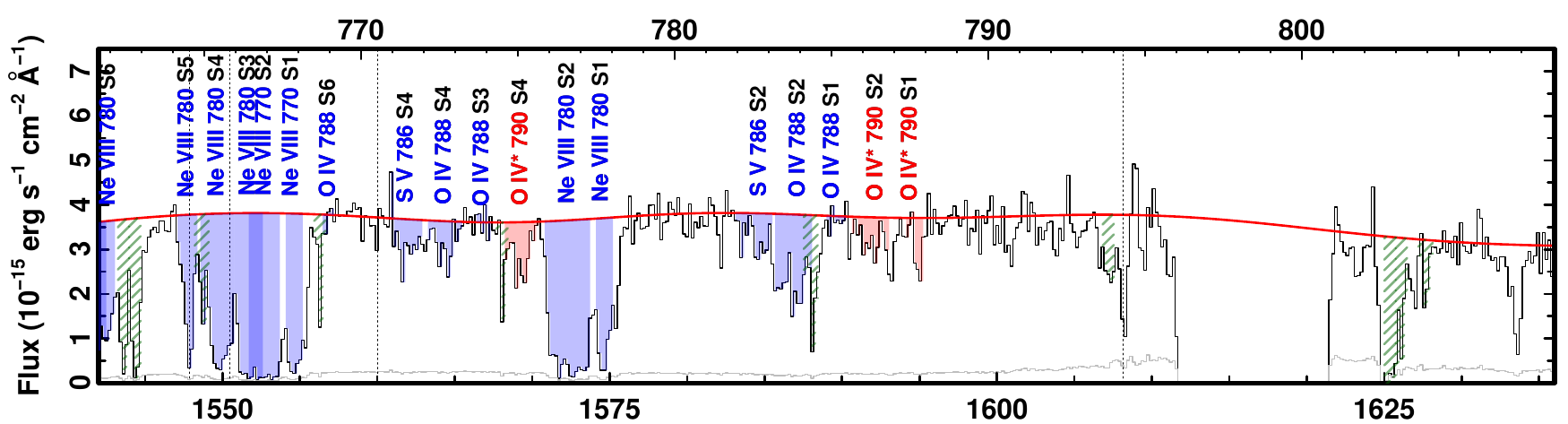}
	\includegraphics[scale=1.0]{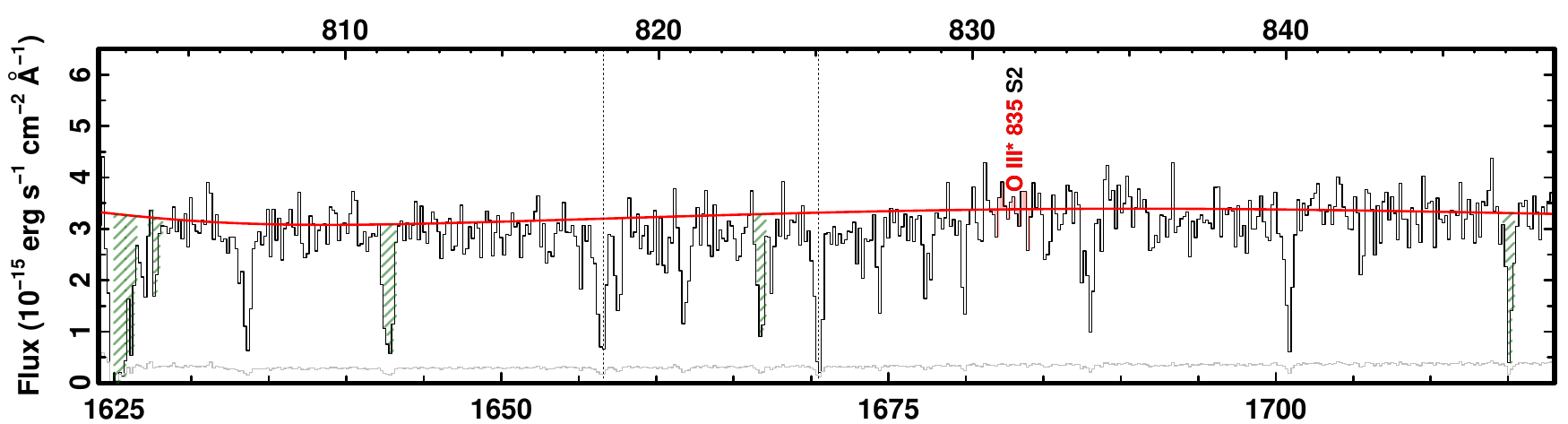}
	\includegraphics[scale=1.0]{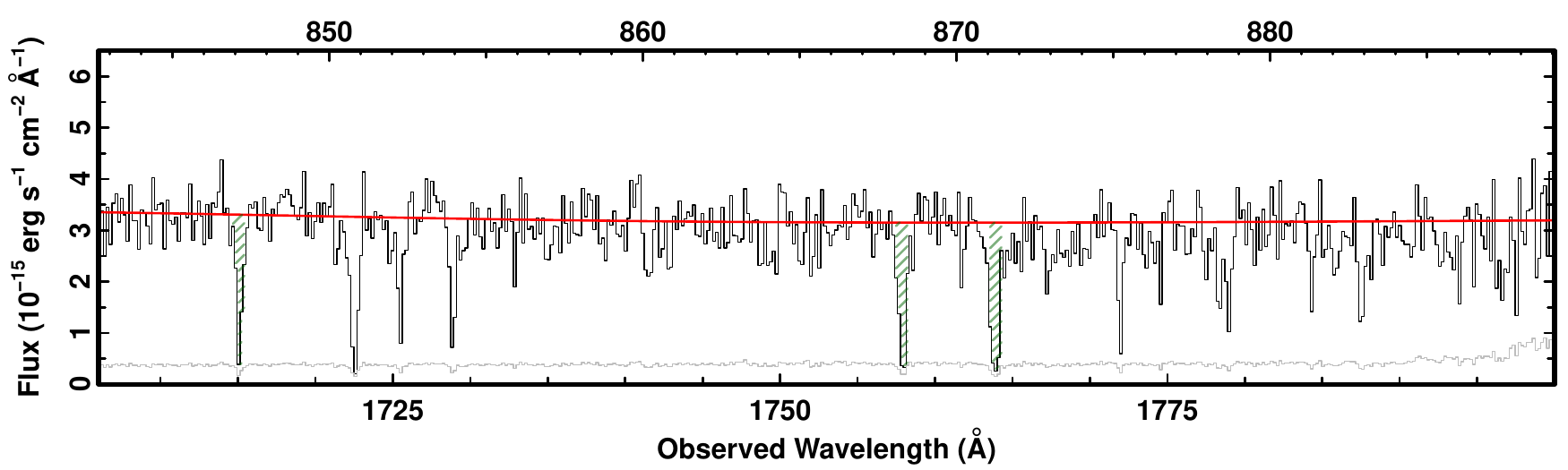}
	\figurenum{1}
	\caption{(Continued.)}
\end{figure*}

\begin{figure*}
	\includegraphics[trim=5mm 0mm 0mm 16mm,clip,scale=0.333]{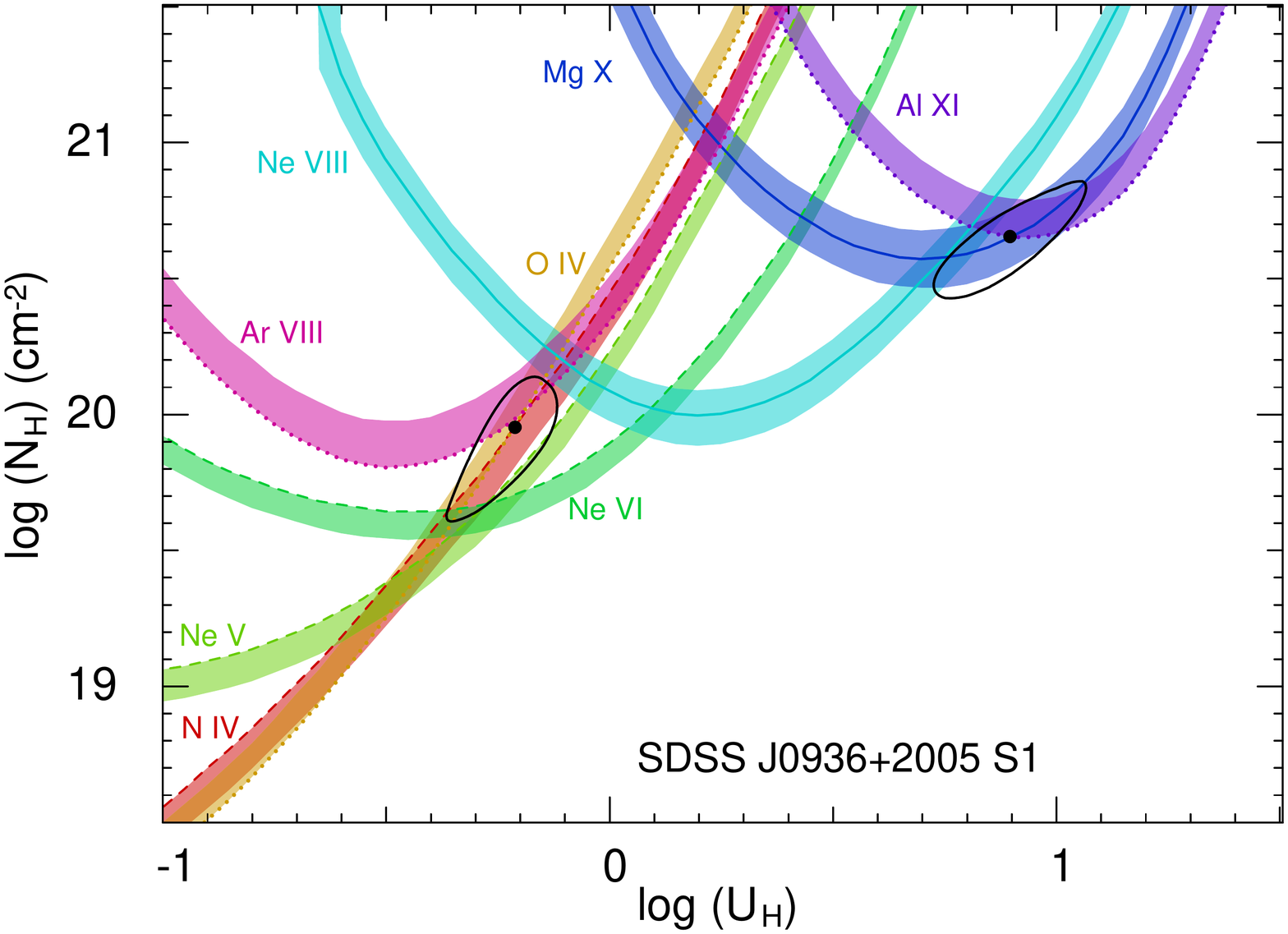}\includegraphics[trim=5mm 0mm 0mm 16mm,clip,scale=0.333]{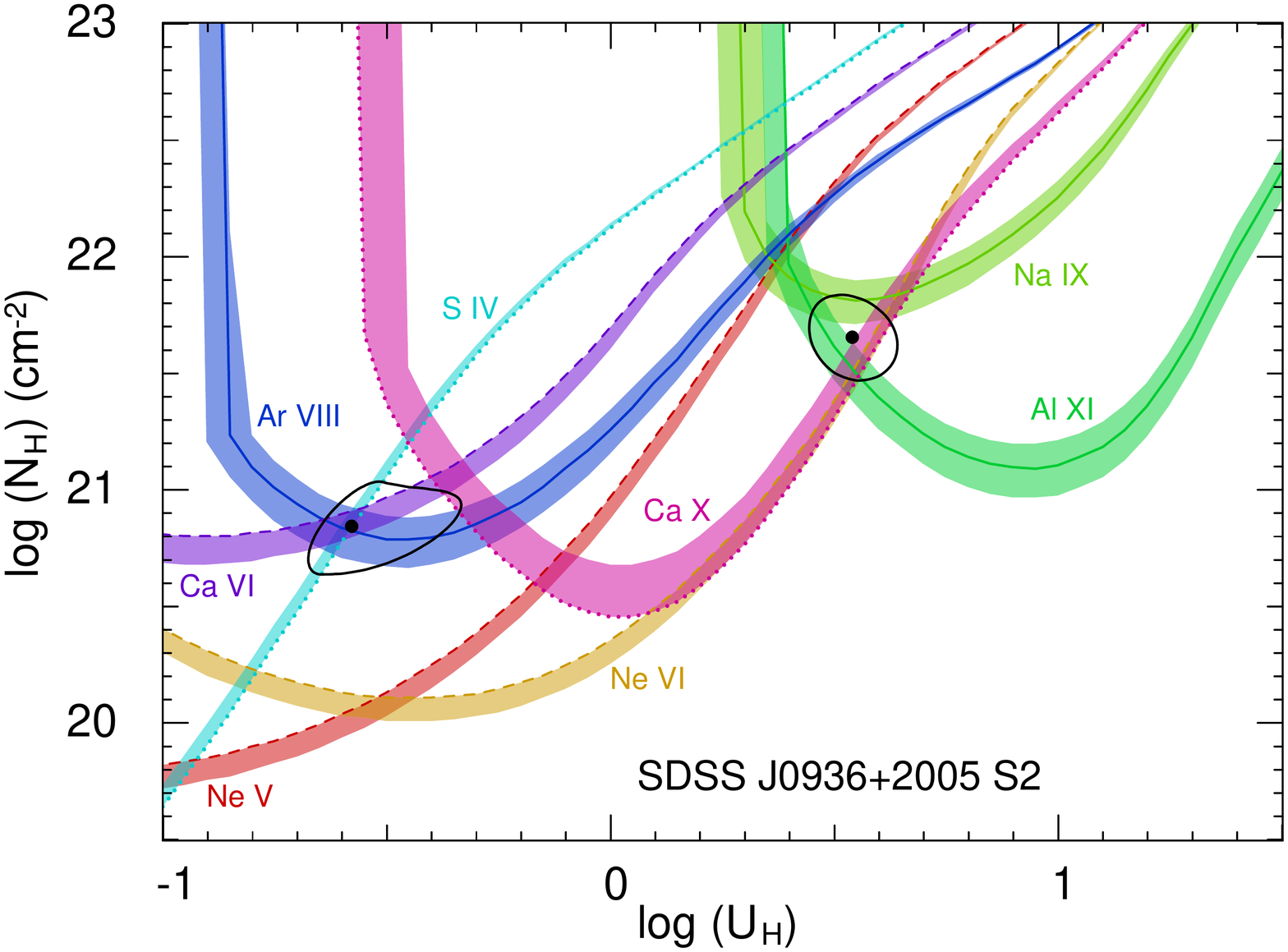}\\
	\includegraphics[trim=5mm 0mm 0mm 16mm,clip,scale=0.333]{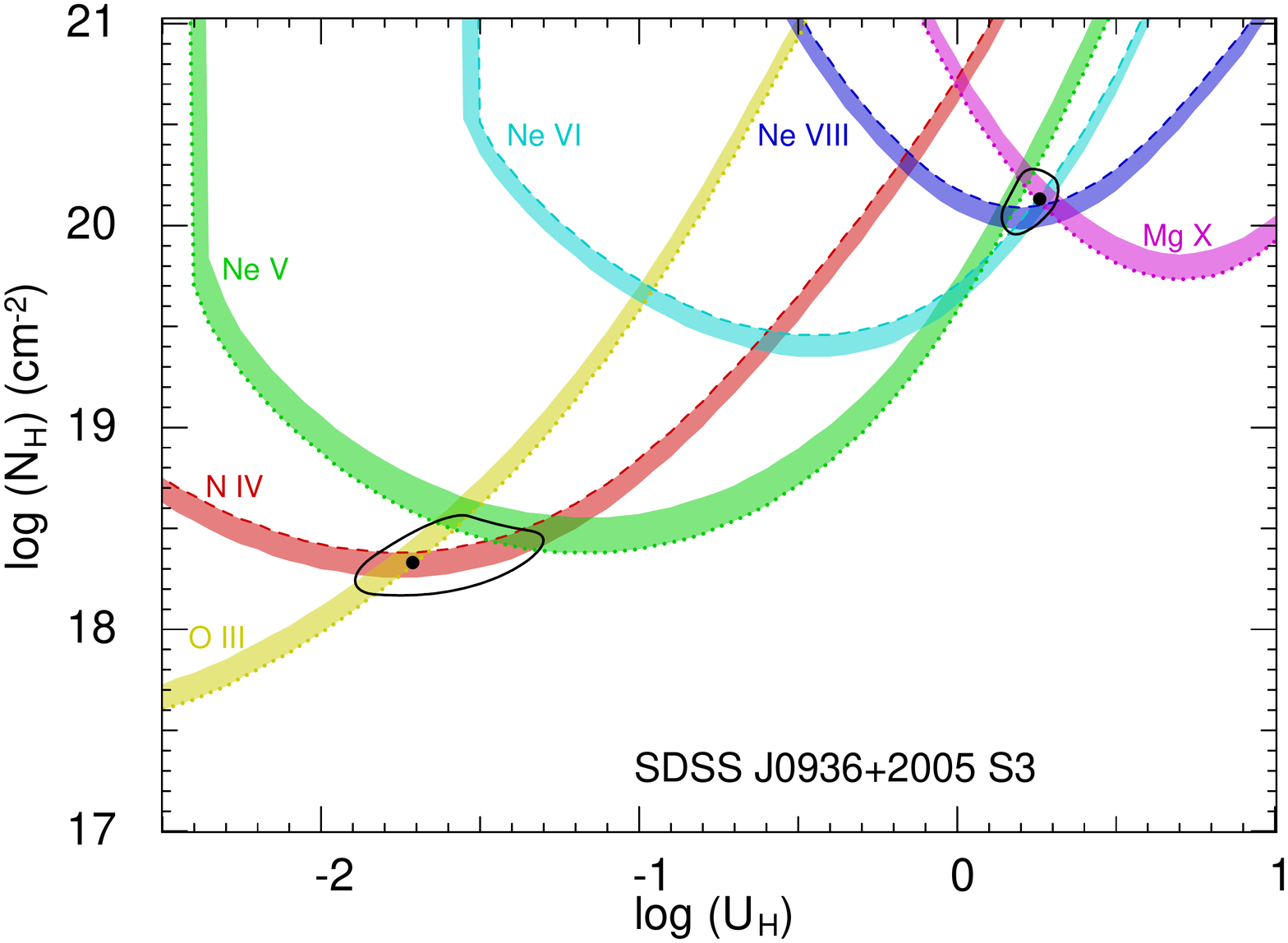}\includegraphics[trim=5mm 0mm 0mm 16mm,clip,scale=0.333]{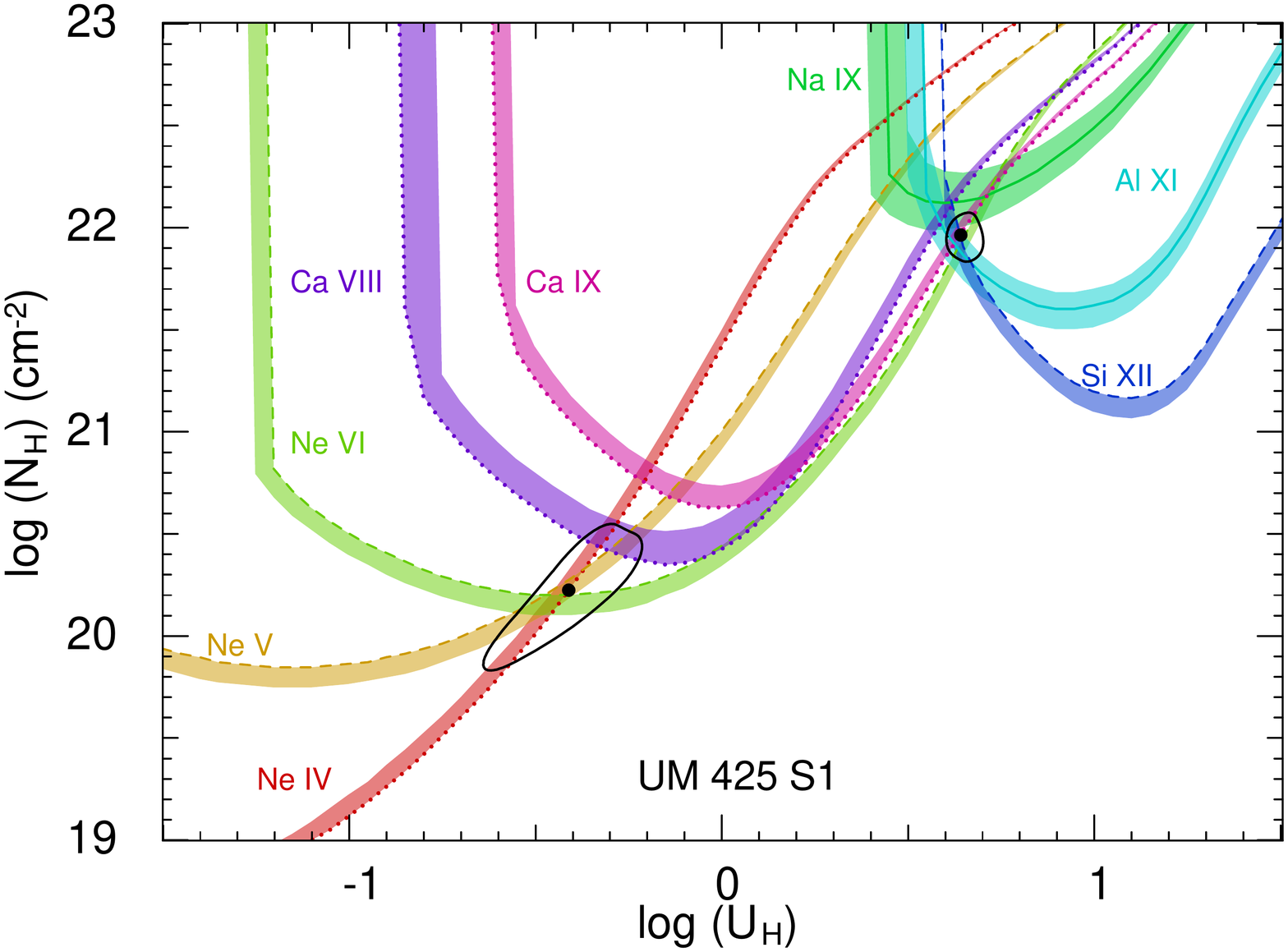}\\
	\includegraphics[trim=5mm 0mm 0mm 16mm,clip,scale=0.333]{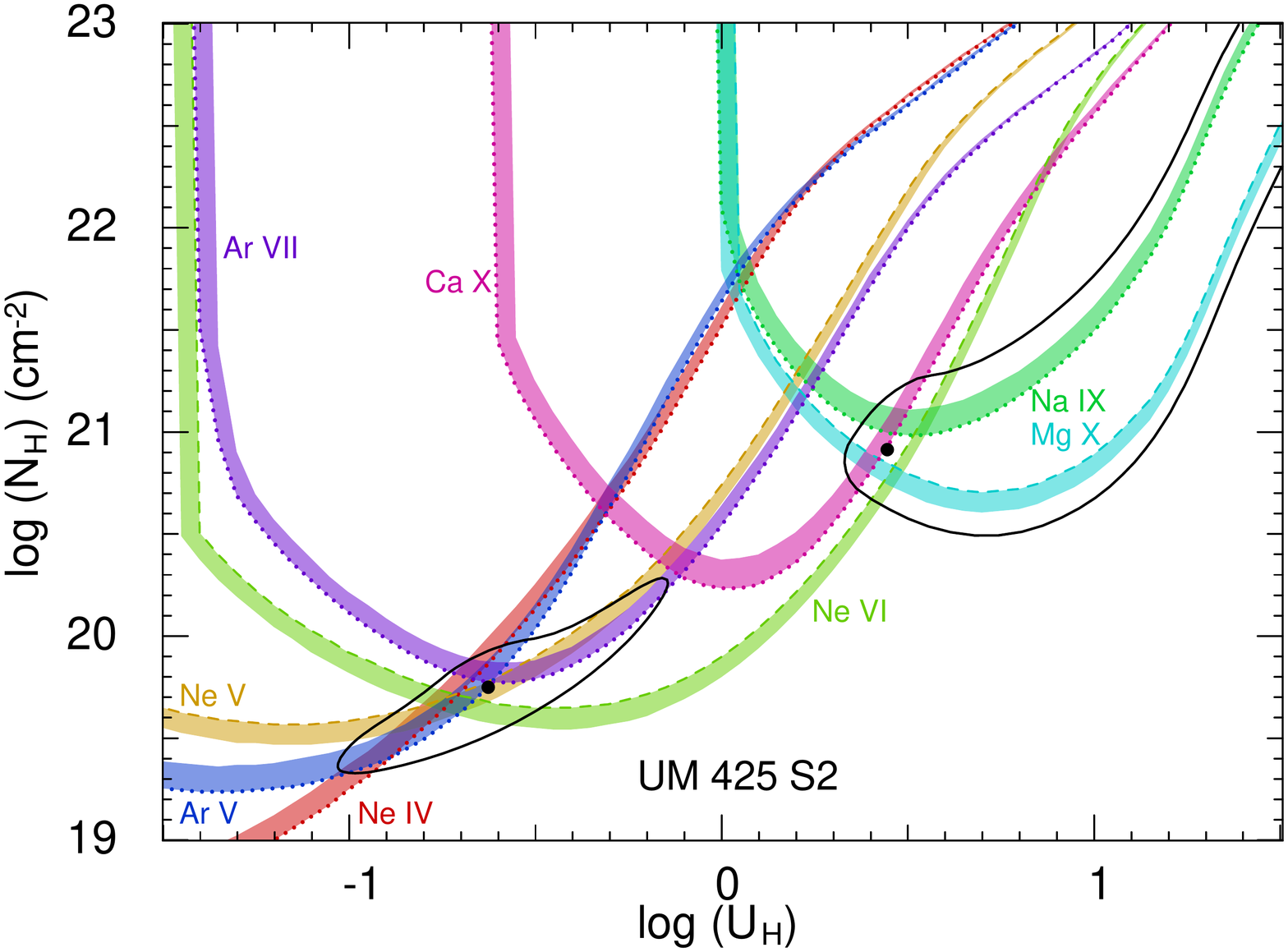}\includegraphics[trim=5mm 0mm 0mm 16mm,clip,scale=0.333]{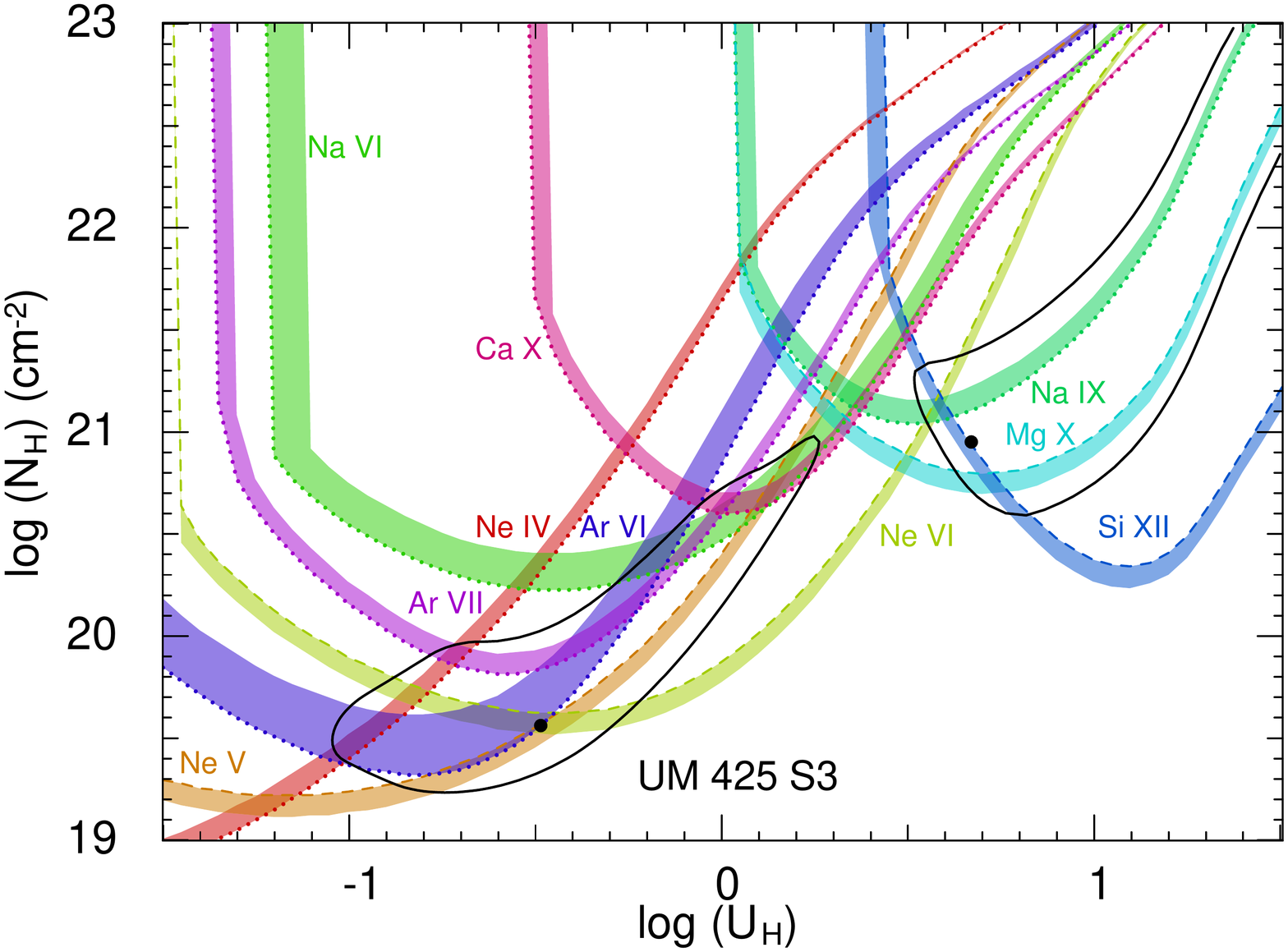}
	\caption{\footnotesize{Photoionization solutions for each outflow system. The colored contours show the model parameters that are consistent with the observed values assuming the HE0238 SED and solar metallicity (for VV2006 J1329+5405, Z = 4.68 Z$_{\astrosun}$; see Table~3 of Paper V). Solid contours represent ionic column densities taken as measurements while dotted and dashed contours are upper and lower limits, respectively. The shaded bands are the 1$\sigma$ uncertainties for each contour (see Table~\ref{tab:col}). The dots are the best $\chi^2$-minimization solutions for each ionization phase and the ellipses encircling them are their 1$\sigma$ uncertainties.}} 
\label{fig:sol}
\end{figure*}
\begin{figure*}
	\includegraphics[trim=5mm 0mm 0mm 16mm,clip,scale=0.333]{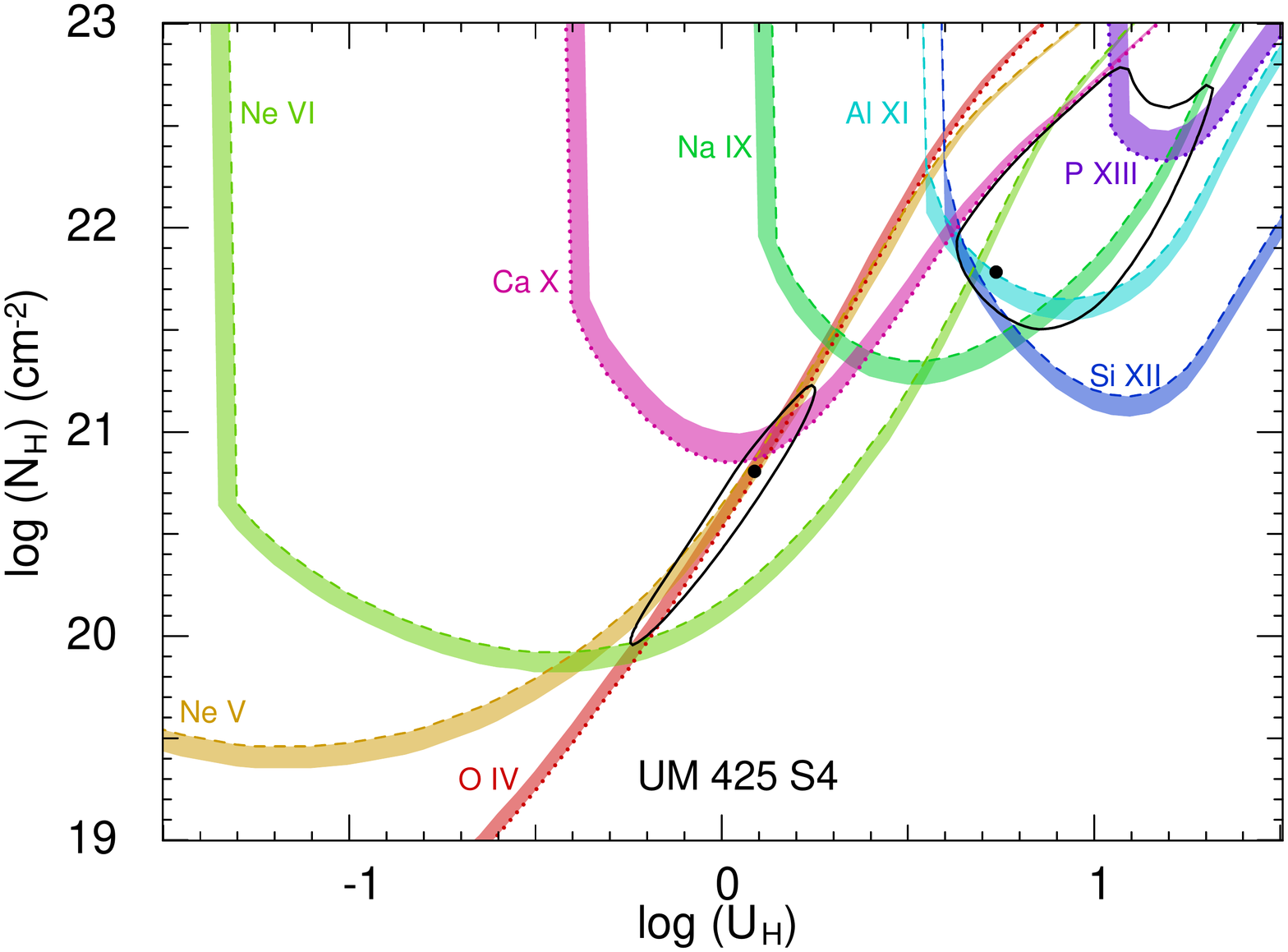}\includegraphics[trim=5mm 0mm 0mm 16mm,clip,scale=0.333]{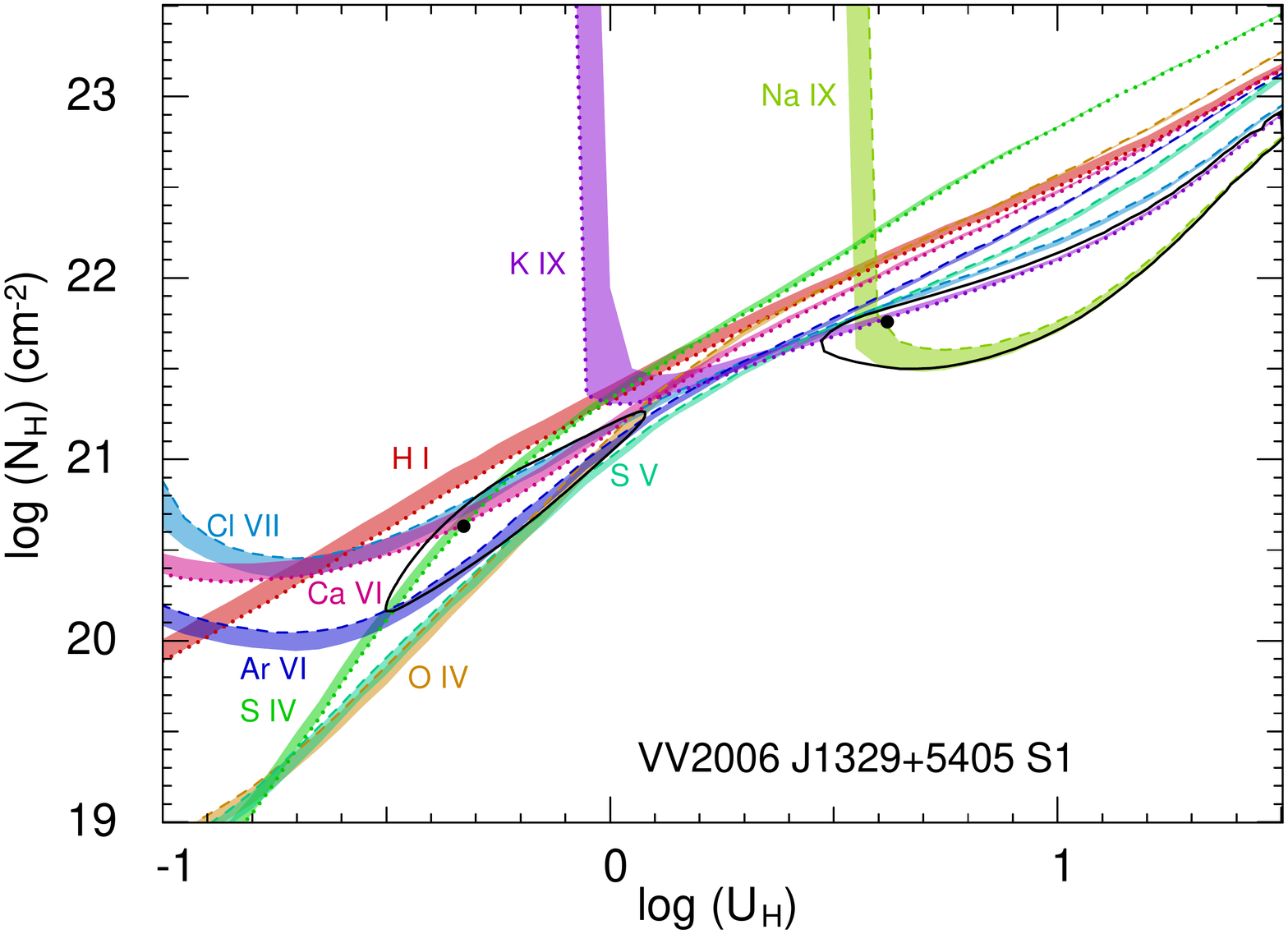}\\
	\includegraphics[trim=5mm 0mm 0mm 16mm,clip,scale=0.333]{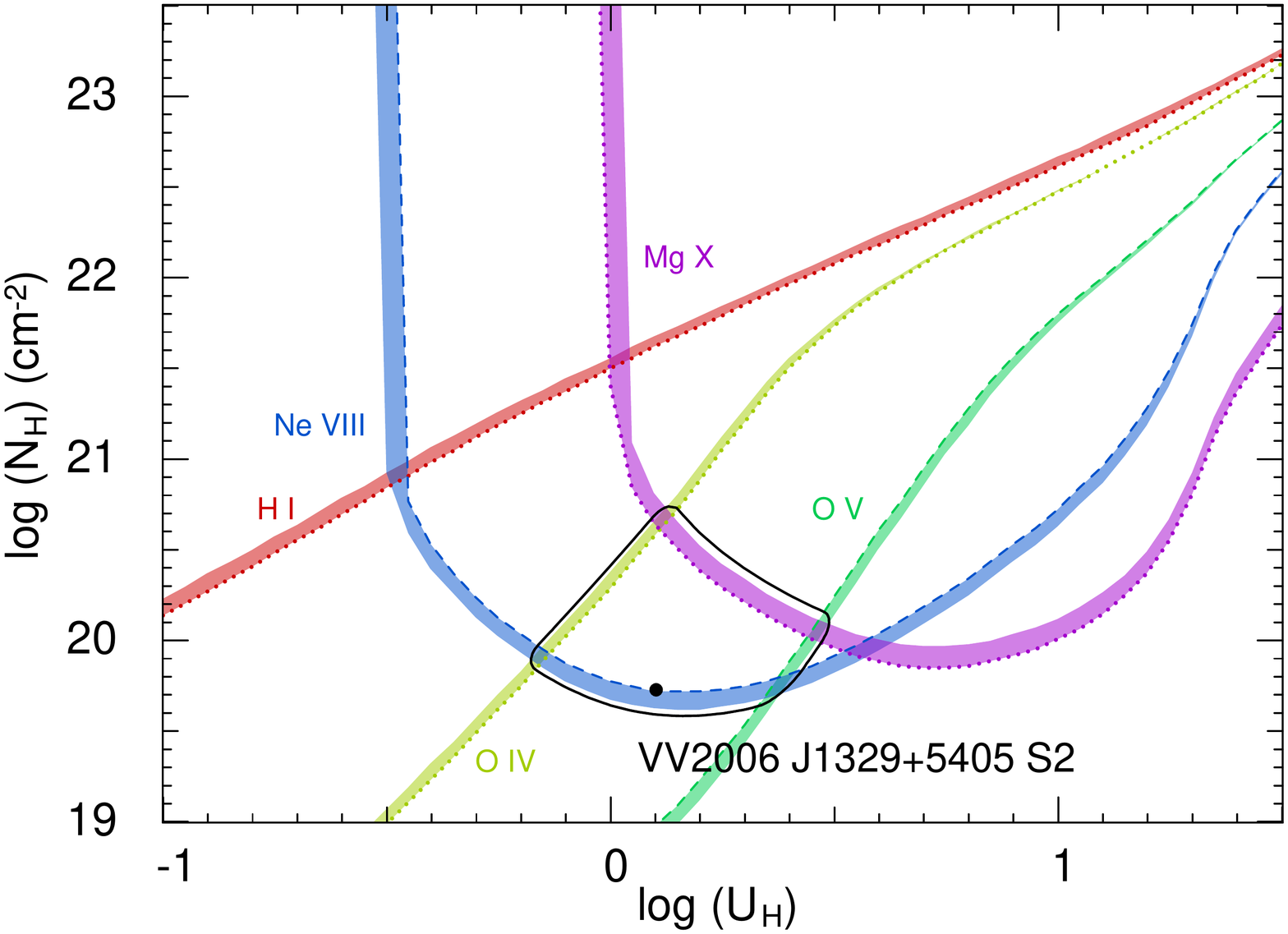}\includegraphics[trim=5mm 0mm 0mm 16mm,clip,scale=0.333]{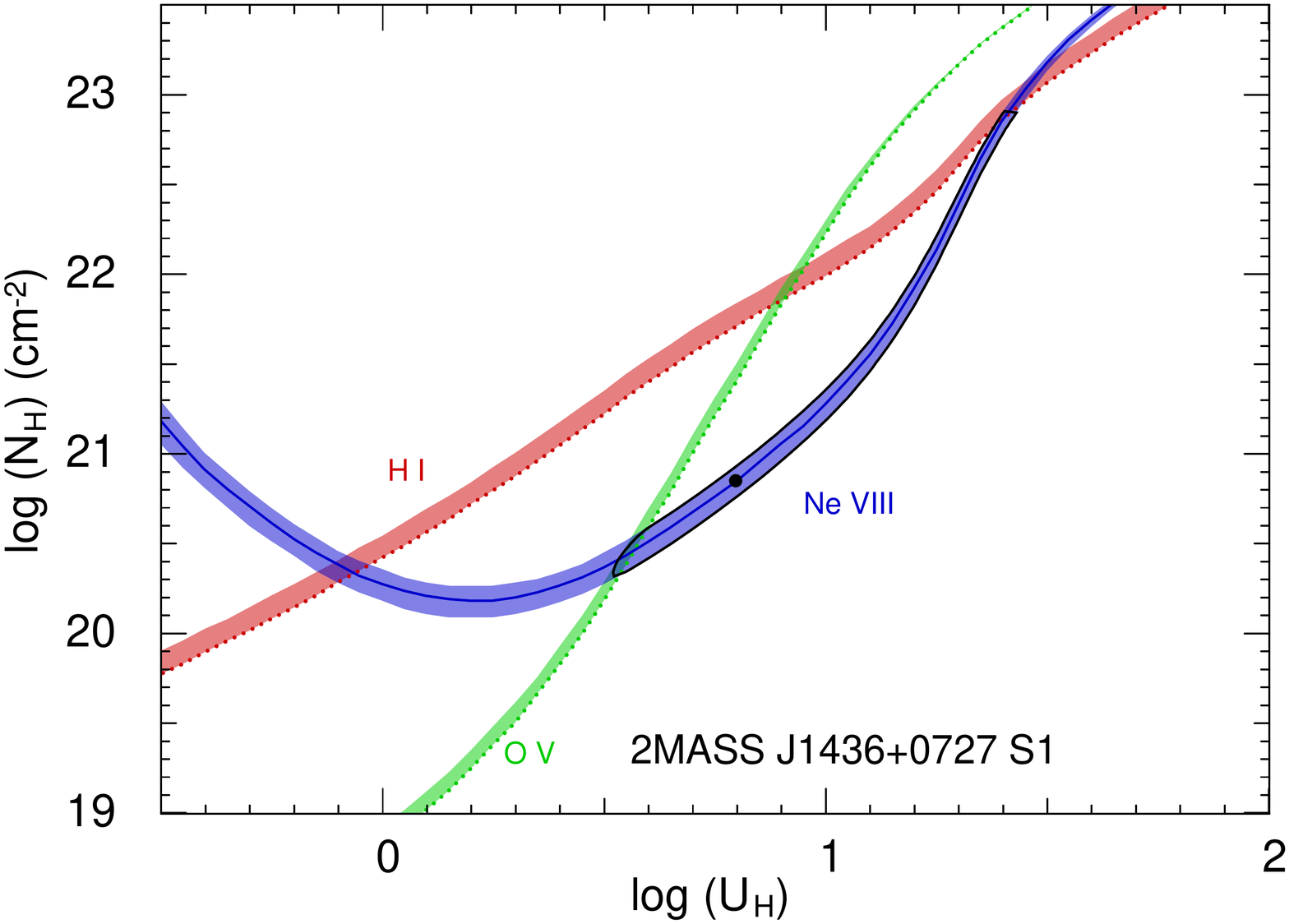}\\
	\includegraphics[trim=5mm 0mm 0mm 16mm,clip,scale=0.333]{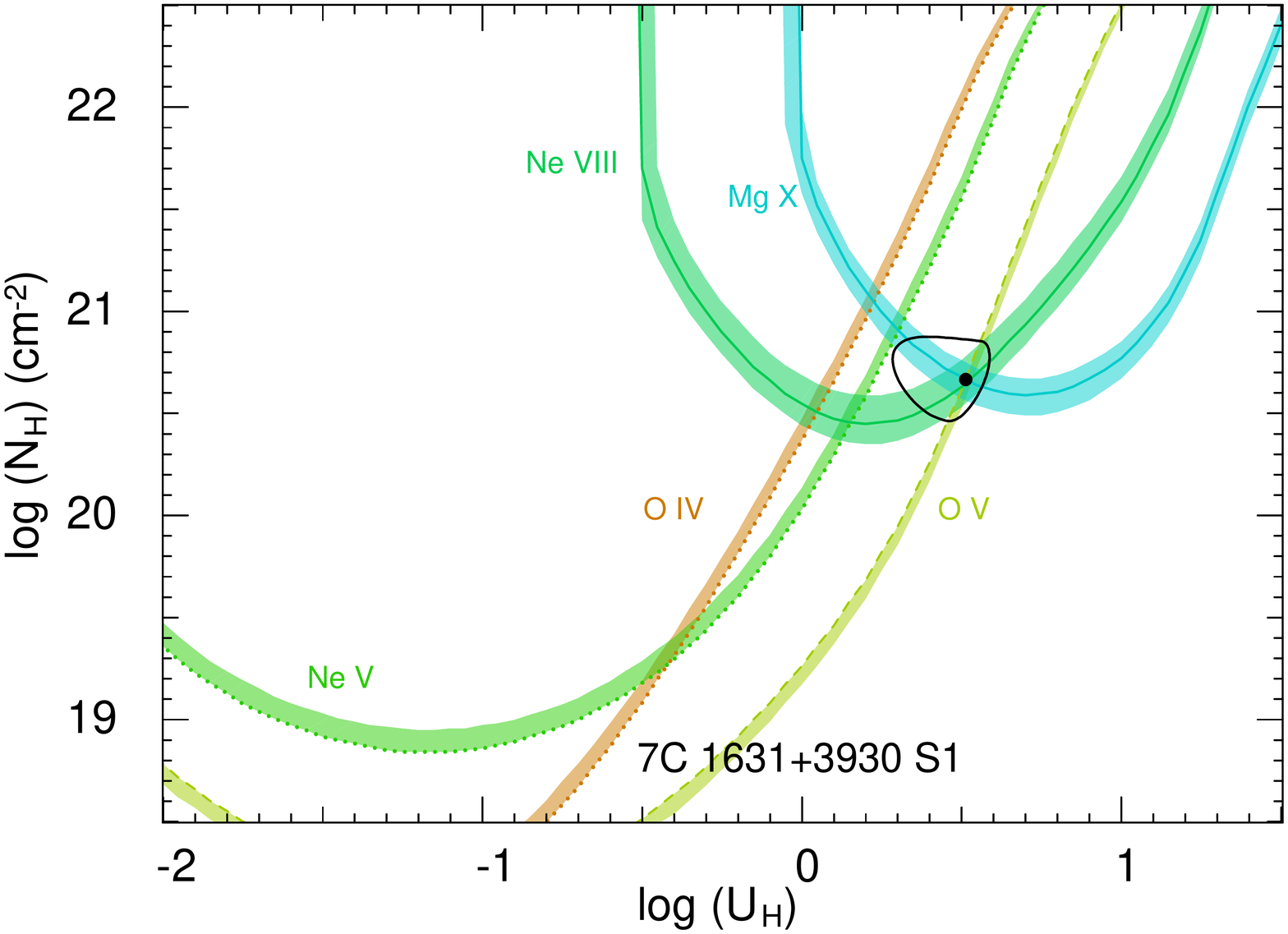}\includegraphics[trim=5mm 0mm 0mm 16mm,clip,scale=0.333]{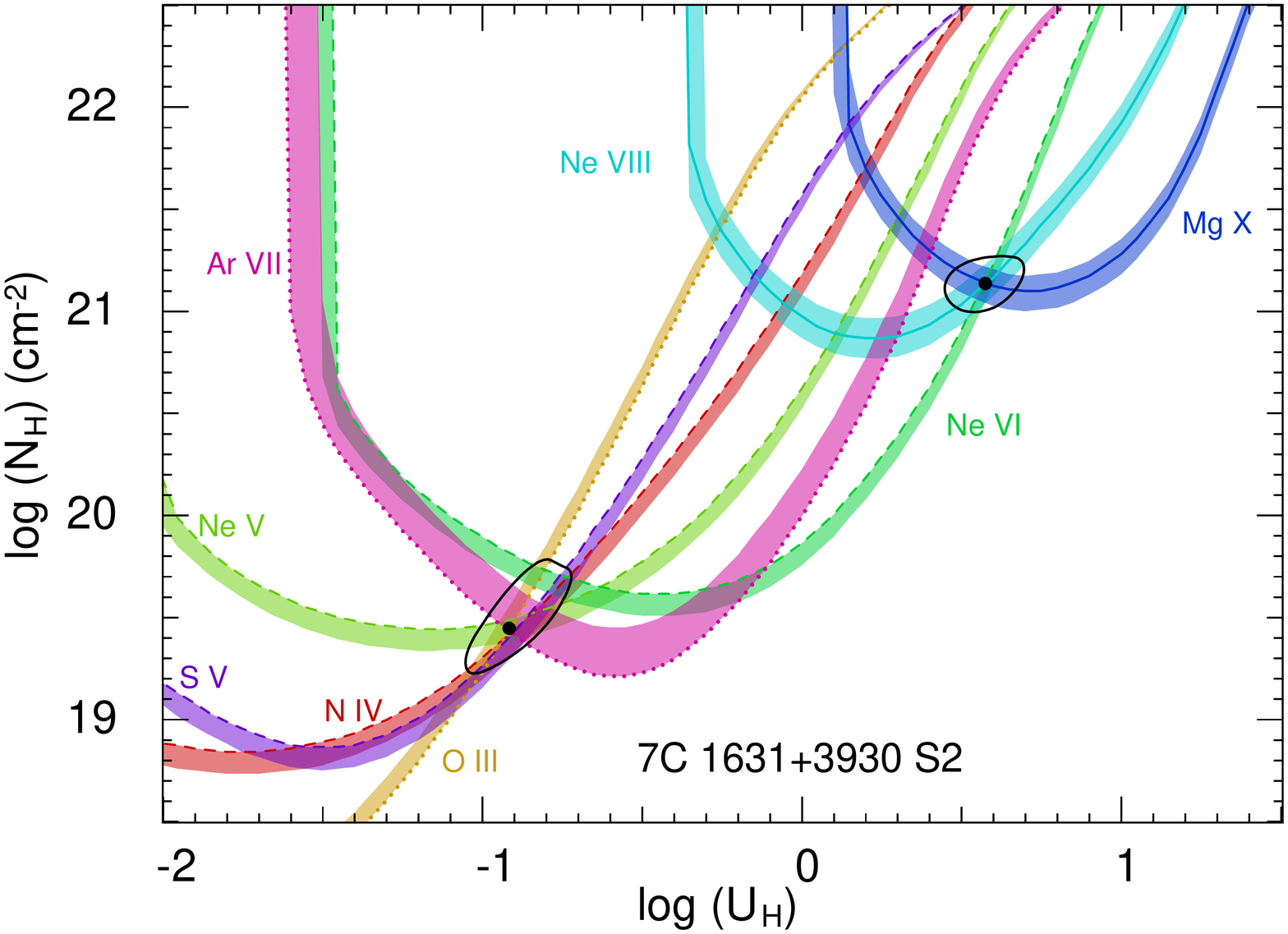}\\
	\figurenum{2}
	\caption{(Continued.)}
\end{figure*}
\begin{figure*}
	\includegraphics[trim=5mm 0mm 0mm 16mm,clip,scale=0.333]{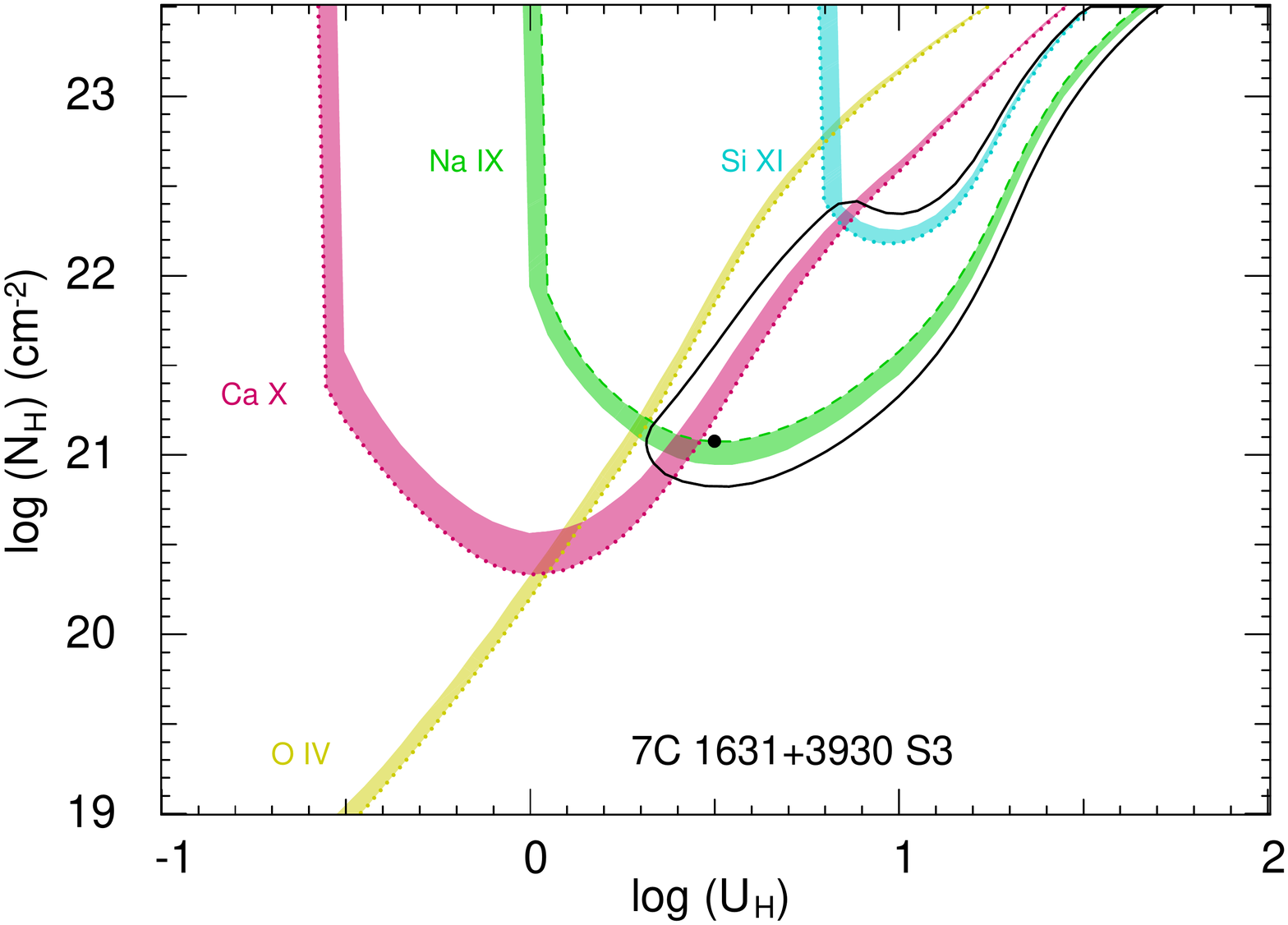}\includegraphics[trim=5mm 0mm 0mm 16mm,clip,scale=0.333]{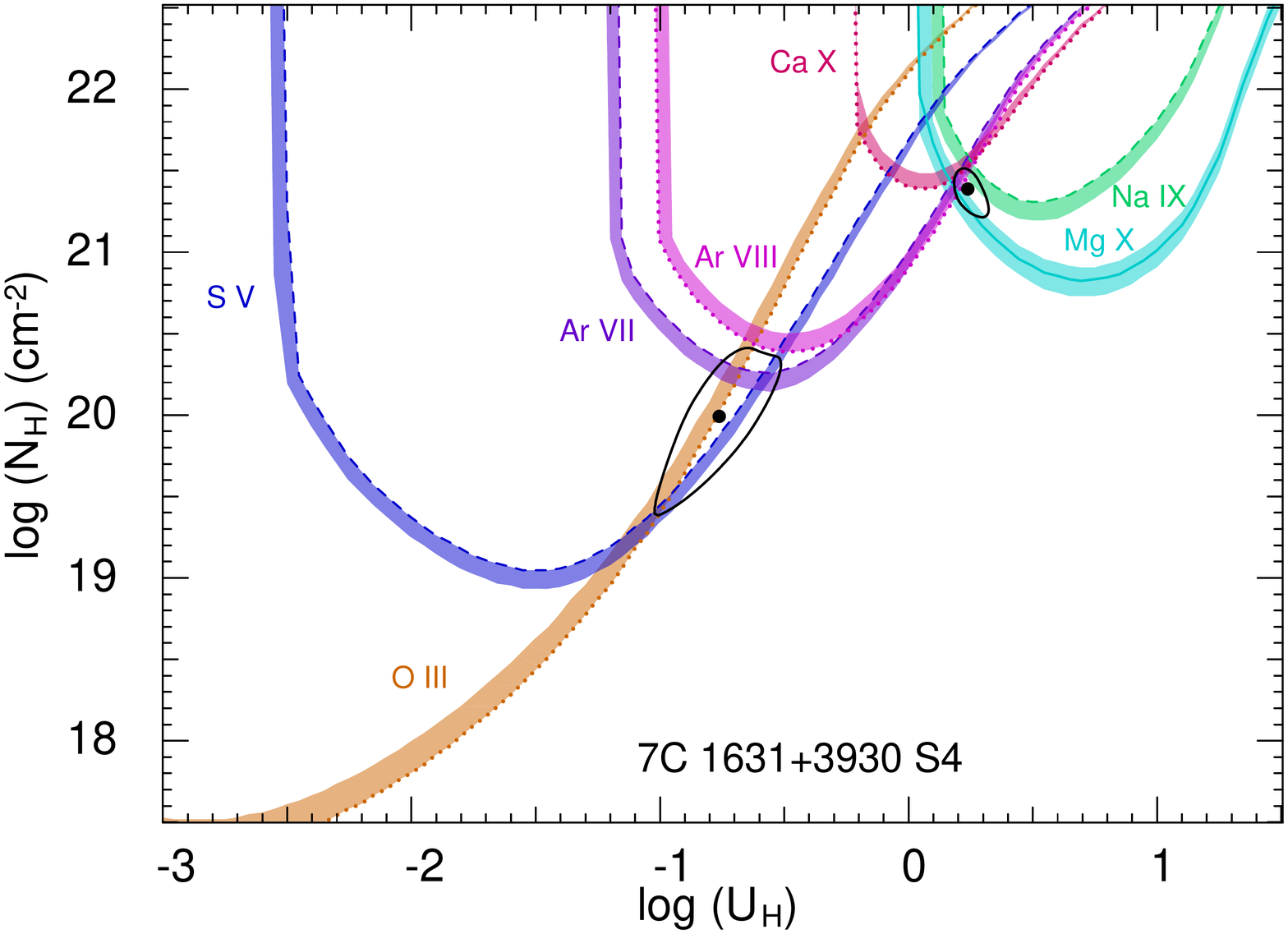}\\
	\includegraphics[trim=5mm 0mm 0mm 16mm,clip,scale=0.333]{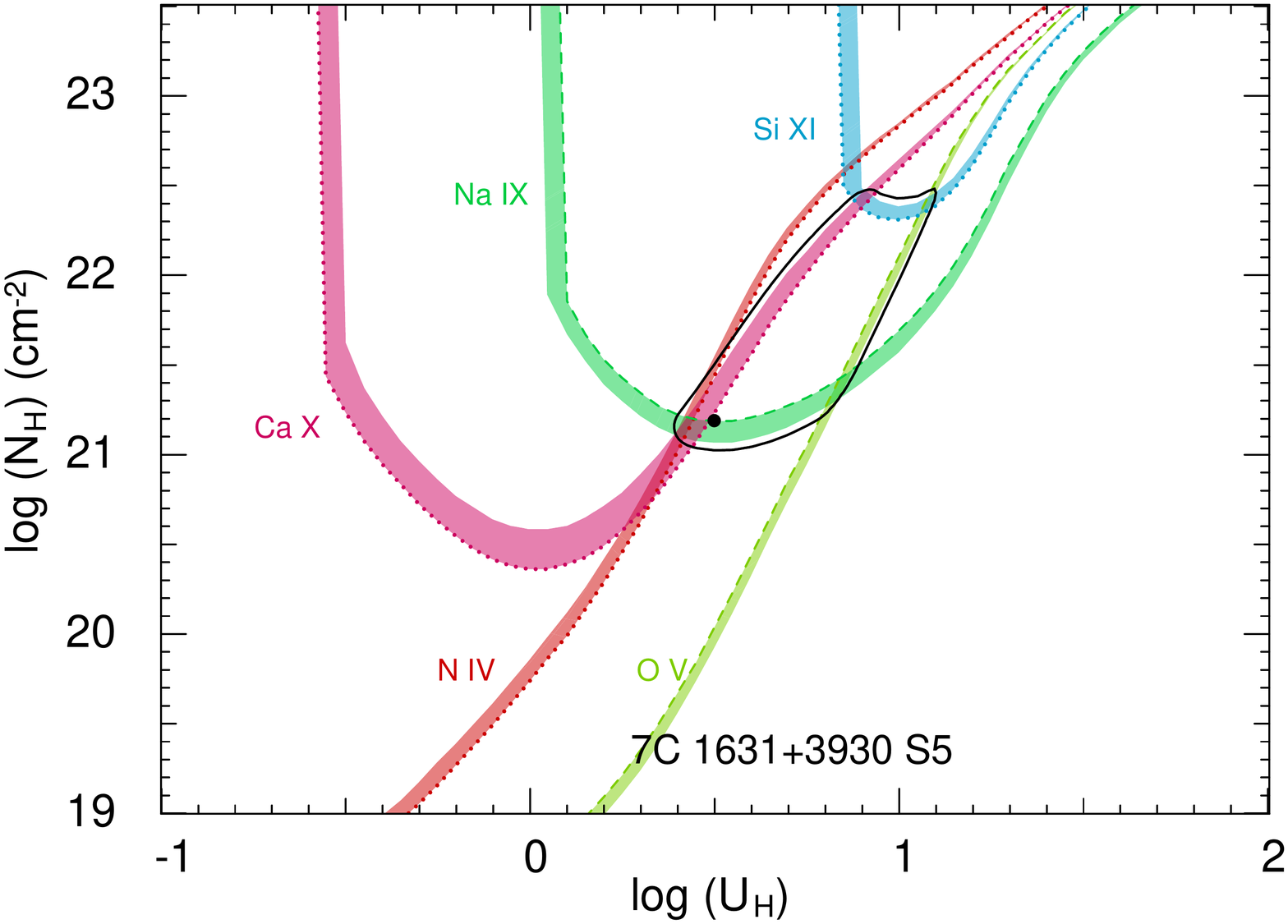}\includegraphics[trim=5mm 0mm 0mm 16mm,clip,scale=0.333]{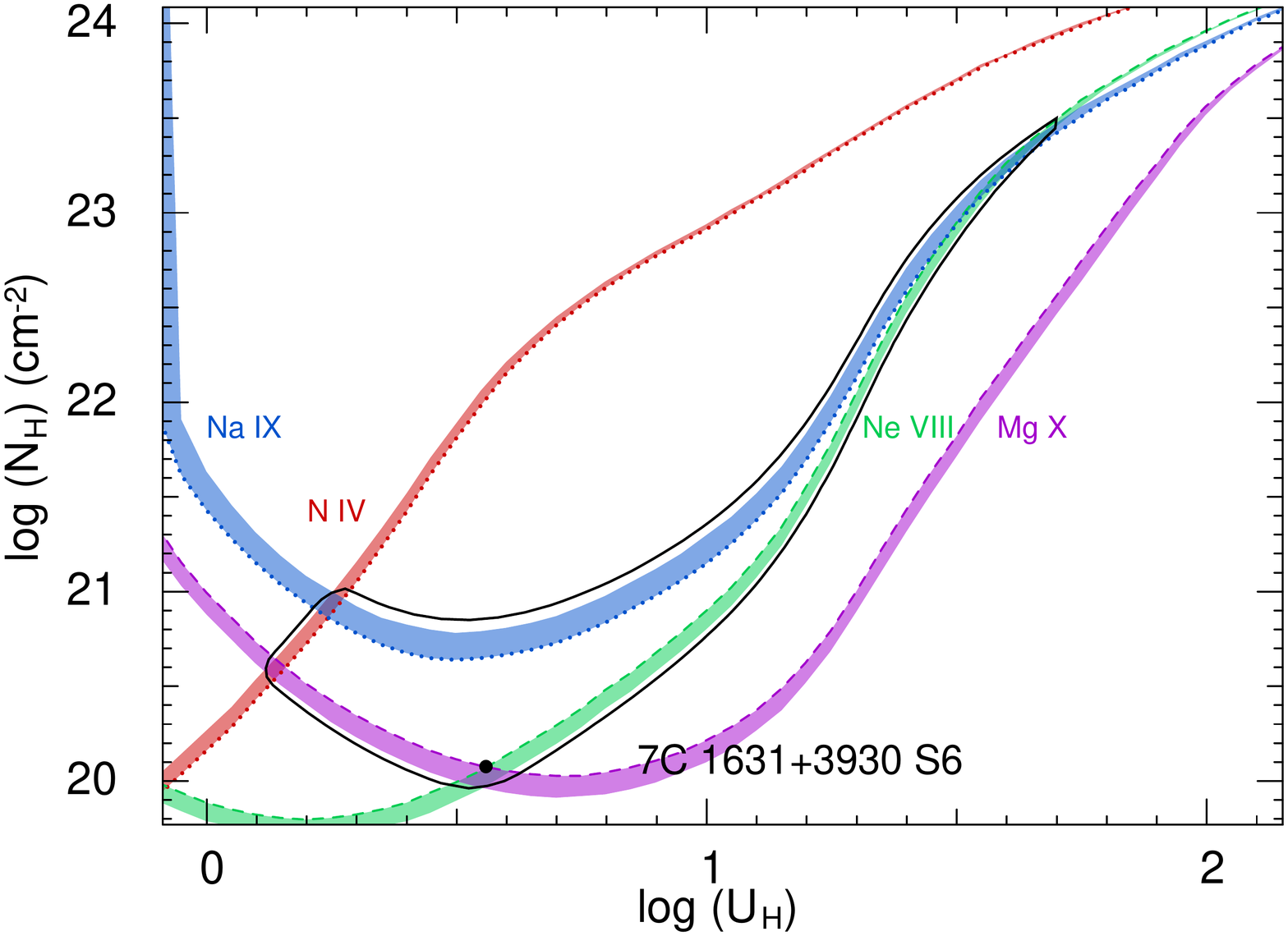}	
	\figurenum{2}
	\caption{(Continued.)}
\end{figure*}

\begin{figure*}
	\includegraphics[trim=6mm 9mm 6mm 5mm,clip,scale=0.37]{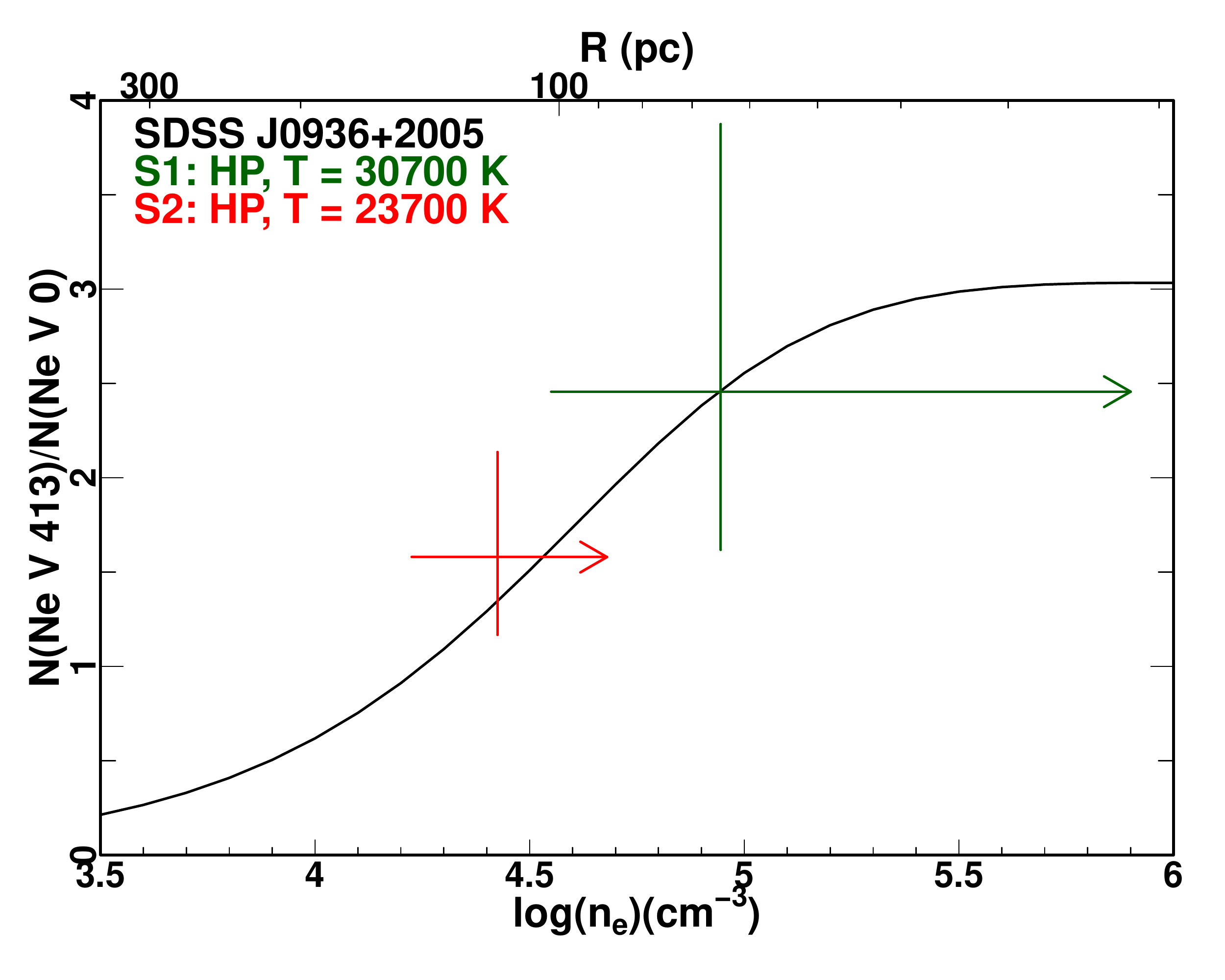}\includegraphics[trim=6mm 9mm 6mm 5mm,clip,scale=0.37]{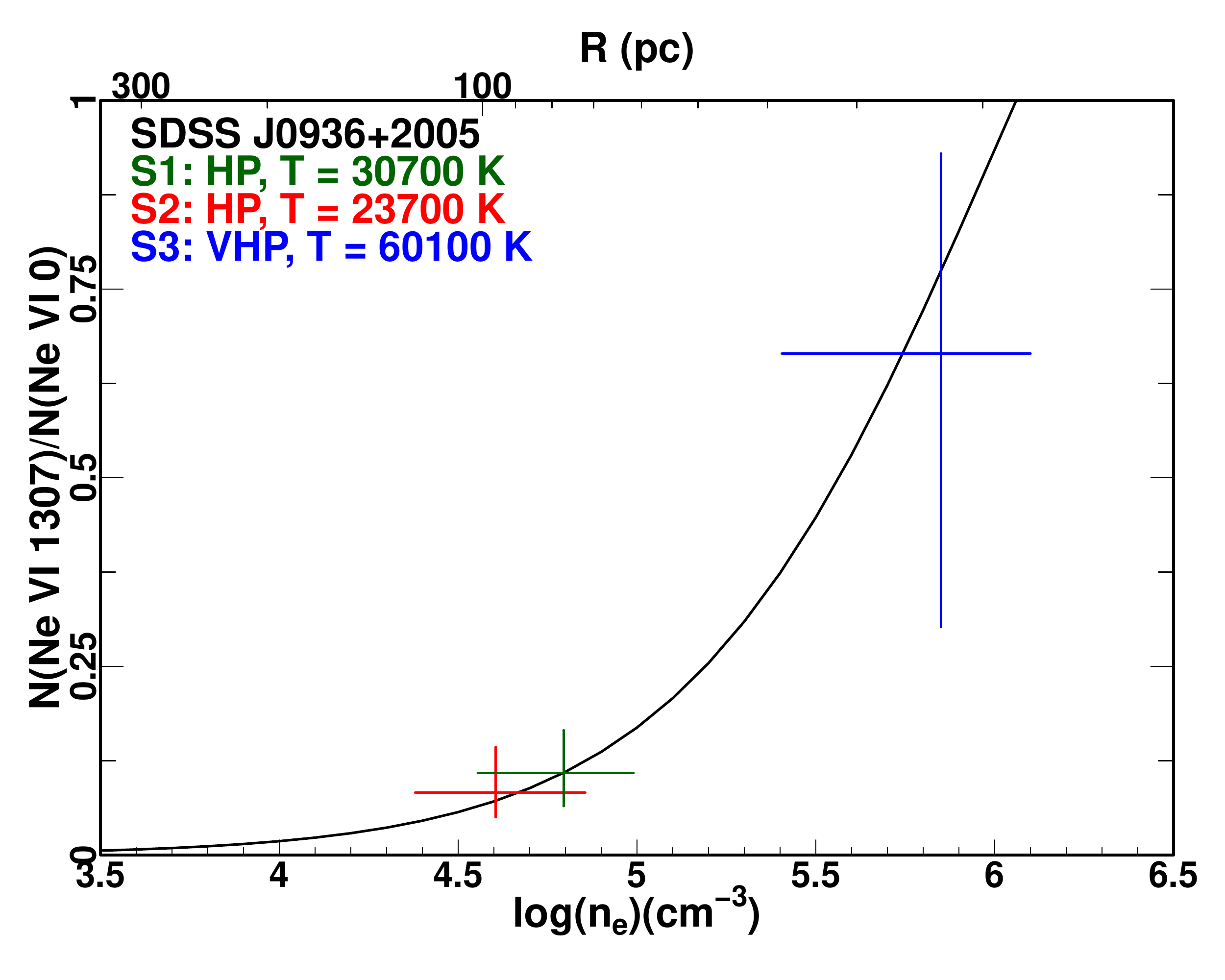}\\
	\includegraphics[trim=6mm 9mm 6mm 5mm,clip,scale=0.37]{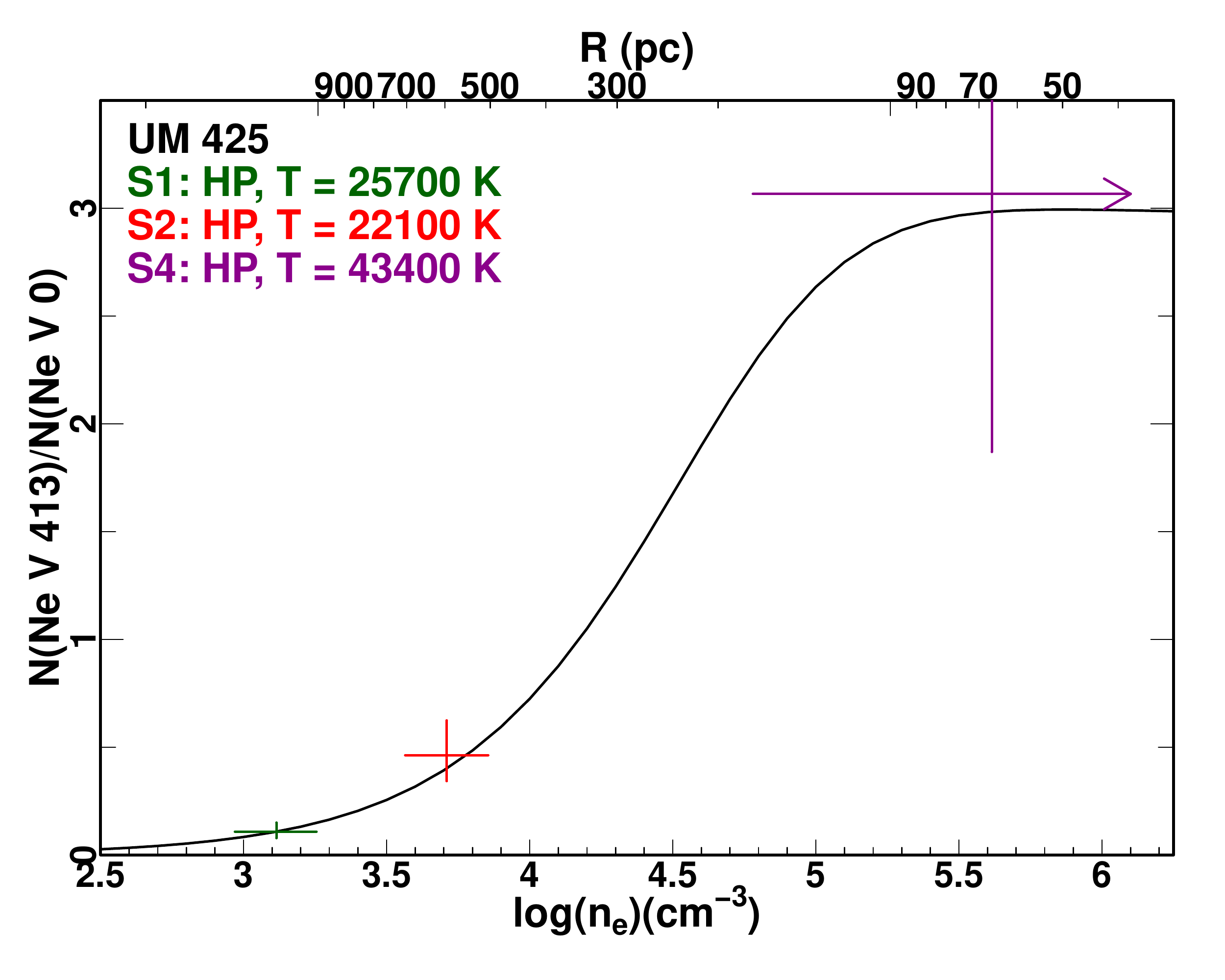}\includegraphics[trim=6mm 9mm 6mm 5mm,clip,scale=0.37]{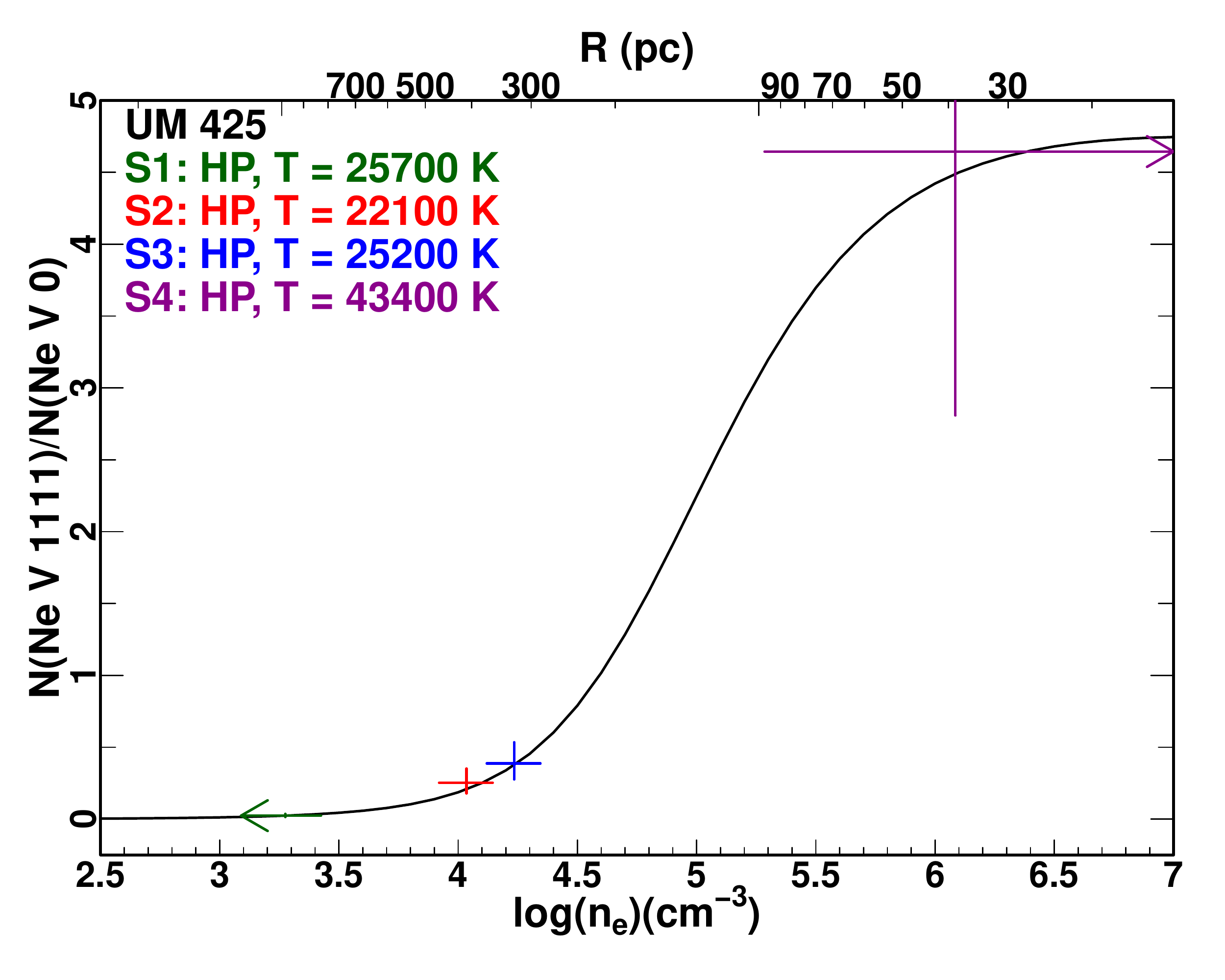}\\
	\includegraphics[trim=6mm 9mm 6mm 5mm,clip,scale=0.37]{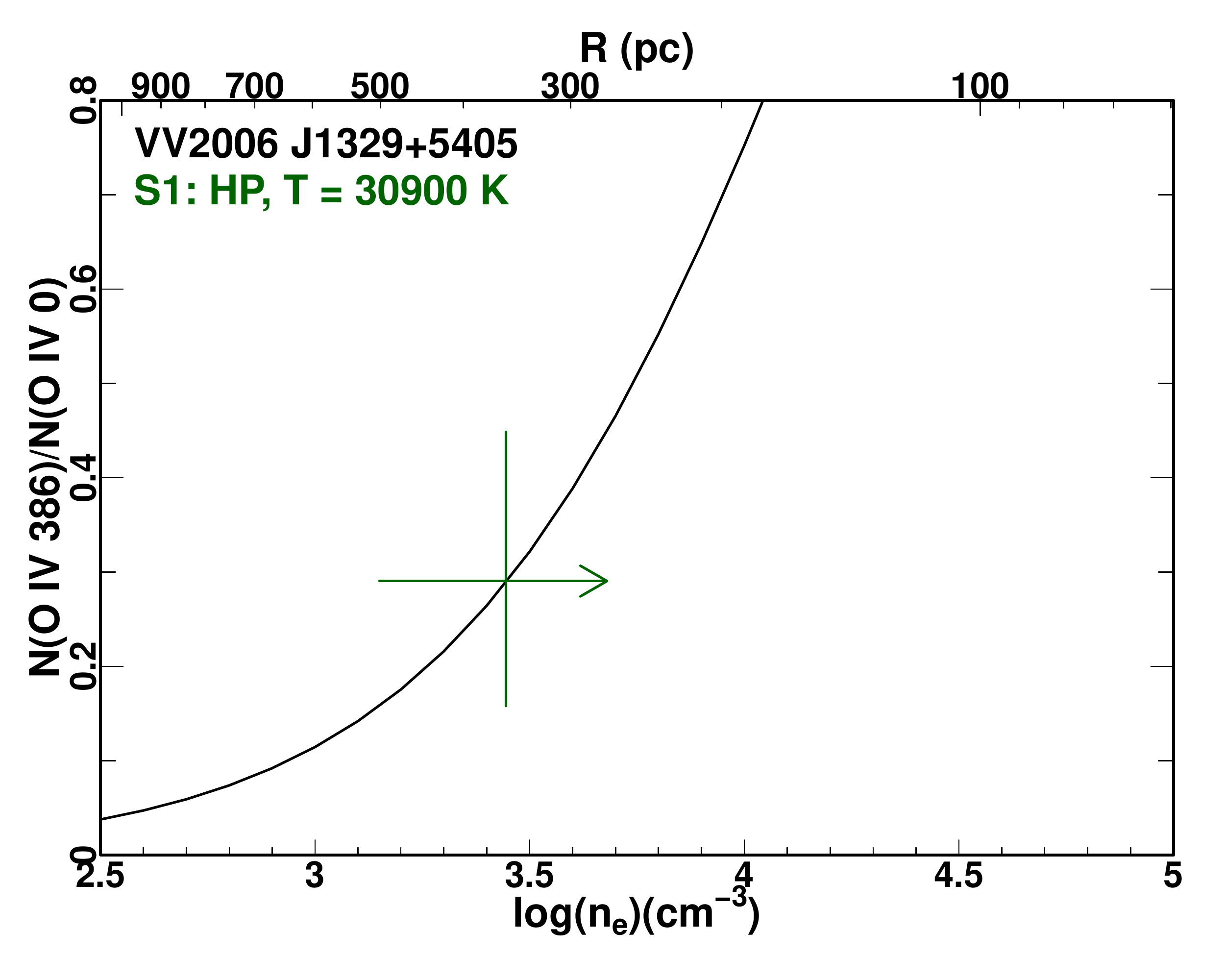}\includegraphics[trim=6mm 9mm 6mm 5mm,clip,scale=0.37]{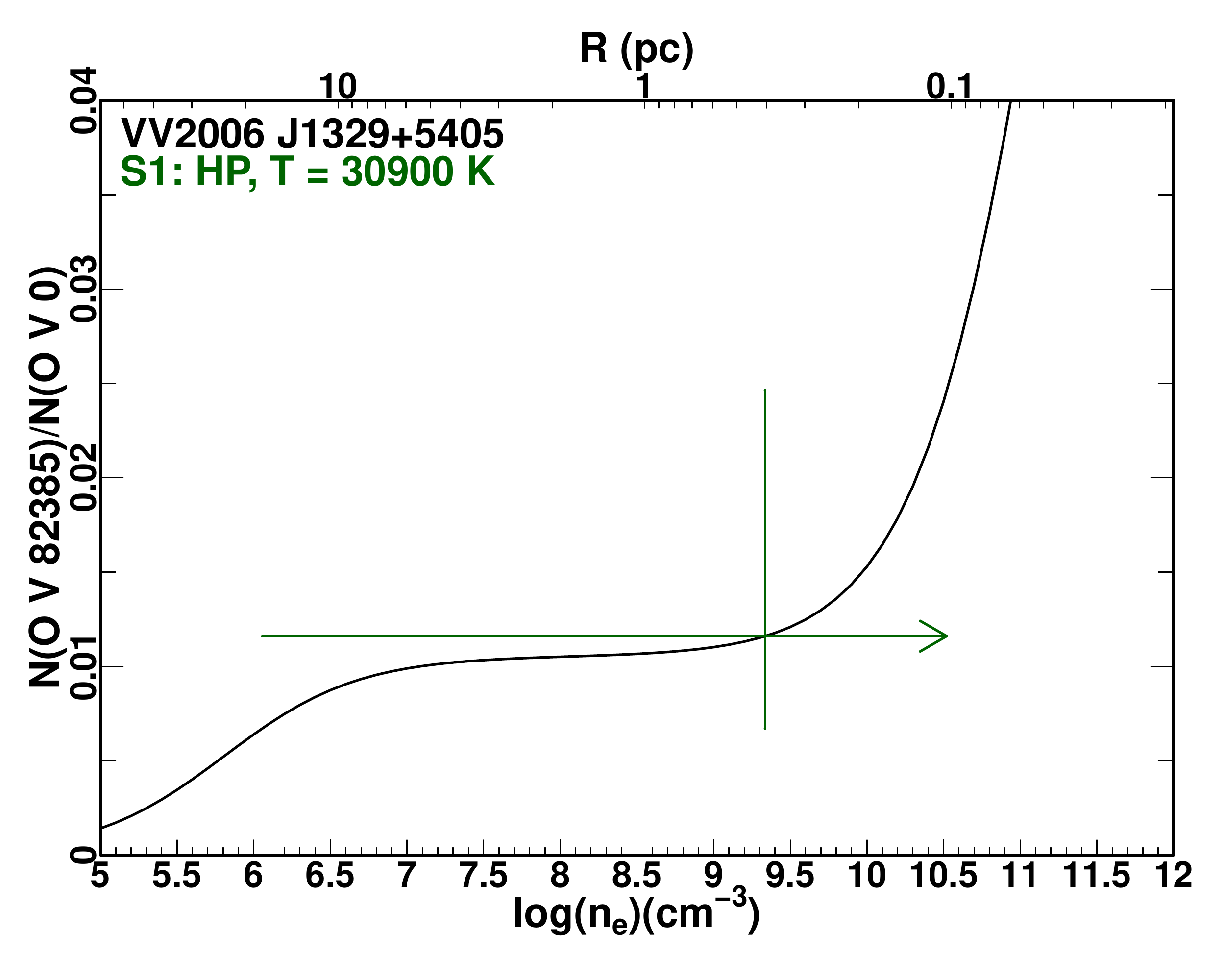}
	\caption{\footnotesize{Electron number density, \sub{n}{e}, of each outflow system based on three population ratios of Ne and one of O. The theoretical predictions from CHIANTI for the population ratios with excited energy levels of \sion{Ne}{v*} 413~cm$^{-1}$, \sion{Ne}{v*} 1111~cm$^{-1}$, \sion{Ne}{vi*} 1307~cm$^{-1}$, \sion{O}{iv*} 386~cm$^{-1}$, \sion{O}{v*} 82385~cm$^{-1}$, or \sion{N}{iv*} 67416~cm$^{-1}$ are overlaid. The curves assume the average electron temperature (T) from the photoionization solution for the corresponding phase of the first listed outflow. The distance, $R$ (from equation \ref{eq:R}), for the first labeled outflow is also shown on the top axis. The offset of the ratios from the shown curves for the other outflows are the result of the different temperatures.}}
\label{fig:dens}	
\end{figure*}
\begin{figure*}
	\includegraphics[trim=6mm 9mm 6mm 5mm,clip,scale=0.37]{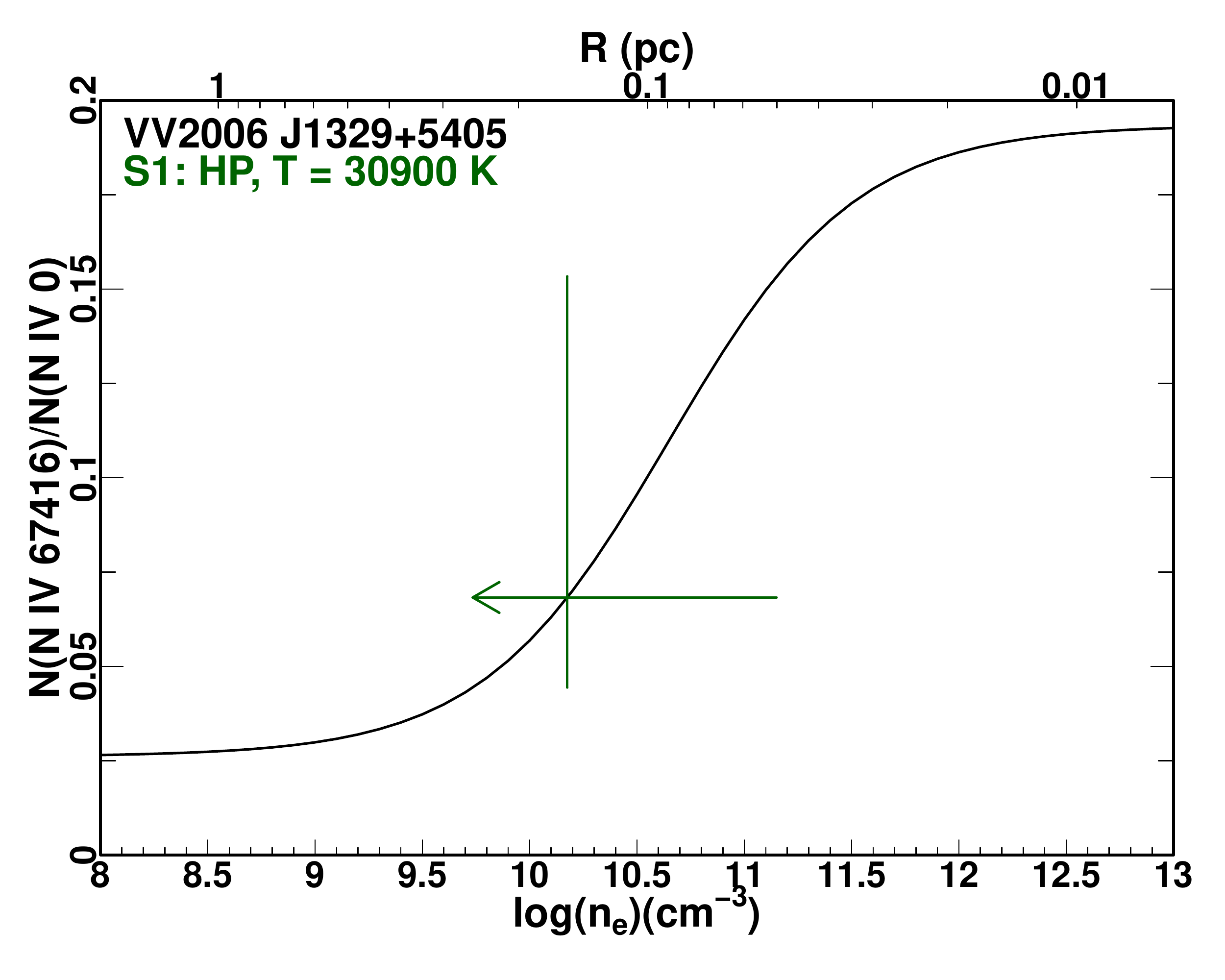}\includegraphics[trim=6mm 9mm 0mm 5mm,clip,scale=0.37]{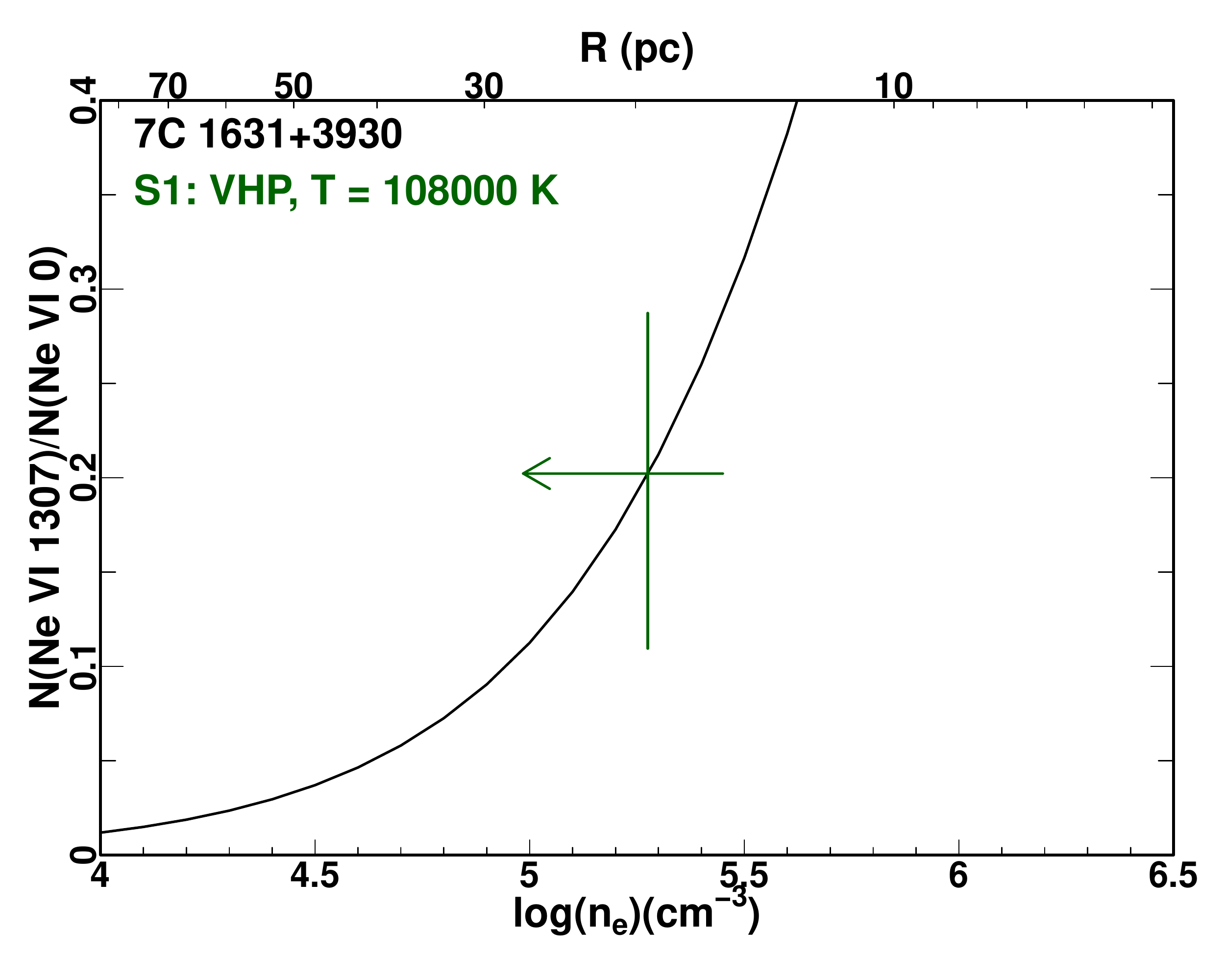}\\
	\includegraphics[trim=6mm 9mm 6mm 5mm,clip,scale=0.37]{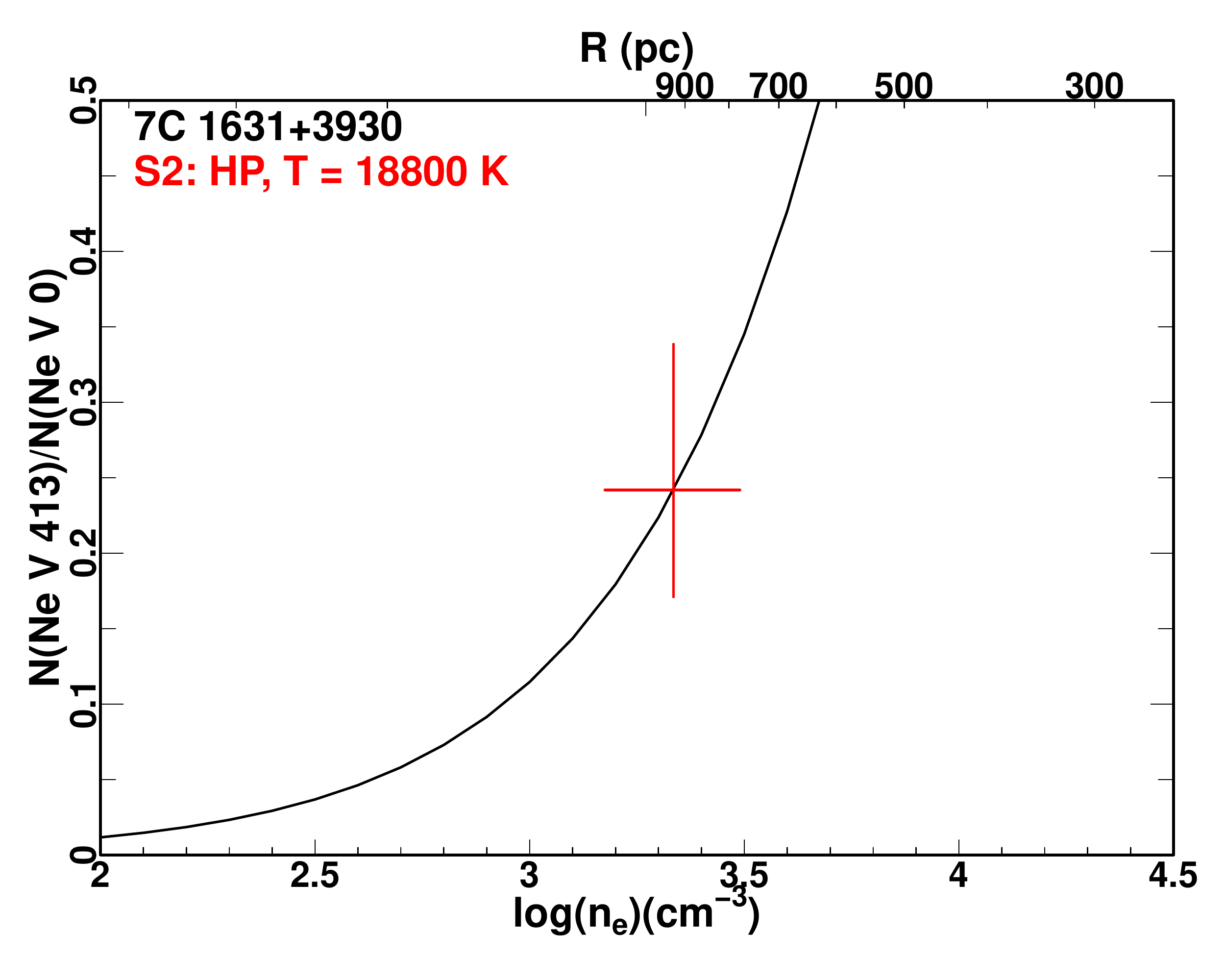}\includegraphics[trim=6mm 9mm 0mm 5mm,clip,scale=0.37]{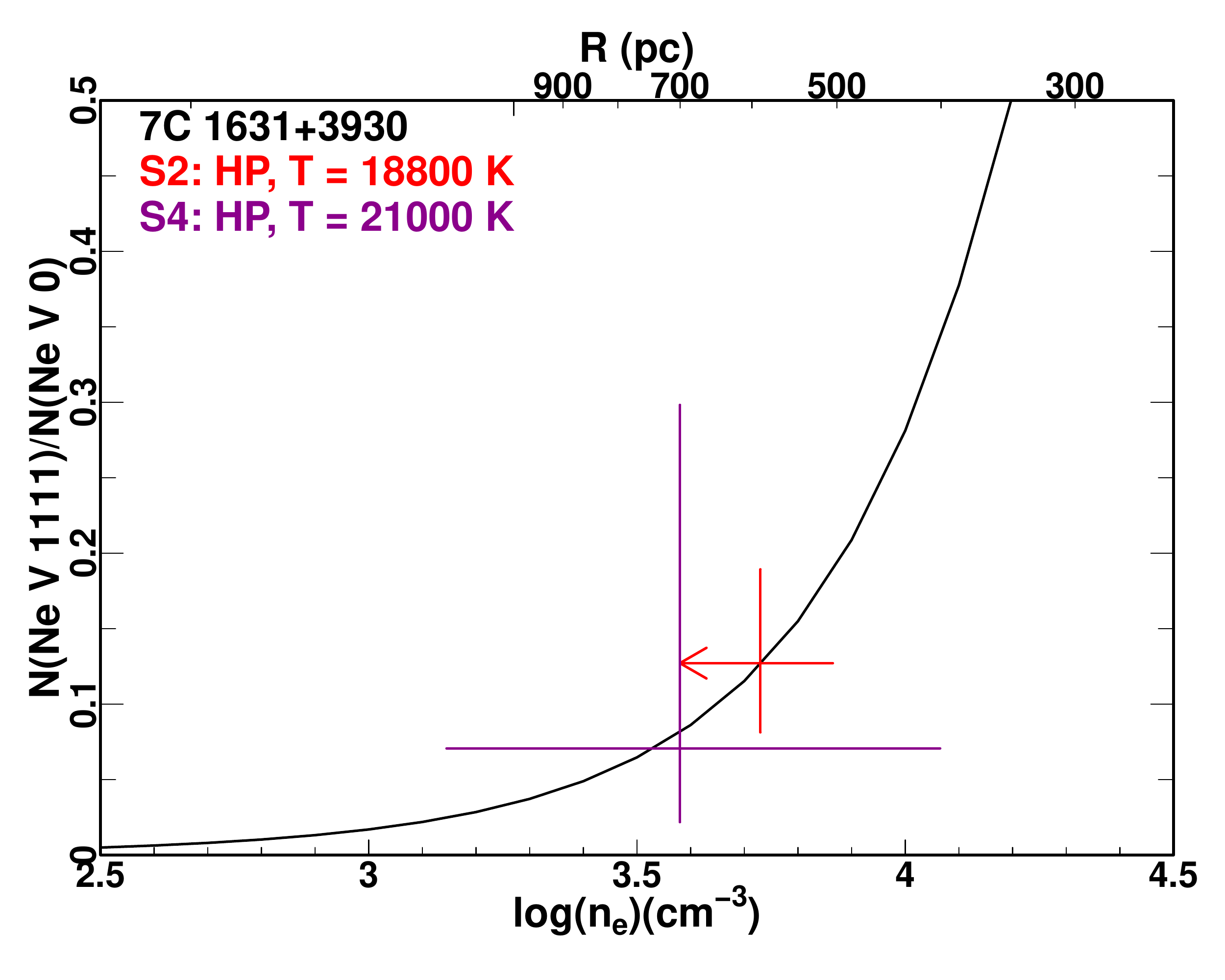}
	\figurenum{3}
	\caption{(Continued.)}
\end{figure*}
\movetabledown=167mm
\begin{rotatetable*}
	\begin{deluxetable}{c c c c c c c c c c c l}
		\tabletypesize{\small}
		\tablecaption{Physical Properties, Distances, and Energetics of Each Outflow system\label{tab:res}}
		\tablehead{	\colhead{System}   & \colhead{ $v$$^{(a)}$} & \colhead{log(\sub{U}{H,HP})} & \colhead{log(\sub{N}{H,HP})} & \colhead{log(\sub{n}{e})}& \colhead{log(\sub{U}{H,VHP})} & \colhead{log(\sub{N}{H,VHP})} & \colhead{log($\ssub{f}{V}$)$^{(b)}$}& \colhead{R} & \colhead{$\dot{M}^{(c)}$}& \colhead{log $\dot{E_{K}}^{(c)}$} & \colhead{ $\dot{E_{K}}/L_{Edd}$$^{(c)}$}	\\	\colhead{}   & \colhead{(km s$^{-1}$)} & \colhead{log}& \colhead{log(cm$^{-2}$)} & \colhead{log(cm$^{-3}$)}& \colhead{log}  & \colhead{log(cm$^{-2}$)} &\colhead{}   & \colhead{pc} & \colhead{($M_{\odot}$ yr$^{-1}$)} & \colhead{log(erg s$^{-1}$)} & \colhead{$\%$}
		}
		
		\startdata
		\multicolumn{11}{l}{\textbf{SDSS J0936+2005, L$_{\text{bol}}$ = 1.3 $\times$ 10$^{47}$ erg s$^{-1}$, L$_{\text{Edd}}$ = 1.4\pme{1.5}{0.7} $\times$ 10$^{47}$ erg s$^{-1}$, Q$_{\text{H}}$ = 7.2 $\times$ 10$^{56}$ s$^{-1}$}} \\
		\hline
		\textbf{1} &\textbf{-7960} &\textbf{-0.2$^{+0.2}_{-0.3}$} &\textbf{20.0$^{+0.4}_{-0.6}$} &\textbf{4.8$^{+0.2}_{-0.2}$} &\textbf{0.8$^{+0.3}_{-0.3}$}&\textbf{20.6$^{+0.4}_{-0.3}$} & \textbf{-1.8$^{+0.6}_{-0.9}$}
		&\textbf{77$^{+40}_{-22}$}&\textbf{17$^{+29}_{-8.5}$} &\textbf{44.5$^{+0.4}_{-0.3}$}&\textbf{0.23$^{+0.59}_{-0.15}$}\\
		\textbf{2} &\textbf{-8200} &\textbf{-0.6$^{+0.3}_{-0.1}$}&\textbf{20.8$^{+0.2}_{-0.2}$}&\textbf{4.6$^{+0.3}_{-0.2}$} &\textbf{0.5$^{+0.1}_{-0.1}$}&\textbf{21.7$^{+0.1}_{-0.2}$} & \textbf{-1.9$^{+0.3}_{-0.4}$}
		&\textbf{150$^{+50}_{-50}$}&\textbf{370$^{+280}_{-180}$} &\textbf{45.9$^{+0.2}_{-0.3}$}&\textbf{5.5$^{+8.5}_{-3.5}$}\\
		\textbf{3} &\textbf{-9300} &\textbf{-1.7$^{+0.4}_{-0.2}$}&\textbf{18.3$^{+0.2}_{-0.1}$}&\textbf{5.9$^{+0.2}_{-0.5}$} &\textbf{0.3$^{+0.02}_{-0.2}$}&\textbf{20.1$^{+0.2}_{-0.1}$} & \textbf{-3.8$^{+0.3}_{-0.5}$}
		&\textbf{14$^{+9}_{-4}$}&\textbf{1$^{+1}_{-0.5}$} &\textbf{43.4$^{+0.3}_{-0.3}$}&\textbf{0.01$^{+0.3}_{-0.005}$}\\
		\hline
		\\ 
		\multicolumn{12}{l}{\textbf{UM 425, L$_{\text{bol}}$ = 3.8 $\times$ 10$^{47}$ erg s$^{-1}$, L$_{\text{Edd}}$ = 6.0\pme{6.3}{3.2} $\times$ 10$^{47}$ erg s$^{-1}$, Q$_{\text{H}}$ = 2.1 $\times$ 10$^{57}$ s$^{-1}$}}\\
		\hline
		\textbf{1} &\textbf{-1640} &\textbf{-0.4$^{+0.2}_{-0.2}$}&\textbf{20.2$^{+0.3}_{-0.4}$}&\textbf{3.1$^{+0.2}_{-0.1}$} &\textbf{0.64$^{+0.06}_{-0.04}$}&\textbf{22.0$^{+0.1}_{-0.2}$} & \textbf{-2.8$^{+0.4}_{-0.5}$}
		&\textbf{1180$^{+430}_{-290}$}&\textbf{1050$^{+680}_{-470}$} &\textbf{44.9$^{+0.3}_{-0.2}$}&\textbf{0.1$^{+0.3}_{-0.03}$}\\
		\textbf{2} &\textbf{-1980} &\textbf{-0.6$^{+0.4}_{-0.4}$}&\textbf{19.8$^{+0.5}_{-0.4}$}&\textbf{3.7$^{+0.2}_{-0.1}$} &\textbf{$>$0.5$_{-0.2}$}&\textbf{$>$20.9$_{-0.4}$} & \textbf{$<$-2.3$^{+0.8}$}
		&\textbf{760$^{+440}_{-320}$}&\textbf{$>$81$_{-55}$} &\textbf{$>$44.0$_{-0.5}$}&\textbf{$>$0.02$_{-0.01}$}\\
		\textbf{3} &\textbf{-2200} &\textbf{-0.4$^{+0.7}_{-0.6}$}&\textbf{19.6$^{+1.4}_{-0.4}$}&\textbf{4.2$^{+0.1}_{-0.1}$} &\textbf{$>$0.7$_{-0.2}$}&\textbf{$>$21.0$_{-0.4}$} & \textbf{$<$-2.5$^{+1.6}$}
		&\textbf{340$^{+370}_{-190}$}&\textbf{$>$43$_{-31}$} &\textbf{$>$43.8$_{-0.5}$}&\textbf{$>$0.011$_{-0.007}$}\\
		\textbf{4} &\textbf{-9420} &\textbf{0.1$^{+0.2}_{-0.3}$}&\textbf{20.8$^{+0.4}_{-0.8}$}&\textbf{$>$6.1$_{-0.8}$} &\textbf{0.7$^{+0.6}_{-0.1}$}&\textbf{21.8$^{+1.0}_{-0.3}$} & \textbf{-1.6$^{+0.5}_{-1.5}$}
		&\textbf{$<$22$^{+37}$}&\textbf{$<$81$^{+850}$} &\textbf{$<$45.4$^{+1.0}$}&\textbf{$<$0.4$^{+4.4}$}\\
		\hline
		\\ 
		\multicolumn{12}{l}{\textbf{VV2006 J1329+5405, L$_{\text{bol}}$ = 8.9 $\times$ 10$^{46}$ erg s$^{-1}$, L$_{\text{Edd}}$ = 1.2\pme{1.5}{0.7} $\times$ 10$^{47}$ erg s$^{-1}$, Q$_{\text{H}}$ = 5.0 $\times$ 10$^{56}$ s$^{-1}$}}\\
		\hline
		\textbf{1} &\textbf{-11600} &\textbf{-0.3\pme{0.4}{0.1}}&\textbf{20.6\pme{0.7}{0.4}}&\textbf{9.3\me{3.2}--10.2\pe{1.0}} &\textbf{$>$0.6$_{-0.1}$}&\textbf{$>$21.8$_{-0.3}$} & \textbf{$<$-2.1$^{+0.7}$}
		&\textbf{0.15\me{0.11}--0.4\pe{17.4}}&\textbf{$>$0.6$_{-0.4}$} &\textbf{$>$43.4$_{-0.6}$}&\textbf{$>$0.022$_{-0.017}$}\\
		\textbf{2} &\textbf{-12900} &\textbf{--}&\textbf{--}&\textbf{--} &\textbf{0.1$^{+0.4}_{-0.3}$}&\textbf{19.7$^{+1.0}_{-0.1}$} & \textbf{--} &\textbf{--}&\textbf{--} &  \textbf{--}&\textbf{--}\\
		\hline
		\\ 
		\multicolumn{12}{l}{\textbf{2MASS J1436+0727, L$_{\text{bol}}$ = 8.3 $\times$ 10$^{46}$ erg s$^{-1}$, L$_{\text{Edd}}$ = 8.5\pme{8.8}{4.5} $\times$ 10$^{46}$ erg s$^{-1}$, Q$_{\text{H}}$ = 4.7 $\times$ 10$^{56}$ s$^{-1}$}}\\
		\hline
		\textbf{1} &\textbf{-14400} &\textbf{--}&\textbf{--}&\textbf{--} &\textbf{0.8$^{+0.6}_{-0.3}$}&\textbf{20.9$^{+2.0}_{-0.6}$} & \textbf{--}
		&\textbf{--}&\textbf{--} &\textbf{--}&\textbf{--}\\
		\hline
		\\ 
		\multicolumn{12}{l}{\textbf{7C 1631+3930, L$_{\text{bol}}$ = 1.2 $\times$ 10$^{47}$ erg s$^{-1}$, L$_{\text{Edd}}$ = 2.3\pme{2.6}{1.4} $\times$ 10$^{47}$ erg s$^{-1}$, Q$_{\text{H}}$ = 6.9 $\times$ 10$^{56}$ s$^{-1}$}}\\
		\hline
		\textbf{1} &\textbf{-1010} &\textbf{--}&\textbf{--}&\textbf{$<$5.3$^{+0.2}$} &\textbf{0.5\pme{0.1}{0.2}} &\textbf{20.7\pme{0.2}{0.2}} & \textbf{--}
		&\textbf{$>$19$_{-4}$}&\textbf{$>$0.5$_{-0.3}$} &\textbf{$>$41.2$_{-0.3}$}&\textbf{$>$0.00007$_{-0.00004}$}\\
		\textbf{2} &\textbf{-1430} &\textbf{-0.9\pme{0.2}{0.1}}&\textbf{19.4\pme{0.3}{0.2}}&\textbf{3.3\pme{0.2}{0.1}} &\textbf{0.6\pme{0.1}{0.2}}&\textbf{21.1\pme{0.2}{0.1}} & \textbf{-3.2$^{+0.4}_{-0.3}$}
		&\textbf{940\pme{260}{230}}&\textbf{110\pme{70}{50}} &\textbf{43.8\pme{0.3}{0.2}}&\textbf{0.03\pme{0.05}{0.02}}\\
		\textbf{3} &\textbf{-5300} &\textbf{--}&\textbf{--}&\textbf{--} &\textbf{$>$0.5$_{-0.2}$}&\textbf{$>$21.1$_{-0.3}$} & \textbf{--}	&\textbf{--}&\textbf{--} &\textbf{--}&\textbf{--}\\
		\textbf{4} &\textbf{-5770} &\textbf{-0.8\pme{0.3}{0.2}}&\textbf{20.0\pme{0.4}{0.6}}&\textbf{3.6\pme{0.5}{0.5}} &\textbf{0.2\pme{0.1}{0.02}}&\textbf{21.4\pme{0.1}{0.2}} & \textbf{-2.4$^{+0.7}_{-0.5}$}
		&\textbf{590\pme{470}{270}}&\textbf{500\pme{520}{290}} &\textbf{45.7\pme{0.3}{0.4}}&\textbf{2.3\pme{4.9}{1.4}}\\
		\textbf{5} &\textbf{-6150} &\textbf{--}&\textbf{--}&\textbf{--} &\textbf{0.5$^{+0.6}_{-0.1}$}&\textbf{21.2$^{+1.3}_{-0.2}$} & \textbf{--}
		&\textbf{--}&\textbf{--} &\textbf{--}&\textbf{--}\\
		\textbf{6} &\textbf{-7210} &\textbf{--}&\textbf{--}&\textbf{--} &\textbf{0.6$^{+1.1}_{-0.5}$}&\textbf{20.1$^{+3.4}_{-0.1}$} & \textbf{--}
		&\textbf{--}&\textbf{--} &\textbf{--}&\textbf{--}\\
		\enddata
		\tablecomments{\\
			(a). The velocity centroid of each outflow system.\\
			(b). The volume filling factor of the outflow's high-ionization phase relative to the very high-ionization phase.\\
			(c). Assuming $\Omega$ = 0.4 and where \sub{N}{H} is the sum of the two ionization phases, if present.\\
		}
	\end{deluxetable}
\end{rotatetable*}

\section{Discussion}
\label{sec:ds}
\subsection{Contribution to AGN Feedback}
The potential for AGN feedback can be assessed by using the criterion set forth by \citet[]{hop10} or \citet[]{sca04} where the kinetic luminosities must exceed 0.5\% or 5\% of the Eddington luminosity (\sub{L}{Edd}), respectively. We use the \sion{Mg}{ii}-based black hole mass equation from \citet[][]{bah19} along with their methodology for measuring the \sion{Mg}{ii} FWHM and the nearby continuum level from Sloan Digital Sky Survey (SDSS) data to estimate the mass of the super massive black hole, which is then used to calculate \sub{L}{Edd} (see Table~\ref{tab:res}; errors include systematics). Outflow S4 in 7C 1631+3930 has a kinetic luminosity above 0.5\%\sub{L}{Edd} while outflow S2 in SDSS J0936+2005 exceeds 5\%\sub{L}{Edd}. Therefore, on average, these two outflows carry enough energy, depending on the method of energy deposition into the surrounding ISM of the host galaxy, to contribute substantially to AGN feedback in galaxies with similar black hole masses as each of these quasars.
\subsection{Photoionization Solution and \sub{n}{e} Accuracy}
The majority of the photoionization solutions shown in Figure~\ref{fig:sol} are constrained only by \sub{N}{ion} upper and lower limits. As such, they are immune to saturation effects. Measurements were only taken for doublet transitions where a partial covering solution could be reliably determined or for single transitions where we could infer that saturation effects would be minimal. The multitude of upper and lower limits along with the few measurements yielded only four out of the 16 outflows with unbounded VHPs. Half of the bounded VHPs have tightly constrained errors to within a factor of two. The photoionization solutions of either the HP (two phase solutions) or VHP (single phase solutions) constrained the total \sub{N}{ion} for \sion{Ne}{v}, \sion{Ne}{vi}, \sion{O}{iv}, \sion{O}{v}, and \sion{N}{iv}, which were used in conjunction with the excited- and ground-state \sub{N}{ion} measured from the data to calculate the population ratios that yielded \sub{n}{e} for each outflow (see sections~\ref{sec:ed}~and~\ref{sec:indobj}). When multiple diagnostics were available for a given outflow, the \sub{n}{e} values were all consistent within errors. This consistency in \sub{n}{e} between multiple diagnostics that arose from constraints from the photoionization solutions shows the results are robust and reliable.   
\subsection{Geometry and Volume Filling Factor}
Comparing outflows within a given quasar, there are two that show a similarity in their respective geometries: S1 and S2 in SDSS J0936+2005. The kinematic similarities of the troughs from the VHP and HP for each outflow suggest that the two phases occupy the same volume. Under the assumption that the VHP occupies the total volume where the outflow resides \cite[i.e., the volume filling factor = 1;][]{ara13}, and since the HP is both denser and has a lower \sub{N}{H} than that of the VHP, the HP would have a small volume filling factor within the VHP. This volume filling factor is given by equation 4 in Section 3.5 of Paper I:   
\begin{equation}
\ssub{f}{V} = \frac{\sub{U}{H,\tiny\textit{HP}}}{\sub{U}{H,\tiny\textit{VHP}}}\times\frac{\sub{N}{H,\tiny\textit{HP}}}{\sub{N}{H,\tiny\textit{VHP}}}
\end{equation}
S1 and S2 in SDSS J0936+2005 have \ssub{f}{V} values that are fully consistent, considering the errors (see Table~\ref{tab:res}). This similarity in geometry, the consistent distances, and small velocity separation (240~km~s$^{-1}$) suggest the two outflows have a common origin. These similarities in geometry, distance, and velocity were also observed for outflows 1a and 1b in Paper II as well as the four outflows in Paper III, implying that it is more than a coincidence. 
\subsection{Comparison to X-Ray Warm Absorbers}
As has been discussed in Papers II, III, and V, X-ray warm absorbers span a wide range of \sub{U}{H}, which can be explained by a continuous \sub{N}{H} absorber as a function of \sub{U}{H}. The single phase and two phase solutions found here are sufficient to explain the data, but we cannot rule out additional ionization phases that are not detected due to the wavelength range and/or data sensitivity limitations. Also, the \sub{U}{H} values in Table~\ref{tab:res} are comparable to the quantity log($\xi$) that is commonly used in X-ray analyses [log($\xi$) $\approxeq$ log(\sub{U}{H})+1.3]. Therefore, the outflows detected here are similar to X-ray warm absorbers.
\subsection{New Absorption Trough Transitions/Ions}
To the best of our knowledge, the detection of \sion{Cl}{vii} 800.64~\AA\ in S1 of VV2006 J1329+5405 is a first in both ion and transition for the astronomical community. Similarly, the \sion{Ne}{v} 480.42~\AA\ and \sion{Ne}{v*} 482.99~\AA\ line transitions in S4 of UM 425 are also first time detections.
\section{Summary and Conclusions}
\label{sec:sc}
This paper presented new HST/COS spectra for five quasars containing 16 outflows. 
Absorption troughs from up to 14 ions yielded column density measurements and lower limits for each outflow. The use of these absorption trough constraints along with additional upper limits and a grid of photoionization models enabled us to determine the best-fitting values of \sub{U}{H} and \sub{N}{H} for each outflow. 

Column density ratios from excited- and ground-state transitions of \sion{Ne}{v}, \sion{Ne}{vi}, \sion{O}{iv}, \sion{O}{v}, and \sion{N}{iv} constrained the electron number densities of 11 outflows. Using those values, the distances of each outflow to the central source were calculated from equation (\ref{eq:R}), leading to the mass flux and kinetic luminosity (equations \ref{eq:M}~and~\ref{eq:E}). Lastly, the potential for AGN feedback was investigated with the final results summarized in Table~\ref{tab:res}.

The main results are as follows:
\begin{enumerate}
\item Of the 16 outflows, 10 require a two phase photoionization solution to simultaneously satisfy the column density measurements from ions spanning over an order of magnitude in ionization potential. Each of the other six contain a VHP.
\item Distances and energetics were determined for 11 outflows with seven having more than one distance indicator, and all of these were consistent within errors. Outflows S2 in SDSS J0936+2005 and S4 in 7C 1631+3930 each have a large enough kinetic luminosity to Eddington luminosity ratio to be major contributors to AGN feedback processes, depending on the energy deposition process. If the electron number density of outflow S1 in VV2006 J1329+5405 is closer to its lower limit, then it too would have enough kinetic energy to be a major contributor to AGN feedback.
\item The outflows S1 and S2 in SDSS J0936+2005 likely originate from the same material at the same distance since the velocity separation between the outflows is small and since the distances and volume filling factors are consistent within the errors.
\item The ion and line transition \sion{Cl}{vii} 800.64~\AA\ as well as the line transitions \sion{Ne}{v} 480.42~\AA\ and \sion{Ne}{v*} 482.99~\AA\ are first time detections.
\end{enumerate}



\acknowledgments

T.M., N.A., and X.X. acknowledge support from NASA  grants HST GO-14777, -14242, -14054, and -14176 as well as HST AR 15786. This support is provided by NASA through a grant from the Space Telescope Science Institute, which is operated by the Association of Universities for Research in Astronomy, Incorporated, under NASA contract NAS5-26555. T.M. and N.A. also acknowledge support from NASA ADAP 48020 and NSF grant AST 1413319. CHIANTI is a collaborative project involving George Mason University (USA), the University of Michigan (USA), and the University of Cambridge (UK).



\end{document}